\documentclass{aa}  

\usepackage{amsmath}
\usepackage{placeins}           

\usepackage{longtable}
\usepackage{graphicx}
\usepackage{txfonts}
\usepackage{afterpage}
\usepackage{amssymb}
\usepackage{adjustbox}

\newcommand{\mh}{$\rm M_{H_2}$}
\newcommand{\mhi}{$\rm M_{HI}$}
\newcommand{\mhiop}{$\rm M_{HI,~R25}$}
\newcommand{\mhitot}{$\rm M_{HI,~tot}$}

\newcommand{\mdust}{$\rm M_{dust}$}
\newcommand{\mstar}{$\rm M_{\star}$}
\newcommand{\msun}{$\rm M_{\odot}$}
\newcommand{\hh}{$\rm H_2$}
\newcommand{\aco}{$\alpha_{\rm CO}$}
\newcommand{\uaco}{$\rm M_{\odot}~pc^{-2}~(K~km~s^{-1})^{-1}$}
\newcommand{\lco}{$\rm L^{\prime}_{CO(1-0)}$}

\newcommand{\as}{$^{\prime\prime}$}
\newcommand{\am}{$^{\prime}$~}
\newcommand{\micron}{$\mu$m~}

\begin{document}

   \title{Exploring the interplay of dust and gas phases in DustPedia star-forming galaxies}

   \subtitle{}

   \author{Francesco Salvestrini
          \inst{1,2}
          \and
          Simone Bianchi\inst{3}
          \and
          Edvige Corbelli\inst{3}
          }

   \institute{
         INAF - Osservatorio Astronomico di Trieste, Via G. Tiepolo 11, 34143 Trieste, Italy
         \email{francesco.salvestrini@inaf.it}
         \and
         IFPU - Institute for Fundamental Physics of the Universe, via Beirut 2, 34151 Trieste, Italy
         \and
         INAF - Osservatorio Astrofisico di Arcetri, Largo E. Fermi 5, 50125 Firenze, Italy 
}
   \date{}

 
  \abstract
{Molecular gas is the key ingredient of the star formation cycle, and tracing its dependencies on other galaxy properties is essential for understanding galaxy evolution.
In this work, we explore the relation between the different phases of the interstellar medium (ISM), namely molecular gas, atomic gas, and dust, and galaxy properties using a sample of nearby late-type galaxies. 

To this goal, we collect CO maps covering
at least 70\% of the optical extent
for 121 galaxies from the DustPedia project, ensuring an accurate determination of \mh, the global molecular gas mass. We investigate
which scaling relations provide the best description of \mh, based on the strength of the correlation and its intrinsic dispersion.

We found that the commonly used correlations between \mh and star formation rate (SFR) and stellar mass (\mstar), respectively, are affected by large scatter, which accounts for galaxies that are experiencing quenching of their star formation activity.
This issue can be partially mitigated by considering a ``fundamental plane'' of star formation, fitting together \mh, \mstar, and SFR.
We confirm previous results from the DustPedia collaboration that the total gas mass has the tightest connection with the dust mass and that the molecular component also establishes a good correlation with dust, once map-based \mh\ estimates are used. Although dust grains are necessary for the formation of hydrogen molecules, the strength of gravitational potential driven by the stellar component plays a key role in driving density enhancements and the atomic-to-molecular phase transition.
By investigating the correlations between the various components of the ISM and monochromatic luminosities at different wavelengths, we propose mid and far-IR luminosities as reliable proxies of \lco\ for those sources lacking dedicated millimeter observations. Luminosities in mid-IR photometric bands collecting PAH emission can be used to trace molecular gas and dust masses.}


   \maketitle
%

\section{Introduction}

Scaling relations are a powerful tool to investigate complex systems like galaxies. 
The study of the correlations between global galaxy properties (e.g., masses, luminosities), as well as between its different components (e.g., stars, ISM gas phases), helps to understand the physical processes that regulate the formation and evolution of galaxies.
Indeed, inferring the internal - or environmental-conditions that favor the formation of new stars in galaxies is not an easy task, due to the complexity of the process.
However, large observational campaigns offer the opportunity to investigate the cycle of star formation in an increasingly larger sample of galaxies, which benefit from a multi-wavelength characterization.
The number of studies dedicated to scaling relations of the interstellar medium (ISM) components is growing (e.g., \citealt{Saintonge11a, Saintonge11b, Corbelli12, Cortese12, Boselli14, Lin19, Lisenfeld19, Sorai19}). These studies have quickly become references and offer constraints for cosmological models of galaxy evolution, which are able to trace the evolution of the different gas phases.
In particular, recent works both at galactic and kpc scales investigated the connection between the molecular gas, which is a key ingredient to form new stars, and the different ISM components, as well as different galaxy properties (e.g., \citealt{Casasola17, Lin19}; \citealt{Ellison20}; \citealt{Baker23a}).
By investigating scaling relations, we can understand which relations are fundamental - i.e., they represent the physical processes that drive the cycle of star formation - and which are by-products.
Scaling relations between different baryonic components can also be used to predict the properties of a galaxy in terms of observed quantities.
The better a scaling relation is  (i.e., with the lower dispersion), the better a certain quantity can be predicted.
This is important because it allows us to retrieve quantities that require time-consuming or unavailable observations.\\
\citet[herafter, C20]{Casasola20} investigated the scaling relations between the molecular gas mass and several galaxy properties for 256 late-type galaxies, drawn from the 875 large and nearby objects included in the DustPedia sample \citep{Davies17,Clark18}. 
Molecular mass estimates were obtained from low-J CO spectroscopic measurements in the literature. However, C20 did not have access to CO observations covering a significant fraction of the galactic disk; instead, in the vast majority of cases, only single pointings of galactic centers were available. Thus, the global CO fluxes (and molecular masses) could only be extrapolated from those measurements, adopting an exponential distribution for the galactic disk and following the methods of \cite{Lisenfeld11} and \cite{Boselli14}. 
Since the work of C20, the results of a few mapping surveys have been published, such as the CO multi-line imaging of nearby galaxies (COMING) survey \citep{Sorai19} and a CO mapping survey of galaxies in the Fornax cluster \citep{MorokumaMatsui22}. 
The availability of these new data for a significant number of DustPedia galaxies allows us to improve the study of scaling relations and to reduce the scatter that is inevitably introduced by general extrapolation techniques.
In the present work, we used these recent databases to estimate the molecular gas mass for DustPedia galaxies, and we expand the investigation on the scaling relations for molecular gas, with two goals: \emph{i)} determine which physical process favors the formation of molecular hydrogen, \emph{ii)} determine which are the best proxies of the molecular gas content in galaxies. 

The paper is organized as follows. In section~\ref{sec:sample_data}, we describe the data collected in this work; in section \ref{sec:methods}, we describe the statistical methods adopted in the analysis and the improvements in mass estimates due to the use of CO maps. In section \ref{sec:results}, we analyze scaling relations involving the molecular gas content and the star formation cycle in our sample. In section \ref{sec:discussion}, we discuss the physical processes that regulate the cycle of star formation and the balance between different phases of the ISM. 
In section~\ref{sec:photometry}, we present an alternative approach to derive the ISM content of galaxies from monochromatic luminosities.
Finally, in section~\ref{sec:conclusions}, we summarize the results of this work.
Throughout the paper, we refer to the $^{12}$C$^{16}$O molecule simply as CO.
Errors are given at a 68\% confidence level.


\section{Sample and data}
\label{sec:sample_data}
\subsection{Sample selection}
\label{sec:sample}
The present sample was drawn from the DustPedia catalog, which includes the largest ($D_{25}>1'$, $D_{25}$ being the major axis isophote at which optical surface brightness falls beneath 25 mag arcsec$^{-2}$) and closest ($d<30~Mpc$) galaxies observed by \emph{Herschel}, for a total of 875 objects \citep{Davies17}.
We focused on late-type galaxies from DustPedia with available observations mapping low-J CO emission lines.
The morphology indicator (or Hubble stage, HT) was obtained from the Hyper-LEDA database \citep{Makarov14}. 
Following C20, we used all objects with Hubble stage in the range $-0.5<HT<10$, thus including all galaxies classified as Sa or later types.
From this point forward, $HT\geq0$ shall denote $HT>-0.5$.
Archival CO observations were collected upon two main criteria: \emph{i)} CO maps covering at least 70$\%$ of the optical extent of the galaxy;  \emph{ii)} Observations must not suffer from short-spacing due to interferometry.
Regarding \emph{i)}, the molecular gas is thought to mostly reside in the central region of the galaxy (e.g., \citealt{Nishiyama01, Kuno07}), following a radial decreasing profile with a scale length which is a fraction of the optical size (e.g., measured in the V band).
In particular, several works from the literature mapping low-J CO emission lines in nearby galaxies found evidence for an exponentially decreasing trend for molecular gas concentration at increasing distances from the center (e.g., \citealt{Lisenfeld11, Boselli14, Casasola17}), with a typical CO scale-radius proportional to the optical one, i.e. $R_{CO}\sim0.2\times R_{D25}$.
\begin{figure}[ht]
        \centering
	\includegraphics[width = 0.5\textwidth, keepaspectratio=True]{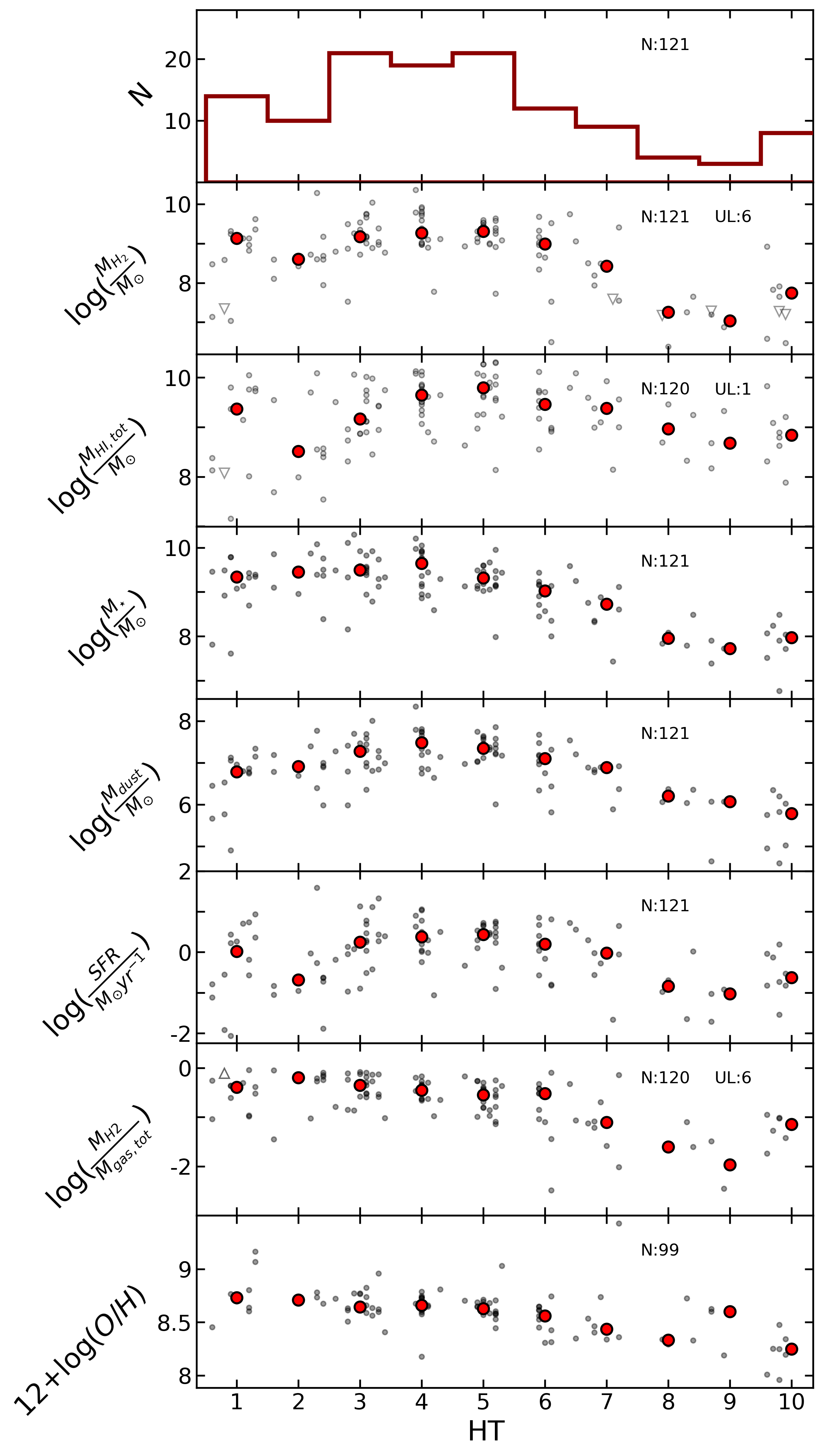}
	\caption{Overview of the properties of our sample as a function of the Hubble stage parameter ($HT$).
	From top to bottom: histogram of the number of galaxies in each bin ($\Delta T=1$); the distribution of the masses of different components, molecular (\mh) and atomic (\mhi) gas, the dust (\mdust) and stars (\mstar), respectively; SFR, molecular-to-total gas fraction, and metallicity.
	Data are represented as gray points, the mean value in each bin of $HT$ is a red circle.
    In each panel, we indicated the number of galaxies with measurements (and upper limits) for each quantity.
	This figure is similar to Fig.~1 by C20.}
	\label{fig:sample_bin}
\end{figure}
Our selection criteria based on collecting low-J CO emission line intensities over the large part of the optical disk ($>70$\% of $R_{25}$, where $R_{25}=D_{25}/2$) ensured us to gather $\gtrsim 90\%$ of the expected total flux based on models (e.g., \citealt{Boselli14}).
Interferometric observations may suffer from short-spacing issues (criterion \emph{ii)}), i.e., the filtering out of the emission from large-scale structures.
For this reason, we excluded archival interferometric observations resolving out emission at scales $\lesssim 2\times R_{CO}$.
Indeed, most of the CO maps considered in this work were obtained with On-The-Fly (OTF) mapping designed to collect the flux from the entire galaxies, satisfying the Nyquist theorem (see next section for further details on the datasets).\\
Given the selection criteria mentioned above, we cross-matched the DustPedia catalog with the following literature works: \cite{Sorai19}; \cite{Brown21}; \cite{Groves15}; \cite{MorokumaMatsui22}; \cite{Chung17}; \cite{Corbelli12}; \cite{Curran01}. 
Literature works are listed in order of the number of molecular gas mass estimates derived by the corresponding paper; details are reported in Table \ref{table:surveys}.
For the targets that appear in multiple literature works, first, we verified the consistency of the measurements, ultimately discovering a consensus within the uncertainties. 
Then, we favored the \mh\ estimate from the catalog that includes a larger number of sources from the DustPedia parent sample, allowing us to build a sample as homogenous as possible.
Furthermore, we also favored the surveys targeting the CO(1--0) transition to mitigate the uncertainties associated with the assumption of a (2--1)/(1--0) emission line ratio correction.
\begin{table*}[ht]
\centering          
\begin{tabular}[0.5\textwidth]{l l l l l l }     
\hline     
    Reference               &   Number  & Non-det. &   Survey name &   CO line &   Mapping techno. \\ 
    (1)                     &   (2)     & (3)      &   (4)         &   (5)     &  (6)\\ 
\hline                  
    \cite{Sorai19}          &   55      & -        &  COMING      &    (1--0) &   OTF \\
    \cite{Brown21}          &   15      & -        &   VERTICO     &    (2--1) &   INT \\
    \cite{Groves15}         &   14      & -        &   HERACLES    &    (2--1) &   OTF \\
    \cite{MorokumaMatsui22} &   13      & 6        &               &    (1--0) &   INT \\
    \cite{Chung17}          &   10      & -        &               &    (1--0) &   OTF \\
    \cite{Corbelli12}       &   9       & -        &   HeViCS      &    (1--0) &   OTF \\
    \cite{Curran01}         &   5       & -        &               &    (1--0) &   OTF \\
\hline                 
    Total                   &   121     & 6        &    \\
\hline                  
\end{tabular}
\caption{Summary of properties and statistics of the data collected from literature works. Columns: (1) Reference; (2) Number of objects for each survey or collection; (3) Number of CO non-detections; (4) Survey name; (5) CO emission line transition considered (see the main text for details; (6) Mapping technique, on-the-fly (OTF) mode with single-dish antenna or mosaic with interferometric (INT) facilities.}
\label{table:surveys}      
\end{table*}

\subsection{Data collection}
\label{sec:co_maps}
Here, we briefly described the properties of the data collected from the literature works reported in Table \ref{table:surveys}.
All the \mh\ measurements collected from literature works were homogenized as follows: i) the original distance was corrected to the one estimated and provided by \cite{Clark18}; ii) a Milky-Way like CO-to-H$_2$ conversion factor ($\alpha_{CO}=3.26$ \uaco) was adopted (as in C20, we neglected the contribution to the gas mass due to elements heavier than molecular hydrogen - mainly helium).
iii) In case the \mh\ estimates in the literature are derived from CO(2--1) emission line intensity maps, we corrected the value for the adoption of a CO(2--1)/CO(1--0) emission line ratio of 0.7, as suggested for nearby galaxies by \cite{denBrok21}.\\
Regarding ii), here we adopted a constant \aco~to maximize the sample size since metallicity measurements are not available for all DustPedia sources. 
In section~\ref{app:gas_mass_Z}, we discussed the impact of this assumption compared to the adoption of a metallicity-dependent \aco.\\
\cite{Sorai19} presents the CO Multiline Imaging of Nearby Galaxies (COMING) survey, targeting molecular lines, including CO(1--0) of 147 nearby spiral galaxies with the 45-meter Nobeyama telescope.
Observations were carried out using the OTF technique, covering the molecular gas emission over 70$\%$ of D25, producing maps with an angular resolution of $\sim$14\as.
In the end, we included in our sample 55 objects from the COMING survey.\\
\cite{Brown21} presented the Virgo Environment Traced in CO (VERTICO) survey, which includes observations of the CO(2--1) for 51 Virgo Cluster galaxies with the ALMA 7-meter array\footnote{The 7m antenna array at the Atacama site is also known as Atacama Compact Array (ACA).} and TP.
The 15 objects collected from \cite{Brown21} have been sampled with an angular resolution of $\sim7-9$\as, and the short-spacing issue can be neglected.
Indeed, the largest-angular-scales recoverable with the observing set-up used by \cite{Brown21} are larger than $1^{\prime}$, i.e. at least as large as the expected $R_{CO}$.
Nevertheless, we investigated the potential flux loss due to the partial short-spacing, by comparing the integrated flux obtained by \cite{Brown21} with other estimates from the literature: we found no evidence for significant flux loss (i.e., larger than the CO integrated flux uncertainty).\\
\cite{Groves15} presents the analysis of the molecular gas properties in the Heterodyne Receiver Array CO Line Extragalactic Survey (HERACLES; \citealt{Leroy09}), carried out at the IRAM telescope and collecting CO(2--1) emission line maps.
Data sets have a native FWHM beam size of 13.3\as\ and large extent, usually covering out beyond the optical radius of the galaxy.
Integrated CO luminosity is extracted over large apertures, which cover up to $\sim$2 optical radii (see also \citealt{Dale12} for further details).
By crossmatching the HERACLES sample with DustPedia galaxies, we retrieved 14 \mh\ measurements.\\
\cite{MorokumaMatsui22} analyzed the ALMA data for 64 galaxies either included in the ALMA Fornax Cluster Survey (AlFoCS) project \citep{Zabel19} or in the vicinity of the Fornax cluster with archival HI data in the HyperLEDA (Makarov et al. 2014).
\cite{MorokumaMatsui22} collected CO(1--0) maps by combining ALMA 12m, 7m, and Total Power (TP) observations, covering the entire optical disk of galaxies, with an average beam of roughly $15^{\prime\prime}\times 8^{\prime\prime}$.
We retrieved the molecular gas mass for 13 sources that are in common with the DustPedia project, among which 6 are upper limits.\\
\cite{Chung17} presented the OTF mapping of CO(1--0) of 28 spiral galaxies from the Virgo cluster with the Five College Radio Astronomy Observatory (FCRAO) 14 m telescope.
The authors performed an OTF mapping of each galaxy with the SEcond QUabbin Optical Imaging Array (SEQUOIA) focal plane array, which consists of 16 horns (4$\times$4 configuration), each with a 45\as\ beam.
The maps cover about 10\am$\times$10\am, way larger than the optical diameter of the galaxies. 
\cite{Chung17} derived the total flux by summing all emission features in the channel maps; 
We retrieved the CO flux for 10 galaxies from \cite{Chung17}, and we then converted the CO flux into \lco\ by using equation (3) from \cite{SolomonVandenBout05}.\\
The Herschel Virgo Cluster Survey (HeViCS) is a magnitude-limited survey of galaxies in the Virgo Cluster.
Out of the 35 galaxies with CO(1--0) detection listed in \cite{Corbelli12}, we retained 9 objects, those that have their optical disk mapped with both new and archival mm observations, at resolutions between 15" and 45" (for further details, see \citealt{Corbelli12,Pappalardo12} ).\\
\cite{Curran01} performed the mapping of the CO(1--0) emission line from 5 nearby Seyfert galaxies with the 15 m antenna of the Swedish-ESO Sub-millimetre Telescope (SEST).
The authors adopted the position-switching method to complete the mapping of the entire galaxy.
We collected the \lco\ measurements at 45" resolution for 5 DustPedia galaxies.\\
In total, we collected global \mh\ estimates for the 121 DustPedia galaxies, including 6 upper limits, and are reported in Table~\ref{table:sample}.
%
%
\subsection{Ancillary data}
\label{sec:ancillary_data}
\cite{Nersesian19} performed the broadband spectral energy distribution (SED) fitting of the entire sample of DustPedia galaxies using the Code Investigating GALaxy Evolution (CIGALE\footnote{\url{https://cigale.lam.fr/}}; \citealt{Boquien19}).
For our purposes, we collected stellar ad dust masses (\mstar~and \mdust, respectively), and SFR estimates resulting from the SED fitting procedure obtained with The Heterogeneous Evolution Model for Interstellar Solids (THEMIS) dust model (\citealt{Jones13, Kohler14}).
We referred to \cite{Nersesian19} for further details on the photometric catalog, libraries, and parameter space included in the SED fitting procedure, as well as the discussion of results.\\
C20 collected the atomic gas mass \mhi~for DustPedia galaxies from the literature.
Since the atomic gas has been observed to extend up to several kpc from the center of a galaxy, its part co-spatial with the molecular component could be only a fraction of the total \mhi.
For this reason, C20 corrected the total mass estimates for the aperture by assuming the model by \cite{Wang20}, thus deriving \mhi~within the optical radius (R25; \mhiop~hereafter), that matches the aperture of the SED fitting analysis by \cite{Nersesian19}.
Atomic gas mass \mhi~from C20 is available for 120 out of 121 galaxies of the sample presented here (NGC 1336 has no \mhi~estimate and only a \mh\ upper limit).
For comparison purposes (see appendix~\ref{sec:motivation}), we also used the original C20 \mh\ estimates, which are available for 103 out of 121 galaxies.
As for distances, we collected other galaxy properties (e.g., optical diameter, D25; optical centers; inclinations) from \cite{Clark18}.
A synthetic presentation of the properties of the sample as a function of HT is shown in Fig.~\ref{fig:sample_bin}, namely, from top to bottom, the histogram of number of source for each HT bin, molecular and total atomic gas masses, stellar and dust masses, SFR, molecular-to-total gas fraction, and oxygen abundance.
In Appendix~\ref{app:data} and Table~\ref{table:sample}, we reported the basic quantity, such as \mstar, SFR, \mh, \mhi, \mdust. 
All the ancillary data we collected are retrievable from the DustPedia online repositories \footnote{\url{https://cdsarc.cds.unistra.fr/viz-bin/cat/J/A+A/609/A37}, \url{http://dustpedia.astro.noa.gr/}}.

\section{Data analysis}
\label{sec:methods}

\subsection{Methods}

In the following sections, scaling relations between two quantities were modeled using the Python package {\it inmix}\footnote{\url{https://github.com/jmeyers314/linmix}} \citep{Kelly07}, a linear regression method based on a Bayesian approach. 
{\it linmix} allowed us to consider errors on both dependent and independent variables and upper limits on the latter.
Furthermore, the code includes a free parameter to account for the intrinsic dispersion about the regression ($\delta_{intr}$, i.e., not related to measurement uncertainties), which has zero means and standard deviation $\delta_{intr}$.\\
In the case of scaling relations involving three quantities, we could not rely on {\it linmix}. 
In this case, best-fit parameters were obtained by minimizing a log-likelihood function with the {\it emcee} package, a pure-Python implementation of Goodman \& Weare's affine invariant Markov chain Monte Carlo (MCMC) ensemble sampler (\citealt{emcee}).
Then, we adopted the following model for the scaling relation:
\begin{equation}
\log(Z) = \alpha \times \log(X) + \beta \times \log(Y) + \gamma.
\label{eq:3d}
\end{equation}
In the log-likelihood function, we also included a free parameter to account for the intrinsic dispersion of the value ($\delta_{intr}$, i.e., not related to measurement uncertainties), which was summed in quadrature to the uncertainties on both X and Y quantities considered in each relation.\\
In addition, we used a set of correlation coefficients (CCs, namely Pearson's CC, and Kendall rank CC) to statistically compare different scaling relations.
Pearson's CC ($P$, hereafter) measures the linear correlation between two sets of data. It is the ratio between the covariance of two variables and the product of their standard deviations.
Kendall's rank CC ($K$, hereafter) is a non-parametric hypothesis test that measures the rank correlation between two variables, i.e., if the two datasets show similar orderings when ranked by each of the quantities. It is less affected by outliers with respect to other rank coefficients, such as the Spearman rank CC.
These two CCs have values ranging between -1 and 1.\\
Upper limits were excluded in the fitting procedure of scaling relations involving three quantities (Eq.~\ref{eq:3d}) or the calculation of CCs.

\subsection{Radial profiles of CO line brightness}
\label{sec:rad_profile}
\begin{figure}[t]
        \centering
	\includegraphics[width = \columnwidth, keepaspectratio=True]{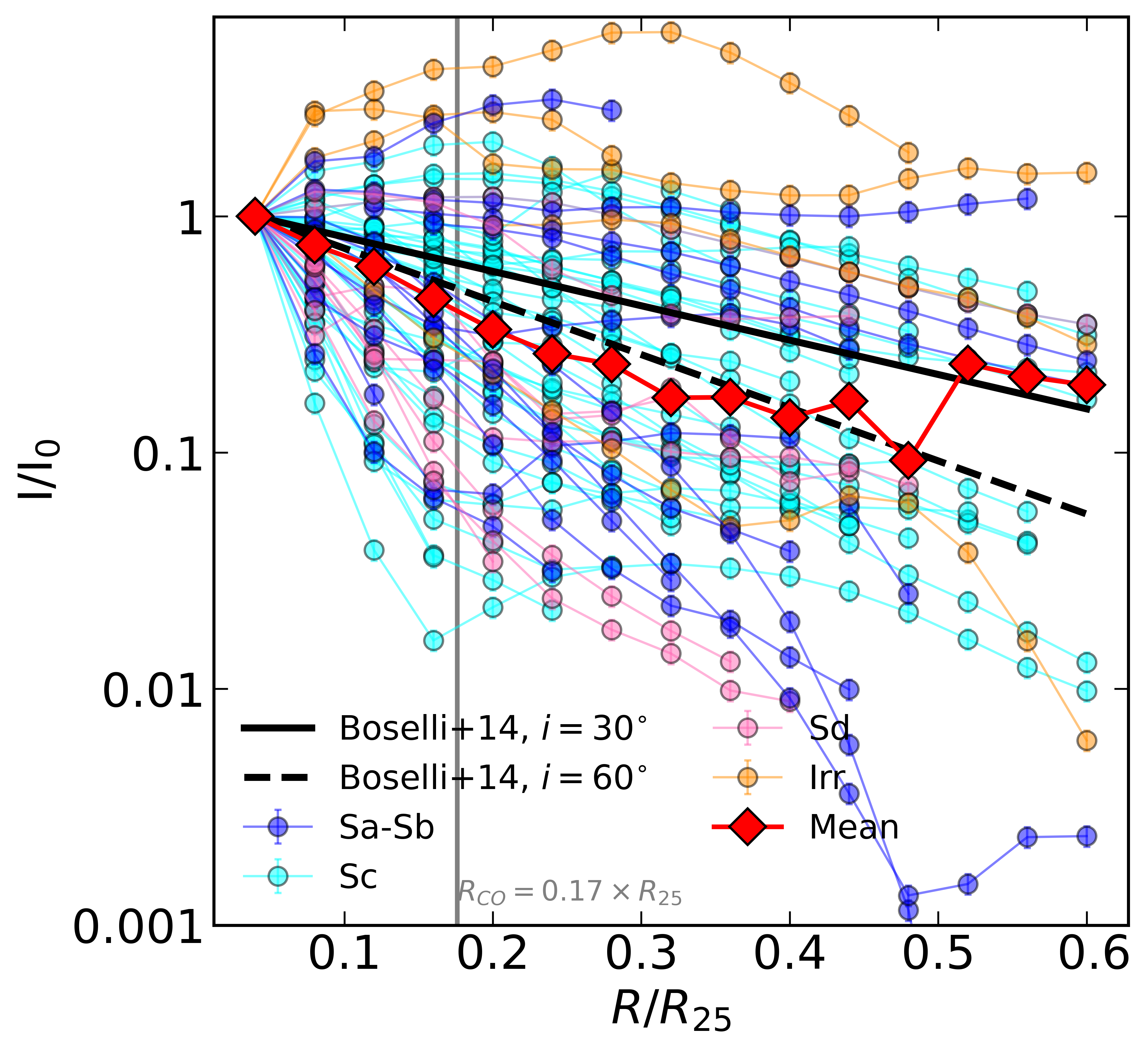}
	\caption{CO emission radial profile, as a function of the optical radius ($R_{25}$). The CO scale radius ($R_{25}$) is reported as a vertical gray line. 
	Objects are color-coded on the basis of morphological classification.
	The median radial profile at each radius is represented with red squares.
	The radial profile predicted by the exponentially decreasing disk by \cite{Boselli14} is shown as solid and dashed lines, when inclination angles of 30$^{\circ}$ and 60$^{\circ}$ are assumed, respectively.}
	\label{fig:radial_profile}
\end{figure}

To accurately study the scaling relations between molecular gas and other galaxy properties, it is necessary to map the gas across the entire galactic disk. Observational evidence suggests that molecular gas is mainly concentrated in the central regions of galaxies, with an exponentially radial decreasing distribution with a scale radius that is a fraction of the optical scale radius \citep[e.g., ][]{Lisenfeld11, Boselli14, Casasola17}.
In Fig.~\ref{fig:radial_profile}, we show the radial intensity profile of 67 out of the 121 galaxies presented in Table \ref{table:sample}. 
The selected galaxies have publicly available CO intensity maps (see references in Table \ref{table:sample} for the link to online repositories).
Radial averaged intensities of CO lines are normalized to the intensity at the center of each galaxy (i.e., corresponding to the optical position, retrieved from the DustPedia archive). Radial profiles were extracted with elliptical apertures projected on the plane of the sky based on the galaxy's inclination and position angle.
In Fig.~\ref{fig:radial_profile}, galaxies are color-coded based on their morphological type, and divided into four bins based on their Hubble stage number (see the legend in Fig.~\ref{fig:radial_profile} for further details).\\
Fig.~\ref{fig:radial_profile} shows that individual radial profiles have a significant scatter with respect to the mean radial profile. Nevertheless, the mean is consistent with the model of an exponentially decreasing gas distribution, within the typical uncertainty (i.e., $\sim15-20\%$, including calibration error; e.g., \citealt{Leroy21a}).
This suggests that the adoption of an average exponential profile 
(\citealt{Lisenfeld11,Boselli14}, C20)
remains a good compromise for estimating total mean gas mass for a sample of galaxies for which only partial observations of the galactic disk are available.
However, the large scatter observed in the profiles implies that each galaxy has some deviations from the mean profile, which does not assure a correct estimate of the total \lco for individual galaxies, and hence of \mh, when the average trend is used (see Appendix~B for more details).
For this reason, in this work, we have collected and used data from the most recent surveys that have mapped the molecular gas over a significant portion of the galaxy's optical disk ($>70\%$; see section~\ref{sec:co_maps}).\\


\section{Scaling relations}
\label{sec:results}
Here, we investigated scaling relations between different ISM components and other galaxy properties (namely, SFR, $\rm M_{\star}$, $\rm M_{dust}$) in a homogeneous sample of late-type ($HT>0$) galaxies, listed in Table~\ref{table:sample}.
In particular, we focused on the analysis of the molecular gas component, the fuel of star formation, to identify possible physical conditions and host-galaxy properties that enhance molecular hydrogen in galaxies. 
To this goal, we characterized a collection of correlations in terms of slope, normalization, and intrinsic dispersion ($\delta_{intr}$; i.e., the dispersion that is not owed to measurement uncertainties), as described in section \ref{sec:methods}.
In the end, we compared and ranked scaling relations between \mh\ and other galaxy properties (or a combination of properties) to determine the relation that has a lower intrinsic scatter.

\begin{figure*}[ht]
        \centering
	\includegraphics[width = 0.95\textwidth, keepaspectratio=True]{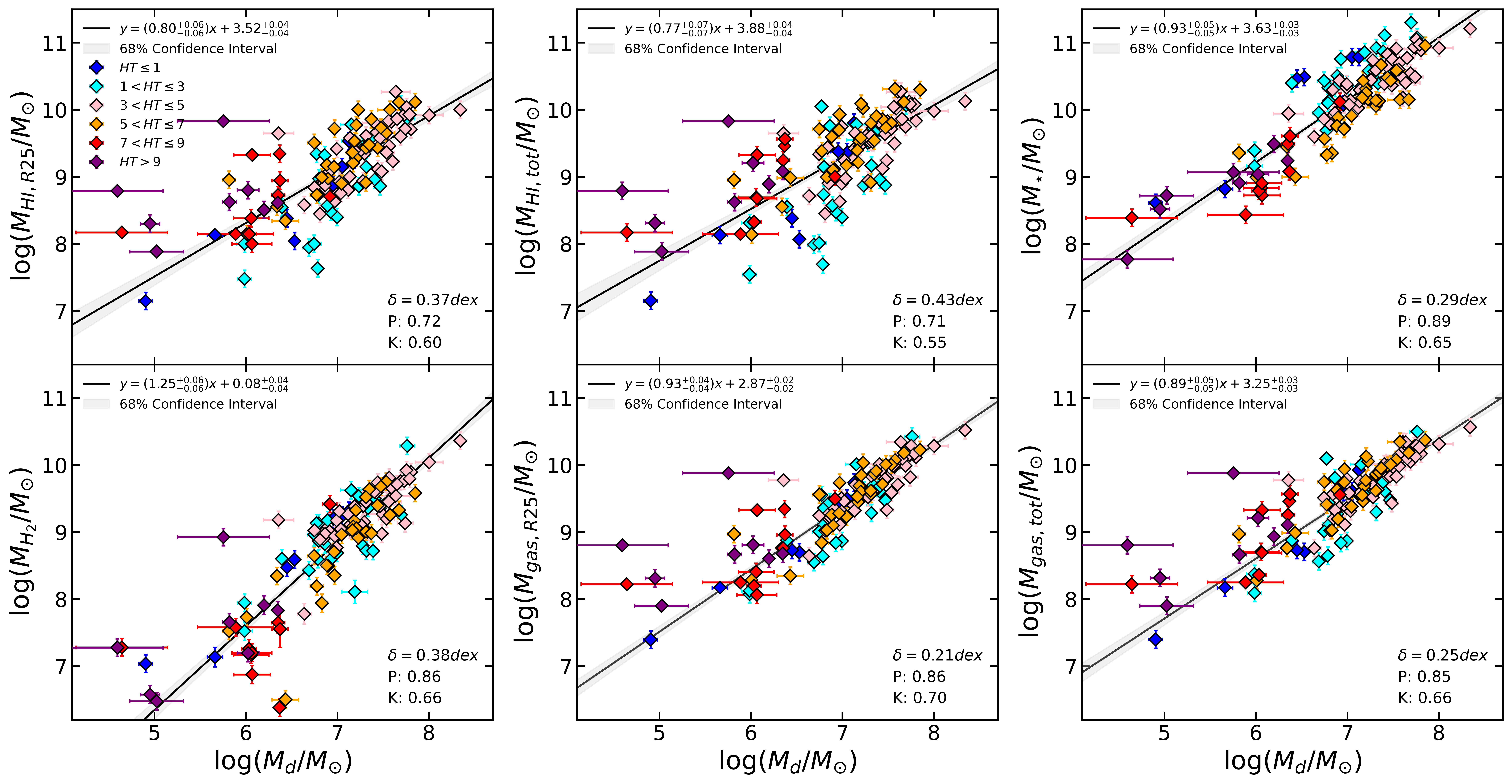}
	\caption{Gas and stellar masses vs dust mass.
    Upper row, atomic gas mass within R25 (left panel), total atomic gas mass (central panel), and stellar mass (right panel) are plotted against the dust mass.
    Bottom row, molecular gas mass (left panel), total gas mass (atomic plus molecular) measured within the optical radius (central panel), and total gas mass measured over the entire galactic disk (right panel) are plotted against the dust mass.
    Data are color-coded as the morphological type ($HT$).
    The best-fit parameters for each scaling relation are reported in the upper left corner, while the best-fit line is represented with the black solid line.
    The 68\% confidence interval of the best-fit parameters is represented as a gray-shaded region.
    The intrinsic dispersion ($\delta$) is reported in the bottom right corner, followed by the correlation coefficients (namely, $P$ and $K$).}    \label{fig:mdust_mgas}
\end{figure*}

%
%
\subsection{Relations between dust and gas}
\label{sec:dust_gas_phases}
In Fig.~\ref{fig:mdust_mgas}, we present the scaling relations between the mass of different gas components, namely the atomic gas within the optical radius ($M_{HI,\ R25}$; upper left panel), the total atomic gas ($M_{HI,\ tot}$; upper central panel), the molecular gas (\mh; bottom left panel), the total gas mass within the optical radius ($M_{gas,\ R25} = M_{H_2} + M_{HI,\ R25}$; bottom central panel) and measured over the entire disk ($M_{gas,\ tot} = M_{H_2} + M_{HI,\ tot}$; bottom right panel) and the dust mass.
In the upper right corner, we also present the relation between \mstar~and \mdust, as stars are one of the key channels to the formation of dust particles \citep{Draine03}.
For a fair comparison, in the six panels of Fig.~\ref{fig:mdust_mgas} we display only 120 out of 121 galaxies of the sample described in section~\ref{sec:sample}, as we excluded NGC1336, which lacks HI measurement (see section \ref{sec:sample}).
We recall that the hydrogen masses presented here have not been corrected for the contribution of heavy elements, from helium onward.
Correcting gas masses for heavy elements would require multiplying masses by a constant factor of 1.36 (e.g., C20). This does not affect the determination of the slope or scatter of the relations analyzed in this work; it merely produces a constant offset in their normalization, which can therefore be accounted for straightforwardly.
Overall, \mdust~shows a strong correlation ($P>0.7$ and $K>0.5$) with all the gas components and \mstar~presented in Fig.~\ref{fig:mdust_mgas}.
Focusing on the atomic gas displayed in the upper row of Fig.~\ref{fig:mdust_mgas}, we observed that the relation \mhiop~vs \mdust, and \mhitot~vs \mdust~are characterized by sub-linear slopes and a relatively large dispersion.
Furthermore, the correlation is stronger (i.e., lower intrinsic dispersion and higher CCs) when considereing only the dust and the co-spatial gas, i.e., measured within the optical extent of the galaxy R25, and not the total atomic gas ($\delta_{intr}=0.37$ and 0.43, respectively).
This may result from the concentration of most of the dust particles within R25, while the cold atomic gas can extend far beyond these areas.\\
Regarding the molecular gas component, using \mh\ estimates based on maps allowed us to detect a tight correlation between dust and molecular gas. This correlation is stronger than the one inferred using \mh\ values extrapolated from pointed observations, and it is also stronger than that between dust and the atomic component (see Appendix~\ref{sec:motivation} for a detailed comparison with C20).
The \mh\ vs \mdust\ relation (MGD, hereafter) shows a scatter ($\delta_{intr}=0.38$) comparable to that for atomic gas, but with higher CCs ( increased by $\sim0.15$) and a super-linear slope ($\alpha=1.25\pm0.05$). 
This suggests that galaxies that are richer in molecular gas are proportionally less rich in dust when compared to galaxies at lower masses.
A significant contribution to the measured intrinsic dispersion is likely driven by the galaxies with HT$>$7, which show a wide dispersion across the best-fit relation, particularly in the low mass end regime ($\log(M_{dust}/M_{\odot})<6.5$).
Among them, NGC7715 stands out in all panels, except for the \mstar~vs \mdust~relation, with a \mdust~that was likely underestimated in the SED fitting procedures due to contrasting fluxes in the far-IR (see \citealt{Nersesian19} and the DustPedia archive\footnote{\url{http://dustpedia.astro.noa.gr/Cigale}}).

Our analysis suggests that massive galaxies are more efficient in converting atomic into molecular gas phase than less massive objects, as the molecular-to-atomic gas mass ratio increases with the stellar mass (see the second panel from the left of Fig.~\ref{fig:gas_phase_trans}), while the dust-to-stellar mass ratio does not -- which is why the MGD relation has a super-linear slope.
The pressure, which is driven by the surface density of stars with an additional contribution by the gas surface density, enhances the gas density, favoring cooling and the formation of molecules \citep{Elmegreen93, Krumholz09}.\\
The stellar mass, \mstar, shows a very good correlation with \mdust~ ($P=0.89$, $K=0.65$), with an intrinsic dispersion which is smaller than the one that is observed for the relation between single gas components and dust ($\delta_{intr}=0.29$).
The slope is just slightly sublinear ($\alpha=0.93\pm0.05$), and holds over almost four orders of magnitude across different mass regimes and morphological types.
Nevertheless, there is a minor clustering in terms of HT in the more massive end ($M_{dust}>10^{6.5}$ \msun) of the relation, with galaxies with lower HT tend to have higher \mstar~for a given \mdust, consistent with previous results from the literature \cite{GonzlezDelgado15}.
Galaxies with lower HT (e.g., S0, Sa types) have more stars and less dust because they have almost depleted their gas (lower \mh-to-\mstar~ratio), having formed more stars when compared to higher HT.
This makes them older, hence dominated by stellar processes, which lead to an unbalanced dust destruction-to-formation rate.
The role of gas pressure in building up larger molecular gas reservoirs will be further discussed in section \ref{sec:MGMS_KS_MGD}.\\
In the bottom central and right panels of  Fig.~\ref{fig:mdust_mgas}, we show the relation between the dust mass and the total (atomic plus molecular) gas mass measured within R25 or over the entire galactic disk, respectively.
The two correlations are slightly sublinear (slopes are $0.93\pm0.04$ and $0.89\pm0.05$, respectively, and they have the lowest intrinsic dispersion among the relations presented in Fig.~\ref{fig:mdust_mgas} ($\delta_{intr}=0.21$, and $0.25$, respectively).
The relation between co-spatial quantities is stronger ($P=0.86$, and $P=0.70$) and has lower $\delta_{intr}$, confirming what has been discussed above for \mhitot.\\
The stronger correlation between global \mdust\ and \mh\ has been confirmed by spatially resolved studies both in our Galaxy (e.g., \citealt{Lee2012, Bialy2015}) and in nearby objects (e.g., \citealt{WongBlitz2002, Leroy08, Morselli20, Casasola22}), where molecular gas and dust are observed to be co-spatial to star-forming regions, dense environments rich in gas and dust. 
However, as found by C20, the relation between the total gas mass, $M_{gas, R25}$, co-spatial to \mdust~has in fact the lowest intrinsic dispersion ($\delta_{intr}=0.21\pm0.02$) and higher CCs.\\
No major clustering in terms of HT was observed in the present sample in any of the scaling relations presented in Fig.~\ref{fig:mdust_mgas}, except for the \mstar~vs \mdust~relation discussed above.\\
\begin{figure*}[ht]
        \centering
	\includegraphics[width = \textwidth, keepaspectratio=True]{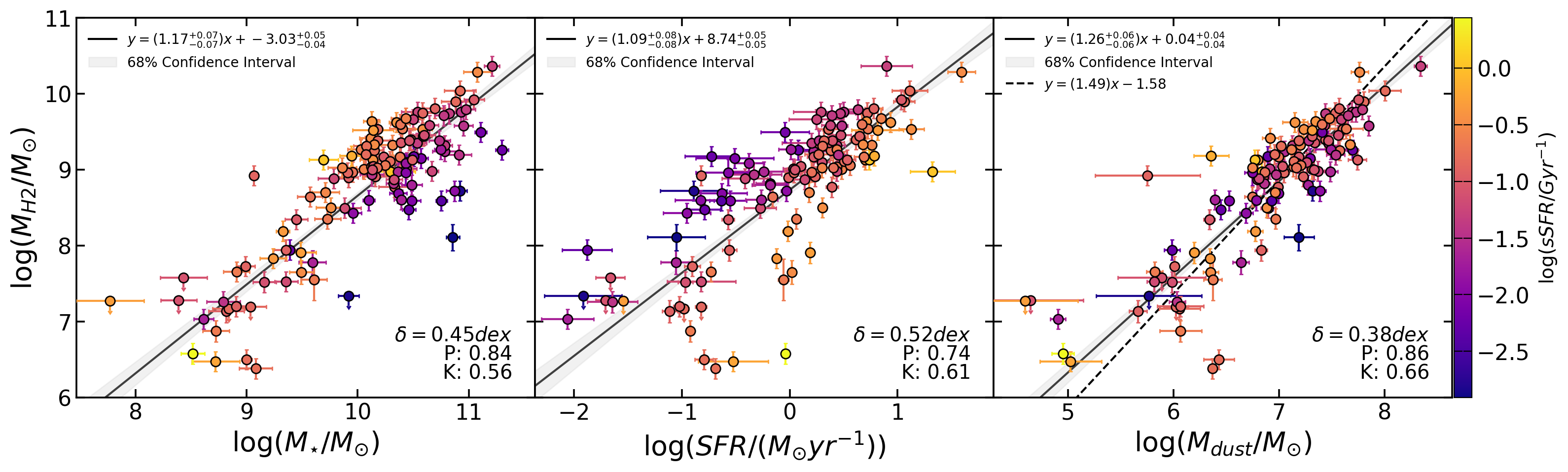}
	\caption{Molecular gas mass scaling relations.
    From left to right: \mh\ vs \mstar, SFR and \mdust.
    Data are color-coded for the sSFR.
    The best-fit parameters for each scaling relation are reported in the upper-left corner, while the best-fit line is represented with the black solid line.
    The intrinsic dispersion ($\delta$) is reported in the lower-right corner, followed by the correlation coefficients (namely, $P$ and $K$).}
    \label{fig:MGMS_KS_MGD}
\end{figure*}
Minor clustering is mostly due to relatively atomic-rich galaxies at later stages ($HT>7$), which deviate more from the best-fit relation, as visible also in the \mdust~versus total gas mass plot (lower right panel of Fig.~\ref{fig:mdust_mgas}).
This segregation is likely driven by the relatively low content of metals usually observed in such types of galaxies (e.g., \citealt{Gallazzi05}).\\
Using the \mdust-\mh\ relation, we also checked that the dispersion is not influenced by the adoption of a common CO-to-H$_2$ conversion factor (see Appendix C) or of a common CO(2--1)-to-(1--0) emission line ratio for the 103 galaxies shown in Fig.~\ref{fig:motivation}.
To this last goal, we divided the galaxies with CO(2--1) observations from those with CO(1--0) measurements (22 and 81 objects, respectively).
The $P$s estimated for the \mh-\mdust~correlation for the two subsamples of galaxies do not differ significantly (R=0.83 and 0.85, respectively).

%
%
\subsection{Molecular Gas Main Sequence and the global Kennicutt-Schmidt relation}
\label{sec:MGMS_KS_MGD}
Understanding what drives the formation of molecular gas in galaxies, for a given \mstar~or SFR, allows us to predict how galaxies evolve, e.g., whether they will continue to grow in mass by forming new stars or become passive.
To this goal, we investigated with our sample the relation between \mh\ and \mstar, also known as Molecular Gas Main Sequence (MGMS, hereafter; e.g., \citealt{Lin19}), and the \mh-SFR relation, commonly referred to as the global -- or galaxy-integrated -- Kennicutt-Schmidt law (gKS, hereafter),
to honor the pioneering works by \cite{Schmidt59} and \cite{Kennicutt89}.
These two relations are presented in the left and central panels of Fig.~\ref{fig:MGMS_KS_MGD}, respectively. 
As discussed in section~\ref {sec:dust_gas_phases}, the molecular gas mass is tightly related to the dust content, which makes the dust a key component to predict \mh, thus we added the MGD relation in the right panel of Fig.~\ref{fig:MGMS_KS_MGD} for comparison.
In all panels of Fig.~\ref{fig:MGMS_KS_MGD}, the 121 galaxies of the entire sample are shown color-coded with their specific star formation rate (sSFR=SFR/\mstar).
Here, the MGD includes NGC 1336, which is not included in Fig.~\ref{fig:mdust_mgas}, due to the missing \mhi~measurement, but the results are fully consistent with those discussed above.
The best-fit parameters obtained from the {\it linmix} procedures are reported in each panel and in Table~\ref{table:corr}.
The gKS is characterized by a close to linear slope, $\alpha=1.09\pm0.08$, while MGMS and MGD are consistently slightly super-linear, $\alpha=1.17\pm0.07 $ and $1.26\pm0.06$, respectively.
The MGMS shows the link between the molecular gas mass and the stellar mass, the latter being the main contributor to the gravitational potential of the disk in the vast majority of the galaxies in our sample, and always dominates in central regions where most of the molecules reside.
Indeed, in recent years, MGMS and gKS have been suggested as fundamental relations for the cycle of star formation (e.g., \citealt{Lin19, Baker23a}, with the linear trend of the gKS relation reflecting the small variations of star formation efficiency.
Consequently, the well-known Main Sequence (MS) of star-forming galaxies (e.g., \citealt{Renzini15}) that relates \mstar~and SFR is considered a by-product of both MGMS and gKS (see also section~\ref{sec:fp_sf}).
The relevance of other physical parameters, such as the stellar mass, in driving the atomic-to-molecular gas conversion can be seen also in the right panel of Fig.~\ref{fig:MGMS_KS_MGD}: the formation of the hydrogen molecules is enhanced as the mass of dust increases, as already shown in Fig.~\ref{fig:mdust_mgas}.
In addition, the gKS relation implies that the increasingly more intense radiation fields produced by the young stars in galaxies with higher SFR may destroy part of the dust particles surrounding the star-forming regions, hence reducing \mdust~in the higher mass regime.\\
Comparing MGMS, gKS, and MGD scaling relations, MGD has the lower intrinsic dispersion ($\delta_{intr}=0.38$) and the higher CCs ($P=0.86$, $K=0.66$) among the relation shown in Fig.~\ref{fig:MGMS_KS_MGD}, suggesting that dust mass can be used as a reliable proxy of the molecular gas reservoir in this sample of galaxies. The MGMS and MGD dispersion shows a weak relation with the sSFR, with galaxies at higher sSFR having possibly a higher molecular gas mass at a given stellar or dust mass (for M$_*> 10^{10}$~M$_\odot$ and for M$_{dust}>5\times 10^6$~M$_\odot$).
The deviations at the high mass end of the MGMS relation, visible for galaxies with a low sSFR, reflect the bimodal distribution usually observed in the SFR-\mstar~(MS; e.g., \citealt{CanoDiaz16, Guo15, Renzini15}).
Galaxies move from the MS to the green valley (e.g., \citealt{Brownson20, Lin22}), and eventually to the so-called red sequence, i.e., passive objects, as the availability of molecular gas gradually decreases (see the discussion in section~\ref{sec:sfe}).
This suggests that \mstar~can not be used alone to accurately predict the amount of molecular gas in a galaxy, and the SFR is required to determine the evolutionary phase of the galaxy accurately.\\
We note that we did not include \mh\ upper limits in the fitting procedures, but they are shown in Fig.~\ref{fig:MGMS_KS_MGD}.
\mh\ upper limits follow a similar trend to that observed for \mh\ detections in the rest of the sample in all scaling relations presented (namely, MGMS, gKS, and MGD), populating the low-\mstar, SFR, and \mdust~regimes. 
\begin{figure*}[ht]
	\includegraphics[width = \textwidth, keepaspectratio=True]{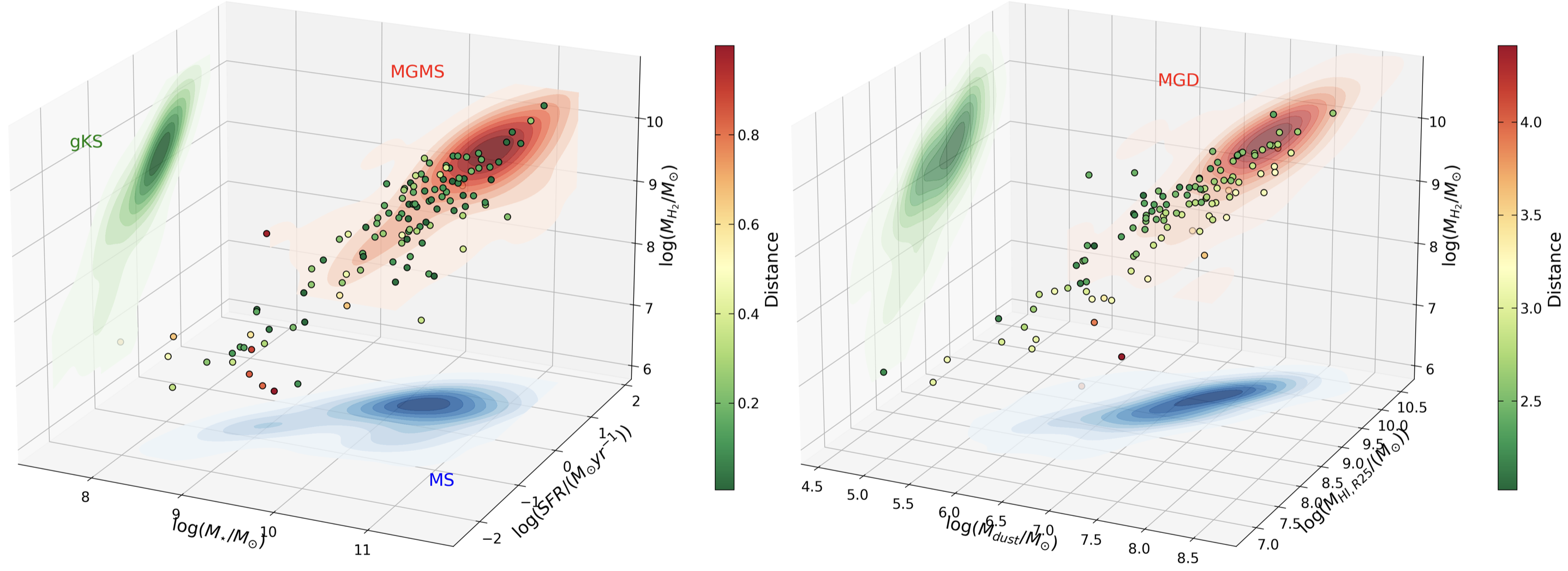}	
	\caption{The fundamental plane of SF and ISM.
    Left panel: \mh\ vs \mstar, and SFR in a three-dimensional projection.
    Data are color-coded as a function of their distance from the best-fit function.
    The green, blue, and red-shaded heat maps represent the density profiles of the data on the gKS, MGMS, and MS planes, respectively.
    Contours are shown to distinguish better the density profile, divided into ten levels.
    The best-fit relation of each 2D panel is color-coded according to the data heat maps, while best-fit parameters are listed in Table\ref{table:corr}.
    Right panel: \mh\ vs \mhiop, and \mdust~in a three-dimensional projection.
    Color coding is the same as in the left panel.
    The green, blue, and red-shaded heat maps represent the density profiles of the data on the \mh-\mhiop, \mdust-\mhiop, and MGD planes, respectively.}
    \label{fig:3d}
\end{figure*}
%
%
\subsection{The fundamental plane for SF}
\label{sec:fp_sf}
Based on the results from gKS and MGMS, we combined \mstar~and SFR to better constrain \mh.
In this case, we fit the relation with equation \ref{eq:3d}, as described in section~\ref{sec:methods}.
We also show the projection of the best-fit three-dimensional plane with each of the axes made by \mstar, SFR, and \mh\ (x, y, and z-axes, respectively).
The best-fit relation is:
\begin{multline}
    \log(M_{H2}/M_{\odot})= 0.79^{+0.07}_{-0.07} \times \log(M_{\star}/M_{\odot}) +  \\ 0.51^{+0.07}_{-0.07} \times \log(SFR/(M_{\odot} yr^{-1})) + 8.86^{+0.03}_{-0.03},
    \label{eq:FP_SF}
\end{multline}
and intrinsic dispersion $\delta_{intr}=0.24^{+0.05}_{-0.05}$.
As mentioned in section~\ref{sec:methods}, we did not include upper limits on \mh\ in this fitting procedure.
The intrinsic dispersion estimated in this case is lower than in the case of MGMS, gKS, and MS relations (see Table~\ref{table:corr}), suggesting that the connection between the three quantities (namely, \mh, \mstar, and SFR) is tighter than when only two are considered at a time.
This way, we removed part of the degeneracy that affects both MGMS and gKS, which comes from the attempt to collapse a multidimensional distribution of galaxy properties onto a linear relation. 
For illustrative purposes, the collapsed distribution of points on each two-dimensional plane (namely, \mh-\mstar, \mh-SFR, and SFR-\mstar) is shown in the left panel of Fig.~\ref{fig:3d}.
To further highlight the relatively low dispersion in the data, we color-coded the points based on their distance from the best-fit plane. 
From a three-dimensional perspective, combining \mh, \mstar, and SFR, we obtained a more complete view of the current evolutionary stage of galaxies, which are thought to move across this plane.
Indeed, several recent studies have suggested the existence of a fundamental plane of SF, such as the one presented in the left panel of Fig.~\ref{fig:3d} (e.g., \citealt{Lin19, Baker23a}).
In this picture, it is clear that only the cycle of star formation relies on the balance between \mh, \mstar, and SFR, as represented by the plane in the parameter space described by the three quantities.
As a galaxy grows in mass following the main sequence, it will have an almost linear proportion between its \mstar, \mh, and SFR.
If at some point, a galaxy cannot replenish its molecular gas reservoir anymore, which can happen because of several physical mechanisms such as sudden feedback-driven gas removal or heating in its interstellar and cirgumgalactic media (e.g., \citealt{Peng15}), these processes will stop its star formation activity within a time which is proportional to the depletion time ($t_{dep}\propto M_{H2}/SFR$).
Once the quenched phase begins, the galaxy will deplete its remaining gas reservoir and then it will evolve passively.
This will make \mstar~to increase proportionally to the availability of cold gas (but this can also depend on galactic morphology and the gas distribution; e.g., \citealt{Peng10, Saintonge16}), but at a lower sSFR (i.e., moving to the region of quenched/passive galaxies in Fig.~\ref{fig:MGMS_KS_MGD}).
Then, the galaxy will evolve passively, following stellar processes, with a low sSFR level.

\subsection{The fundamental plane for the ISM}
\label{sec:fp_ism}
Similarly, we fit equation~\ref{fig:3d} to the data, which include \mdust~on the x-axes, \mhi~on the y-axis, and \mh\ on the z-axis.
In the right panel of Fig.~\ref{fig:3d}, we show the projection of the best-fit three-dimensional plane, with best-fit relation:
\begin{multline}
    \log(M_{H2}/M_{\odot})= 1.38^{+0.10}_{-0.10}\times \log(M_{dust}/M_{\odot}) - \\ 0.24^{+0.10}_{-0.10} \times \log(M_{HI, R25}/M_{\odot}) + 1.38^{+0.54}_{-0.53},
    \label{eq:FP_ISM}
\end{multline}
and intrinsic dispersion $\delta_{intr}=0.34^{+0.03}_{-0.03}$.
First, $\delta_{intr}$ is larger with respect to that measured in the cases of the MGD and the dust-versus-total gas relation.
This suggests that the physical connection between dust and molecular gas is stronger than that between the three main components of the ISM\footnote{Here, we did not include the ionized gas phase, which has a negligible contribution to the total ISM mass; e.g., \citealt{Draine11}.}.
Then, the atomic gas trend is almost constant (the corresponding slope is consistent with zero within $\sim2-2.5\sigma$) when both dust and molecular gas content increase.
This is likely driven by the large scatter that characterizes the distribution of \mhiop~for a given \mdust~and \mh.
\begin{figure*}[ht]
        \centering
	\includegraphics[width = 1.0\textwidth, keepaspectratio=True]{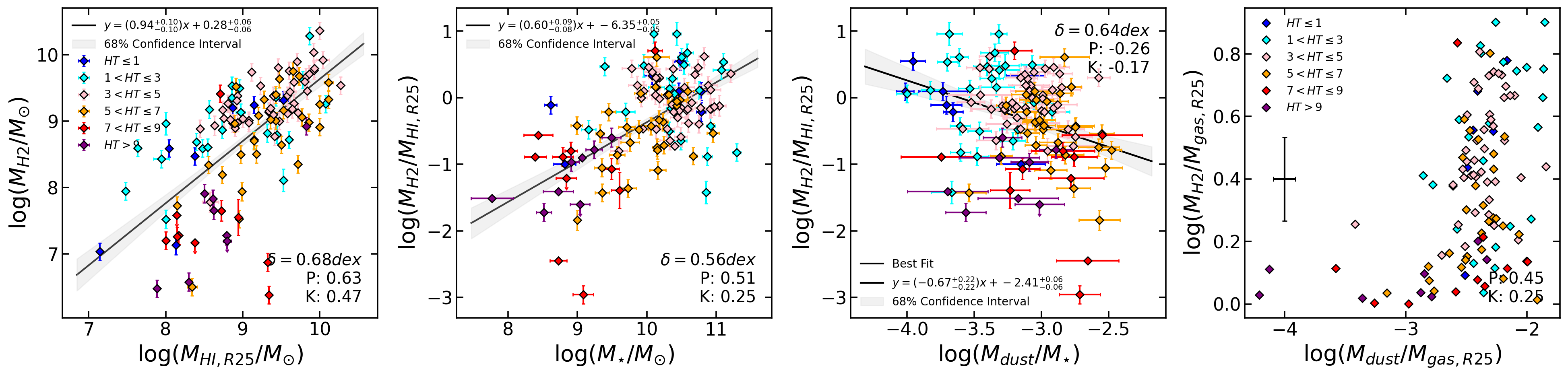}
	\caption{Left panel: Molecular gas mass vs. atomic gas mass.
    Data are color-coded with the morphological type as in Fig.~\ref{fig:mdust_mgas}.
    The best-fit line is represented with the black solid line, and the gray-shaded region represents the 68\% confidence interval of the posterior distribution of the parameters.
    The intrinsic dispersion ($\delta_{intr}$) and correlation coefficients (namely, $P$ and $K$) are reported in the lower part of each panel.
    Central left panel: Molecular gas fraction vs. stellar mass.
    Central right panel: Molecular gas fraction vs. dust-to-stellar mass ratio.
    Right panel: Molecular-to-total gas fraction vs DTGR. The black cross represents the median error on both x and y-axis.}
    \label{fig:gas_phase_trans}
\end{figure*}
We found no significant difference considering the total atomic gas mass (\mhitot) over the atomic gas phase co-spatial to the molecular one (\mhiop).
Given this trend, we concluded that it is not the global cold atomic gas richness in galaxies that drives the molecular gas abundance, but rather the ability of a galaxy to make the gas dense enough to favor the atomic-to-molecular transition (see section~\ref{sec:gas_phase_transition}).
In this regard, dust is crucial to prevent the cold molecular gas from dissociating, shielding the gas from ionizing radiation \cite{Draine03}.
The best-fit slope associated with \mdust~in equation~\ref{eq:FP_ISM} is super linear, while in MGD it was sublinear, suggesting that the dust content dominates over \mhiop~to determine the mass of the molecular gas reservoir and can be used to trace  \mh.
To conclude, it is worth noticing that the accurate determination of \mdust~strongly depends on the modeling of the far-IR emission and is limited to galaxies, mostly in the nearby Universe, with their rest-frame far-IR regime (i.e., $\sim40-1000\mu m$) properly sampled (e.g., \citealt{Clark18}; \citealt{Galliano18}). 
For this reason, the application of MGD to predict \mh\ to large samples of galaxies is limited by the availability of dust masses.
However, this problem may be (at least partially) bypassed by using photometric observations that trace the bulk of the dust emission, as shown in the next section. 
\section{Discussion}
\label{sec:discussion}
\subsection{Gas phase transitions}
\label{sec:gas_phase_transition}
As discussed in section~\ref{sec:dust_gas_phases}, the atomic and molecular phases of gas are closely linked to dust in our sample.
Here, we discuss which are the physical conditions that favor molecular gas enrichment.
To this goal, in the left panel of Fig.~\ref{fig:gas_phase_trans} we present the \mh\ vs the co-spatial \mhi~relation, color-coded as a function of the Hubble type.
The plot includes all 120 galaxies presented in Fig.~\ref{fig:mdust_mgas}.
CCs are $P=0.63$ and $K=0.47$, respectively, which tells us that there is a moderate-to-strong correlation between the two co-spatial gas phases, yet there is also a significant scatter.
Later-type galaxies are those that are deviating most from the main trend, with a larger atomic-to-molecular gas ratio.
However, their relatively low statistics did not impact the global trend of the sample and the CCs ($P$ and $K$ are 0.68 and 0.4,8, limiting the sample to HT$\leq$7).
This indicates that the dispersion is likely driven by individual galaxy characteristics, such as differences in morphology, stellar mass, or accretion history.
When dividing our sample between galaxies in clusters (e.g., the Virgo Cluster; \citealt{Corbelli12, Brown21}) and field galaxies, the environment does not seem to impact this relation, as cluster and field objects show similar trends.\\
In the central left panel of Fig.~\ref{fig:gas_phase_trans}, we show the molecular-to-atomic gas mass ratio (H2-to-HI) as a function of the stellar mass.
The $P$ index suggests a moderate correlation (=0.5), but the scatter is quite large ($\delta_{intr}=0.56$), as highlighted by the value of $K$ (=0.25).
However, as discussed for the MGMS, we suggest that the positive trend of this correlation is a direct consequence of the gravitational potential driven by the stars (see discussion in section~\ref{sec:MGMS_KS_MGD}) that favors the formation of the cold gas reservoir.\\
In the central right panel of Fig.~\ref{fig:gas_phase_trans}, the segregation of galaxies according to their HT is likely driven by \mstar, which increases when HT decreases (see Fig.~\ref{fig:sample_bin} for a binned view of this trend). The H2-to-HI ratio increases from galaxies with higher HT (and lower \mstar) to earlier-type (and more massive) ones.
In the right panel of Fig.~\ref{fig:gas_phase_trans}, we plot the molecular gas fraction as a function of the dust-to-total-gas ratio (DTGR). 
Galaxies with  $HT>7$ show a low molecular fraction for a variety of DTGR ratios, while galaxies with $HT\leq7$ are distributed across the whole range of molecular fractions and have $0.0016\lesssim DGR\lesssim 0.016$, although the distribution peaks at DTGR$\simeq 0.004$. 
The galaxy's ability to increase its dust mass depends on the balance between stellar processes that create new dust grains and those that destroy them. The sublinear slope of the MGD suggests that, at higher masses, dust might be destroyed more efficiently than at lower masses, and this also explains the marginal dependence of the molecular fraction on the DTGR and the marginal dependence of the DTGR on the galaxy Hubble-type.\\
Indeed, despite the strong correlation and relatively low intrinsic dispersion found for the MGD relation in section~\ref{sec:MGMS_KS_MGD}, Fig.~\ref{fig:gas_phase_trans} highlights the crucial role of \mstar through the local gravity in driving the formation of molecules.
The anticorrelation we found between the molecular gas fraction and the dust-to-stellar mass ratio for our sample confirms that the process of atomic-to-molecular gas conversion is more efficient for high stellar masses rather than for high dust masses as soon as atomic gas is available.\\
\begin{figure*}[ht]
        \centering
	\includegraphics[width = 0.9\textwidth, keepaspectratio=True]{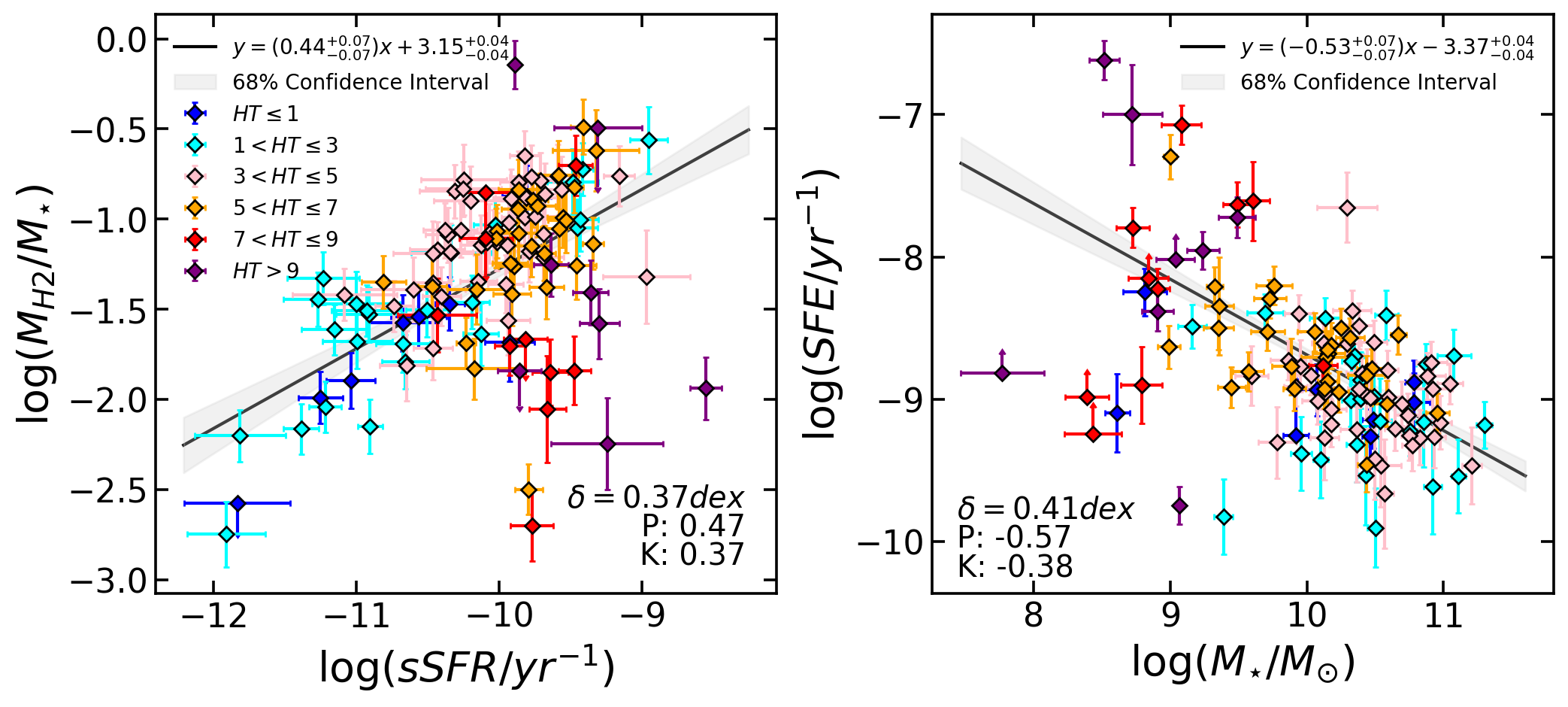}
	\caption{Left panel: Molecular-to-stellar mass ratio vs. sSFR.
    Data are color-coded with the morphological type, as in Fig.~\ref{fig:mdust_mgas}.
    The best-fit line is represented with the black solid line, and the gray-shaded region represents the 68\% confidence interval of the posterior distribution of the parameters.
    The intrinsic dispersion ($\delta_{intr}$) and correlation coefficients (namely, $P$ and $K$) are reported in the lower (upper) right corner of the left (right) panel.
    Right panel: Star formation efficiency vs. stellar mass.}
    \label{fig:fgas_sfe}
\end{figure*}
\subsection{Molecular gas fraction and star formation efficiency}
\label{sec:sfe}
As discussed above, classical scaling relations between \mh, \mstar, and SFR suffer from the contamination of galaxies facing different evolutionary phases.
Specifically, in section~\ref{sec:MGMS_KS_MGD}, we discuss the presence of passive galaxies in the MGMS as those galaxies having a relatively low sSFR due to a relatively low molecular gas-to-star ratio ($M_{H2}/M_{\star}$). 
To this goal, we plot the $M_{H2}/M_{\star}$ as a function of the sSFR in the left panel of Fig.~\ref{fig:fgas_sfe}.
The ratio between \mh\ and \mstar~indicates the amount of molecular gas that is available to sustain the growth of the galaxy, given the mass of stars already formed.
The sSFR is an important parameter to describe how fast a galaxy has grown since 1/sSFR indicates the timescale that the galaxy would have needed to build the current mass at the current SFR.
Therefore, the diagram shown in the left panel of Fig.~\ref{fig:fgas_sfe}, allows us to distinguish quiescent (or soon-to-be quiescent) galaxies from those that are in a relatively active build-up phase.
Indeed, earlier-type galaxies from our sample (HT$\leq$3) show relatively low molecular gas fraction and low sSFR ($<10^{-10}$ yr$^{-1}$).
Conversely, later type galaxies, HT$>7$, are mostly molecular gas-poor, but they are growing at a sustained rate ($sSFR>10^{-10}$ yr$^{-1}$).
This subsample may also include galaxies that are classified as irregular, which can be a consequence of a tidal interaction with a close companion galaxy.
Even if mergers are not frequent in the local Universe ($\sim 2$\%; \citealt{Casteels14}), past interactions with nearby galaxies can induce instabilities in the disks of galaxies, which can enhance the star formation even in relatively gas-poor objects. 
In between these two classes, there is the bulk of our sample (71 out of 121 galaxies) that has a median \mh-to-\mstar~ratio of about 0.1.\\
The distinction between two types of molecular gas-poor galaxies is evident in the right panel of Fig.~\ref{fig:fgas_sfe}, where later type galaxies are relatively efficient at forming stars, with a mean star formation efficiencies ($SFE=SFR/M_{H2}<10^{-8}$ yr$^{-1}$) higher than the bulk of the sample, although with a large scatter, while early types with HT$<7$ have SFE similar to later type galaxies with similar stellar mass (the median value being $\sim 10^{-9}$ yr$^{-1}$) irrespective of their \mh-to-\mstar~ratio).
From an evolutionary perspective, the HT$>$7 galaxies in our sample represent those galaxies still in a growth phase, with the highest SFE. 
Their future evolution depends on their ability to convert their atomic gas reservoir into molecular gas, sustaining their star formation.
These galaxies exhibit some of the highest atomic-to-molecular gas ratios in our sample (see Fig.~\ref{fig:gas_phase_trans} and the discussion below).
Conversely, the relatively low SFE of galaxies with HT$\leq$7 suggests they could have entered a quenching phase.
We underline that there is no evidence of chemical evolution between 3$<$HT$\leq$7 and HT$\leq$3 objects since they show similar metallicity and similar DTGR, as shown in Fig.~\ref{fig:gas_phase_trans}.
To conclude, in this analysis, we underline that we did not find differences between field and cluster galaxies (e.g., from the Virgo cluster; \citealt{Corbelli12, Brown21}), since both classes of sources share similar properties (e.g., sSFR, SFE) when paired for morphological types.
\section{An alternative approach to trace gas reservoirs}
\label{sec:photometry}
\subsection{The molecular gas}
To avoid limitations of using physical quantities derived from galaxy models, such as \mdust, \mstar, and SFR,  we tested the reliability of photometric measurements as a proxy of \mh and \mhiop. We collected photometric data from the UV to far-IR wavelengths from the DustPedia database (see \citealt{Clark18} for details), considering all the photometric bands from \cite{Clark18}, except for \emph{Spitzer} IRAC 1 and 2 bands (because these cover similar wavelength ranges of WISE band 1 and 2, and all objects targeted by the IRAC instrument already had WISE observations). Similarly, we did not consider \emph{Spitzer} MIPS 70 and 160 bands, since \emph{Herschel} PACS 70 and 160 cover the same wavelength regimes.
We complemented the limited number of photometric observations from PACS 70 (available for 64 out of 121 galaxies), using \emph{Spitzer} MIPS 70 data for 21 galaxies that lack \emph{Herschel} PACS 70 data.
This way, we had 85 galaxies with a monochromatic luminosity at 70~$\mu$m.\\
In Appendix~F in Fig.~\ref{fig:photometry}, we showed the relations between \lco\ (i.e., \mh\ divided by the constant \aco=3.26 \uaco, and converted into units of solar luminosity, L$_{\odot}$) and the luminosity measured within each photometric filter.
Since the photometric coverage is not uniform for the entire sample of 121 objects, we tested the correlation between \lco\ and the luminosity in each band for two different sample selections: \emph{i)} by crossmatching all photometric measurements, we selected a subsample of 47 objects that have a measurement in all the bands (sample A); \emph{ii)} we considered all galaxies having a photometric measurement in each band (sample B). 
This allowed us to compare the results of the correlation between \lco\ and the luminosity of the galaxy at different wavelengths on a common sample (sample A), but also to explore each correlation using better statistics (sample B). Both best-fit lines are represented in Fig.~\ref{fig:photometry} and the subsamples are color-coded accordingly. In Table~\ref{table:corr}, we provided the best-fit parameters and CCs for each correlation between \lco\ and $\nu L_{\nu}$.\\
The best-fit relations obtained for samples A and B are in general agreement in each band, with deviations that become significant only in GALEX bands, given the significantly higher number of galaxies in sample B than in sample A.
The intrinsic dispersion measured for sample A is almost systematically smaller than that measured for sample B, likely due to a combination of lower statistics and a greater homogeneity in the former sample.
Based on the CCs shown in each panel of Fig.~\ref{fig:photometry} and $\delta_{intr}$ reported in Table \ref{table:corr}, \lco\ better correlates with the luminosities measured in mid-IR and far-IR bands, with $P$ and $K$ larger than 0.7.
In particular, the tight relation between \lco\ and the luminosity in WISE band 3 and \emph{Spitzer} IRAC 8.0 $\mu$ m in the mid-IR is likely driven by the presence of polycyclic aromatic hydrocarbons (PAH) features that dominate the emission in those bands (e.g., \citealt{Whitcomb23}).
Indeed, PAH are emission features that are tightly related to star formation activity \citep[e.g., ][]{Cortzen19, Zanchettin24}, and they are usually observed in actively star-forming galaxies. 
However, given that similar but steeper correlations hold also in early-type galaxies \citep{Gao25} because old stars also contribute to the mid-IR emission, and because of the inference of AGN activity on PAH emission (e.g., \citealt{GarciaBernete22, Salvestrini22}), future studies are needed to better understand the use of 12~$\mu$m emission as a proxy for \lco.
All mid-IR bands have lower values of CCs than FIR bands because hot dust does not represent the bulk of dust mass available in the dense ISM.\\
The tight correlation between far-IR bands and \lco\  reflects that observed between \mh\ and \mdust, already discussed in section \ref{sec:dust_gas_phases} and \ref{sec:MGMS_KS_MGD}.
All luminosities measured in the $100-500\ \mu$m range are good proxies for the molecular gas luminosity, with $P\geq0.85$, with \emph{Herschel} SPIRE 250 having the highest CCs among all bands (e.g., $P\sim0.92$; see Fig.~\ref{fig:photometry} and Table~\ref{table:corr} in Appendix~\ref{app:corr_coeff} for CCs and best-fit parameters).\\
Concerning the correlations between UV bands (namely, GALEX FUV and NUV filters) and \lco, we found a large dispersion and low CCs.
UV bands are commonly used to derive the SFR estimates, which roughly cover the star formation activity that occurred over the last $\sim$100-300~Myr \citep[see also the discussion on SFR recipes in Appendix~\ref{app:gKS_multiSFR}]{KennicuttEvans12}.
However, UV bands only trace the unobscured star formation and need to be corrected for the emission absorbed by dust.
This can also significantly affect optical observations at shorter wavelengths, as in the case of the SDSS u filter.
Dust obscuration is also dependent on the dust geometry, as well as the inclination of the galaxy with respect to the line of sight.\\
While stellar mass is a relatively good proxy of the molecular gas mass, individual photometric bands dominated by the stellar emission (namely, SDSS and 2MASS filters) show relatively low CCs ($P=0.66-0.8$,  $K\sim0.5$) compared to the mid-IR ones, and cannot be considered a good proxy for \lco in our sample. This limitation might arise because of the small range of luminosities covered by the bulk of our sample. with only a few galaxies extending to low luminosities.
For this reason, we suggest using the \lco-$\nu L_{\nu}$ scaling relations to predict the molecular gas emission with optical to near-IR photometric measurement only for $\log(\nu L_{\nu}/L_{\odot})>9.5$.\\
To conclude, we suggest that far-IR emission at wavelengths $\geq100$ $\mu$m is the best proxy of \lco\ over three orders of magnitude, providing a reliable approach to measure the molecular gas mass when (sub-)mm observations are not available. 
In addition, the emission arising due to more prominent PAH features at 7.7 and 11.3$\mu$m is a relatively reliable tracer of \mh, still covering a similar dynamical range in \lco.
%
%
\subsection{The atomic and total gas}
We also tested if any monochromatic luminosity can be used as a proxy for the atomic gas and total (H$_{\rm 2}$+HI) gas masses using subsamples A and B.
Best-fit parameters are listed in Table~\ref{table:corr_HIr25} and~\ref{table:corr_gasr25}, while best-fit lines are shown in Fig.~\ref{fig:hi_photometry} and~\ref{fig:gas_photometry}.
In general, the atomic gas mass, \mhiop~showsa  lower level of correlation (in terms of CCs) and a higher dispersion (i.e., larger $\delta_{intr}$) with monochromatic luminosities, when compared to \lco.
At the longest far-IR wavelengths (namely, \emph{Herschel} SPIRE observations), \mhiop~shows the tightest relations in terms of intrinsic dispersion ($\delta_{intr}=0.38$~dex for SPIRE 500) and highest CCs ($P\sim0.8$ and $K>0.6$ for SPIRE 500).
This is in general agreement with the result presented in section~\ref{sec:dust_gas_phases}.
Moreover, the atomic phase shows better correlation parameters with the emission that arises from the cold diffuse dust (SPIRE 500), and the CCs (intrinsic dispersion) decrease (increase) monotonically going at short wavelengths down to WISE4 and \emph{Spitzer} 24 bands.
This is the first difference between \mhiop~and \lco, with the latter that correlates best with monochromatic emission close to the peak of the far-IR SED.
Furthermore, we found that \mhiop~shows a relatively good correlation with UV/optical luminosities (namely, GALEX FUV, NUV, and SDSS u bands), only slightly less significant than with FIR luminosities, in contrast with the results for \lco.\\
As expected, the atomic phase of gas is not directly involved in recent star formation events, and \mhiop\ does not strongly correlate with SFR tracers (e.g., PAH features, warm dust emission; see \cite{KennicuttEvans12}). However, the good correlation between \mhiop~and NUV suggests that the atomic gas reservoir is needed to sustain the galaxy's ability to form new stars, it provides the gas fuel to sustain star formation over a long time interval, 300-500~Myr, for several lifecycles of molecular clouds (15~Myr, \citealt{Corbelli17}). Over this period, star formation can be traced by emission in the NUV band. \\
The HI gas also establishes a good correlation with 500$\mu$m emission,  a good proxy of the diffuse dust component. 
Regarding the total gas mass within R25, we found an overall improvement of CCs when analyzing the correlations with monochromatic luminosities in the optical bands.
As for \lco,  the strongest correlations for $M_{gas, R25}$ are established with the far-IR bands ($P>0.85$ and $K\gtrsim0.7$). Slopes are in this case close to linear, slightly flatter than for \lco. 
\subsection{Dust content}
Similarly to previous sections, we investigated the best proxies of the dust mass among monochromatic luminosities.
Looking at Fig.~\ref{fig:dust_photometry}, as expected the best proxies for \mdust~are the monochromatic luminosities at wavelengths longer than 160~\micron,
(from whose fit \mdust is derived) which show strong correlations ($P>80$ and $K\geq70$) and relatively low dispersion ($\delta_{intr}\leq0.1$~dex at 500~\micron). However, if FIR observations are not available
strong correlations ($P>0.8$ and $K\geq0.7$) can be found between \mdust~and the luminosities from optical to mid-IR wavelengths, up to WISE3 (i.e., $\sim8$~\micron).
The tight connection between dust mass and optical or near-IR emission is likely a consequence of the tight correlation between \mdust~and \mstar~(e.g., \citealt{McGaugh14}), already shown in section~\ref{sec:dust_gas_phases}, with a scaling close to linear.
The intensity of PAH emission features falling within the WISE1, WISE2, and WISE3 bands suggests the use of luminosities in these bands as a proxy for \mdust. The scaling is however, non-linear.
The reason for this is not yet clear because how PAH molecules form is still a matter of debate, with some dust evolution models that predict PAHs formation by fragmentation of large carbonaceous grains \citep[e.g., ][]{Seok14}, while observations on the Small Magellanic Clouds support models with PAHs that forms in molecular clouds\citep[e.g., ][]{Sandstrom10}.

Our conclusions for the best reliable proxies of the different phases of the ISM are the following:
$i)$ emission at far-IR wavelengths (particularly from 250 to 500~\micron) can be used as a proxy of all cold gas phases and dust in the ISM.
$ii)$ If observations in the FIR are unavailable, the monochromatic luminosity measured in mid-IR bands where PAH features fall is also a very good tracer for the molecular gas and the dust mass ($P>0.8$, $K\geq0.7$, and $\delta_{intr}<0.3$~dex).
$iii)$ The dust content is also traced with good approximation ($\delta_{intr}<0.3$~dex) by the optical and near-IR luminosities. 
$iv)$ The NUV luminosity can be used to estimate \mhiop, but the intrinsic scatter of the relation would allow us to estimate \mhiop~with a $\sim0.4$~dex uncertainty.

\section{Conclusions}
\label{sec:conclusions}
In this work, we selected a large sample of late-type galaxies (121) in the nearby Universe, which have been fully mapped in CO and at FIR luminosities and therefore have reliable estimates of molecular gas and dust masses. Using the DustPedia database, we complemented these data with estimates of other physical properties, such as the SFR, HI, and stellar masses, to investigate the physical processes that drive the star formation cycle as galaxies evolve.
The main results can be summarized as follows:
   \begin{enumerate}
      \item The global dust mass correlates better with the molecular rather than with the atomic gas mass, in agreement with resolved observations. The use of measurements of CO emission over the large part of the optical extent ($>70\%$ R25) of a galaxy has been a key element to reach this conclusion. 
      \item 
      We confirmed the important role of stellar populations in driving the formation of molecules because their gravity in the disk enhances the gas density.
      \item Determining \mh, \mstar, and SFR is necessary to properly understand the evolutionary stage of a galaxy and the emergence of a quenching phase. We determined the fundamental plane of star formation for late-type galaxies characterized by a lower intrinsic dispersion than MGMS and gKS relations. 
      \item \mdust~is the quantity that best correlates with the gas mass in late-type galaxies (i.e., it has the lowest intrinsic dispersion), whether it is molecular or total gas mass. 
      Dust is co-spatial with molecular gas and plays a crucial role in the formation of molecules themselves. However, dust production and destruction processes limit the growth of dust masses, and galaxies in the highest stellar mass regime are more efficient in converting atomic into molecular ga,s even with a lower dust content.
      \item Monochromatic luminosities measured in the FIR at 250-500 $\mu$m are the best proxies of atomic and molecular gas masses. In the absence of FIR data, luminosities in mid-IR photometric bands collecting PAH emission can be used to trace molecular gas and dust masses.
      \end{enumerate}

Finally, we would like to underline the importance of the availability of CO maps for a reliable estimate of the global molecular content of late-type galaxies, as shown in Section 3.2 and Appendix~B. This is due to the large scatter shown by the radial profile of the CO surface brightness. In fact, although the model of an exponentially decreasing molecular gas mass surface density, often adopted in the literature (e.g., \citealt{Boselli14}), is a good proxy for the mean molecular gas radial profile of a galaxy sample, individual galaxies can show strong departures from it. This, in turn, implies large uncertainties in molecular mass estimates if CO maps are not available. With future radio interferometers, we also hope to increase the availability of 21-cm resolved gas maps and improve the estimates of atomic gas in the star-forming regions of late-type galaxies. These maps are also necessary to fully understand the baryonic cycle in resolved regions of galaxies, from the center to the galaxy outskirts, where the ISM is mostly atomic. 


\begin{acknowledgements}
We acknowledge V. Casasola for helpful discussions that improved the presentation of results and E. Liuzzo for her support in defining the sample.
This work benefits from the financial support by the National Operative Program (Programma Operativo Nazionale-PON) of the Italian Ministry of University and Research “Research and Innovation 2014-2020”, Project Proposals CIR01$\_$00010.
F. S. acknowledges financial support from the PRIN MUR 2022 2022TKPB2P - BIG-z, Ricerca Fondamentale INAF 2023 Data Analysis grant ``ARCHIE ARchive Cosmic HI \& ISM  Evolution'', Ricerca Fondamentale INAF 2024 under project "ECHOS" MINI-GRANTS RSN1.
S. B., E. C. acknowledge funding from the grant PRIN MIUR 2017 – 20173ML3WW\_001 and from INAF minigrant 2023 SHAPES. 
F. S., S. B., E. C. acknowledge funding from the INAF mainstream 2018 program ``Gas-DustPedia: A definitive view of the ISM in the Local Universe''.
F. S., S. B. acknowledge funding from the INAF minigrant 2022 program ``Face-to-face with the Local Universe: ISM's Empowerment (LOCAL)''.
This work made use of HERACLES, ``The HERA CO-Line Extragalactic Survey'' \citep{Leroy09}.
This work made use of THINGS, ``The HI Nearby Galaxy Survey'' \citep{Walter08}.
This publication made use of data from COMING, CO Multi-line Imaging of Nearby Galaxies, a legacy project of the Nobeyama 45-m radio telescope.
\emph{Software:} {\sc Astropy} \citep{astropy1}; {\sc emcee} \citep{emcee1}; {\sc linmix} \citep{linmix}; {\sc Matplotlib} \citep{matplotlib}; {\sc NumPy} \citep{numpy}; {\sc Scipy} \citep{scipy}.


\end{acknowledgements}

%
%

\bibliographystyle{aa}
\bibliography{MAIN.bib} 
%
%

\begin{appendix}

\section{Data}
\label{app:data}
The table here will be available only online at CDS
\FloatBarrier 

\begin{table*}[hb]
\caption{Sample properties} 
\label{table:sample}
\begin{adjustbox}{width=\textwidth,center}

\begin{tabular}{lllllllll}

\hline
Name & Sample & $HT$ & $\log(M_{\star}/M_{\odot})$ & $\log(SFR/M_{\odot}/yr)$ & $\log(M_{dust}/M_{\odot})$ & $\log(M_{H2}/M_{\odot})$ & $\log(M_{HI, R25}/M_{\odot})$ & $\log(M_{HI, tot}/M_{\odot})$ \\
\hline
ESO097-013 & Curran+01 & $3.3$ & $10.29 \pm 0.22$ & $1.32 \pm 0.21$ & $6.84 \pm 0.09$ & $8.97 \pm 0.01$ & $8.93 \pm 0.09$ & $9.44 \pm 0.07$ \\
ESO358-015 & MM+22 & $8.7$ & $8.39 \pm 0.16$ & $-1.71 \pm 0.09$ & $4.64 \pm 0.50$ & $<7.28$ & $8.17 \pm 0.05$ & $8.17 \pm 0.05$ \\
ESO358-016 & MM+22 & $7.1$ & $8.43 \pm 0.21$ & $-1.66 \pm 0.14$ & $5.89 \pm 0.42$ & $<7.58$ & $8.14 \pm 0.05$ & $8.14 \pm 0.05$ \\
ESO358-051 & MM+22 & $0.6$ & $8.82 \pm 0.16$ & $-1.11 \pm 0.07$ & $5.66 \pm 0.09$ & $7.14 \pm 0.07$ & $8.13 \pm 0.12$ & $8.13 \pm 0.12$ \\
ESO358-054 & MM+22 & $7.9$ & $8.84 \pm 0.15$ & $-0.98 \pm 0.04$ & $6.06 \pm 0.19$ & $<7.17$ & $8.38 \pm 0.04$ & $8.69 \pm 0.04$ \\
ESO358-060 & MM+22 & $9.8$ & $7.77 \pm 0.31$ & $-1.54 \pm 0.03$ & $4.59 \pm 0.50$ & $<7.28$ & $8.79 \pm 0.01$ & $8.79 \pm 0.01$ \\
ESO358-063 & MM+22 & $6.9$ & $9.89 \pm 0.14$ & $-0.27 \pm 0.15$ & $6.90 \pm 0.07$ & $8.50 \pm 0.02$ & $8.97 \pm 0.04$ & $9.10 \pm 0.04$ \\
IC0010 & COMING & $9.9$ & $8.72 \pm 0.22$ & $-0.52 \pm 0.33$ & $5.02 \pm 0.29$ & $6.48 \pm 0.01$ & $7.88 \pm 0.03$ & $7.88 \pm 0.03$ \\
IC2574 & HERACLES & $8.9$ & $8.72 \pm 0.12$ & $-0.92 \pm 0.03$ & $6.07 \pm 0.20$ & $6.87 \pm 0.04$ & $9.33 \pm 0.09$ & $9.33 \pm 0.09$ \\
IC3392 & VertiCO & $2.4$ & $9.39 \pm 0.07$ & $-1.88 \pm 0.23$ & $5.98 \pm 0.07$ & $7.95 \pm 0.02$ & $7.48 \pm 0.03$ & $7.54 \pm 0.03$ \\
NGC0150 & COMING & $3.4$ & $10.33 \pm 0.14$ & $0.39 \pm 0.06$ & $6.99 \pm 0.03$ & $8.77 \pm 0.01$ & $9.41 \pm 0.04$ & $9.75 \pm 0.04$ \\
NGC0337 & COMING & $6.7$ & $9.76 \pm 0.13$ & $0.30 \pm 0.03$ & $6.88 \pm 0.03$ & $8.50 \pm 0.01$ & $9.18 \pm 0.02$ & $9.60 \pm 0.02$ \\
NGC0520 & COMING & $1.3$ & $10.35 \pm 0.07$ & $0.93 \pm 0.04$ & $7.15 \pm 0.12$ & $9.62 \pm 0.02$ & $9.70 \pm 0.05$ & $9.78 \pm 0.05$ \\
NGC0613 & COMING & $4.0$ & $10.88 \pm 0.13$ & $1.06 \pm 0.10$ & $7.80 \pm 0.06$ & $9.90 \pm 0.03$ & $9.71 \pm 0.03$ & $9.84 \pm 0.03$ \\
NGC0628 & COMING & $5.2$ & $10.15 \pm 0.07$ & $0.38 \pm 0.08$ & $7.58 \pm 0.14$ & $9.25 \pm 0.01$ & $10.00 \pm 0.07$ & $10.31 \pm 0.05$ \\
NGC0891 & COMING & $3.1$ & $10.54 \pm 0.15$ & $0.29 \pm 0.26$ & $7.69 \pm 0.02$ & $9.76 \pm 0.01$ & $9.60 \pm 0.04$ & $9.74 \pm 0.04$ \\
NGC0925 & HERACLES & $7.0$ & $9.73 \pm 0.12$ & $0.06 \pm 0.02$ & $6.97 \pm 0.04$ & $8.35 \pm 0.02$ & $9.72 \pm 0.06$ & $9.93 \pm 0.04$ \\
NGC1068 & Curran+01 & $3.0$ & $10.58 \pm 0.09$ & $1.13 \pm 0.12$ & $7.23 \pm 0.06$ & $9.54 \pm 0.09$ & $8.86 \pm 0.03$ & $8.87 \pm 0.02$ \\
NGC1317 & MM+22 & $0.8$ & $10.49 \pm 0.08$ & $-0.55 \pm 0.15$ & $6.53 \pm 0.04$ & $8.59 \pm 0.01$ & <$8.04$ & $<8.07$ \\
NGC1336 & MM+22 & $0.8$ & $9.92 \pm 0.09$ & $-1.92 \pm 0.36$ & $5.76 \pm 0.50$ & $<7.34$ &  &  \\
NGC1351A & MM+22 & $4.2$ & $9.59 \pm 0.12$ & $-1.06 \pm 0.14$ & $6.64 \pm 0.08$ & $7.78 \pm 0.08$ & $8.58 \pm 0.11$ & $8.71 \pm 0.11$ \\
NGC1365 & MM+22 & $3.2$ & $10.92 \pm 0.14$ & $1.11 \pm 0.17$ & $8.00 \pm 0.15$ & $10.04 \pm 0.01$ & $9.92 \pm 0.03$ & $9.98 \pm 0.03$ \\
NGC1427A & MM+22 & $9.9$ & $9.04 \pm 0.14$ & $-0.82 \pm 0.05$ & $6.02 \pm 0.12$ & $<7.19$ & $8.80 \pm 0.05$ & $9.21 \pm 0.05$ \\
NGC1436 & MM+22 & $2.0$ & $9.96 \pm 0.08$ & $-0.95 \pm 0.22$ & $6.69 \pm 0.06$ & $8.43 \pm 0.02$ & $7.94 \pm 0.10$ & $7.99 \pm 0.10$ \\
NGC1437B & MM+22 & $8.3$ & $8.79 \pm 0.15$ & $-1.64 \pm 0.23$ & $6.03 \pm 0.08$ & $7.26 \pm 0.05$ & $8.15 \pm 0.03$ & $8.33 \pm 0.03$ \\
NGC1569 & COMING & $9.6$ & $8.52 \pm 0.11$ & $-0.04 \pm 0.03$ & $4.95 \pm 0.11$ & $6.58 \pm 0.04$ & $8.30 \pm 0.02$ & $8.31 \pm 0.02$ \\
NGC2146 & COMING & $2.3$ & $11.08 \pm 0.12$ & $1.59 \pm 0.13$ & $7.76 \pm 0.08$ & $10.28 \pm 0.02$ & $9.87 \pm 0.03$ & $10.09 \pm 0.03$ \\
NGC2273 & COMING & $0.9$ & $10.79 \pm 0.11$ & $0.22 \pm 0.16$ & $7.05 \pm 0.14$ & $9.24 \pm 0.01$ & $9.15 \pm 0.02$ & $9.37 \pm 0.02$ \\
NGC2798 & COMING & $1.1$ & $10.14 \pm 0.12$ & $0.70 \pm 0.05$ & $6.80 \pm 0.06$ & $9.13 \pm 0.01$ & $8.97 \pm 0.06$ & $9.14 \pm 0.06$ \\
NGC2841 & COMING & $2.9$ & $11.30 \pm 0.06$ & $0.08 \pm 0.10$ & $7.70 \pm 0.05$ & $9.26 \pm 0.02$ & $10.08 \pm 0.02$ & $10.06 \pm 0.02$ \\
NGC2976 & COMING & $5.2$ & $8.99 \pm 0.08$ & $-0.90 \pm 0.07$ & $6.01 \pm 0.02$ & $7.73 \pm 0.01$ & $8.15 \pm 0.02$ & $8.14 \pm 0.02$ \\
NGC2992 & COMING & $0.9$ & $10.78 \pm 0.07$ & $0.44 \pm 0.08$ & $7.13 \pm 0.03$ & $9.32 \pm 0.01$ & $9.53 \pm 0.06$ & $9.81 \pm 0.06$ \\
NGC2993 & COMING & $1.2$ & $9.69 \pm 0.13$ & $0.74 \pm 0.02$ & $6.77 \pm 0.02$ & $9.13 \pm 0.01$ & $9.34 \pm 0.03$ & $10.05 \pm 0.03$ \\
NGC3034 & COMING & $7.2$ & $10.12 \pm 0.10$ & $0.65 \pm 0.06$ & $6.92 \pm 0.08$ & $9.42 \pm 0.02$ & $8.71 \pm 0.05$ & $9.00 \pm 0.04$ \\
NGC3077 & COMING & $6.1$ & $9.36 \pm 0.11$ & $-0.82 \pm 0.32$ & $5.82 \pm 0.07$ & $7.53 \pm 0.01$ & $8.95 \pm 0.02$ & $8.95 \pm 0.02$ \\
NGC3079 & COMING & $6.4$ & $10.59 \pm 0.10$ & $0.72 \pm 0.10$ & $7.53 \pm 0.09$ & $9.75 \pm 0.02$ & $9.65 \pm 0.04$ & $9.80 \pm 0.05$ \\
NGC3147 & COMING & $3.9$ & $11.21 \pm 0.07$ & $0.90 \pm 0.24$ & $8.34 \pm 0.06$ & $10.36 \pm 0.04$ & $10.00 \pm 0.02$ & $10.13 \pm 0.02$ \\
NGC3184 & HERACLES & $5.9$ & $10.23 \pm 0.07$ & $0.22 \pm 0.06$ & $7.19 \pm 0.06$ & $9.17 \pm 0.05$ & $9.53 \pm 0.02$ & $9.53 \pm 0.02$ \\
NGC3198 & COMING & $5.2$ & $10.16 \pm 0.12$ & $0.23 \pm 0.04$ & $7.23 \pm 0.02$ & $8.91 \pm 0.02$ & $10.00 \pm 0.02$ & $10.02 \pm 0.02$ \\
NGC3338 & COMING & $5.0$ & $10.18 \pm 0.05$ & $0.36 \pm 0.10$ & $7.61 \pm 0.07$ & $9.53 \pm 0.02$ & $9.23 \pm 0.06$ & $10.27 \pm 0.05$ \\
NGC3351 & COMING & $3.1$ & $10.44 \pm 0.09$ & $0.03 \pm 0.10$ & $6.91 \pm 0.04$ & $9.01 \pm 0.01$ & $9.08 \pm 0.05$ & $9.11 \pm 0.04$ \\
NGC3370 & COMING & $5.1$ & $10.06 \pm 0.13$ & $0.48 \pm 0.03$ & $7.31 \pm 0.05$ & $9.01 \pm 0.03$ & $9.43 \pm 0.08$ & $9.80 \pm 0.07$ \\
NGC3437 & COMING & $5.2$ & $10.32 \pm 0.11$ & $0.76 \pm 0.03$ & $7.20 \pm 0.03$ & $9.33 \pm 0.02$ & $9.28 \pm 0.03$ & $9.58 \pm 0.03$ \\
NGC3521 & COMING & $4.0$ & $10.93 \pm 0.08$ & $0.50 \pm 0.17$ & $7.73 \pm 0.03$ & $9.76 \pm 0.03$ & $9.92 \pm 0.03$ & $10.12 \pm 0.02$ \\
NGC3627 & COMING & $3.1$ & $10.83 \pm 0.08$ & $0.47 \pm 0.14$ & $7.44 \pm 0.02$ & $9.74 \pm 0.01$ & $9.11 \pm 0.03$ & $9.12 \pm 0.03$ \\
NGC3628 & COMING & $3.1$ & $10.50 \pm 0.08$ & $0.25 \pm 0.25$ & $7.59 \pm 0.02$ & $9.67 \pm 0.01$ & $9.82 \pm 0.02$ & $10.02 \pm 0.02$ \\
NGC3655 & COMING & $5.0$ & $10.03 \pm 0.12$ & $0.43 \pm 0.04$ & $7.12 \pm 0.02$ & $9.27 \pm 0.02$ & $8.81 \pm 0.03$ & $9.26 \pm 0.02$ \\
NGC3686 & COMING & $4.1$ & $9.92 \pm 0.08$ & $-0.01 \pm 0.04$ & $6.85 \pm 0.04$ & $8.90 \pm 0.03$ & $8.90 \pm 0.02$ & $8.90 \pm 0.02$ \\
NGC3729 & Chung+17 & $1.2$ & $10.32 \pm 0.07$ & $-0.18 \pm 0.13$ & $6.86 \pm 0.05$ & $8.82 \pm 0.04$ & $9.32 \pm 0.02$ & $9.76 \pm 0.02$ \\
NGC3813 & COMING & $3.3$ & $10.12 \pm 0.09$ & $0.43 \pm 0.05$ & $7.13 \pm 0.06$ & $9.04 \pm 0.01$ & $9.00 \pm 0.05$ & $9.42 \pm 0.05$ \\
NGC3938 & COMING & $5.2$ & $10.48 \pm 0.11$ & $0.61 \pm 0.05$ & $7.44 \pm 0.06$ & $9.40 \pm 0.02$ & $9.43 \pm 0.05$ & $9.86 \pm 0.05$ \\
NGC3953 & Chung+17 & $4.0$ & $10.36 \pm 0.12$ & $-0.24 \pm 0.34$ & $7.49 \pm 0.07$ & $8.97 \pm 0.03$ & $9.08 \pm 0.02$ & $9.24 \pm 0.02$ \\
NGC4013 & Chung+17 & $3.1$ & $10.57 \pm 0.06$ & $-0.51 \pm 0.36$ & $7.30 \pm 0.02$ & $9.15 \pm 0.02$ & $9.34 \pm 0.05$ & $9.53 \pm 0.05$ \\
NGC4030 & COMING & $4.0$ & $11.05 \pm 0.09$ & $1.03 \pm 0.05$ & $7.75 \pm 0.05$ & $9.92 \pm 0.02$ & $10.04 \pm 0.08$ & $10.05 \pm 0.08$ \\
\hline
\end{tabular}
\end{adjustbox}
\footnotesize {{\bf Notes.} ID of the target; Survey as in Table~\ref{table:surveys}; $HT$; logarithm of \mstar, $SFR$, $M_{dust}$, \mh, \mhiop, \mhitot.
The uncertainty on \mh\ does not include the systematic uncertainty of $\sim30\%$ due to the adoption of an \aco~\cite{Bolatto13}.}
\end{table*}

\FloatBarrier 

\begin{table*}
\begin{adjustbox}{width=\textwidth,center}
\begin{tabular}{lllllllll}

\hline
Name & Sample & $HT$ & $\log(M_{\star}/M_{\odot})$ & $\log(SFR/M_{\odot}/yr)$ & $\log(M_{dust}/M_{\odot})$ & $\log(M_{H2}/M_{\odot})$ & $\log(M_{HI, R25}/M_{\odot})$ & $\log(M_{HI, tot}/M_{\odot})$ \\
\hline
NGC4045 & COMING & $1.3$ & $10.39 \pm 0.05$ & $0.36 \pm 0.14$ & $7.34 \pm 0.06$ & $9.36 \pm 0.01$ & $9.43 \pm 0.01$ & $9.73 \pm 0.01$ \\
NGC4051 & Chung+17 & $4.0$ & $9.95 \pm 0.08$ & $0.20 \pm 0.13$ & $6.87 \pm 0.11$ & $8.97 \pm 0.02$ & $8.87 \pm 0.04$ & $9.07 \pm 0.04$ \\
NGC4100 & Chung+17 & $4.1$ & $10.44 \pm 0.10$ & $0.30 \pm 0.07$ & $7.26 \pm 0.02$ & $9.10 \pm 0.02$ & $9.40 \pm 0.02$ & $9.61 \pm 0.02$ \\
NGC4102 & Chung+17 & $3.1$ & $10.38 \pm 0.11$ & $0.69 \pm 0.09$ & $6.99 \pm 0.06$ & $9.17 \pm 0.03$ & $8.74 \pm 0.03$ & $8.90 \pm 0.03$ \\
NGC4189 & HeViCS & $5.9$ & $9.91 \pm 0.08$ & $0.04 \pm 0.08$ & $7.05 \pm 0.05$ & $8.96 \pm 0.05$ & $8.90 \pm 0.02$ & $9.17 \pm 0.01$ \\
NGC4192 & HeViCS & $2.6$ & $10.49 \pm 0.09$ & $-0.18 \pm 0.17$ & $7.28 \pm 0.02$ & $8.80 \pm 0.04$ & $9.45 \pm 0.02$ & $9.51 \pm 0.01$ \\
NGC4212 & HeViCS & $4.9$ & $10.14 \pm 0.07$ & $0.10 \pm 0.04$ & $7.03 \pm 0.02$ & $9.05 \pm 0.03$ & $8.76 \pm 0.02$ & $8.97 \pm 0.02$ \\
NGC4214 & COMING & $9.8$ & $8.91 \pm 0.12$ & $-0.73 \pm 0.03$ & $5.82 \pm 0.08$ & $7.65 \pm 0.02$ & $8.62 \pm 0.02$ & $8.62 \pm 0.02$ \\
NGC4216 & VertiCO & $3.0$ & $10.92 \pm 0.06$ & $-0.89 \pm 0.31$ & $7.32 \pm 0.04$ & $8.72 \pm 0.02$ & $9.15 \pm 0.02$ & $9.17 \pm 0.02$ \\
NGC4217 & Chung+17 & $3.0$ & $10.54 \pm 0.08$ & $0.20 \pm 0.27$ & $7.47 \pm 0.02$ & $9.35 \pm 0.02$ & $8.86 \pm 0.04$ & $8.87 \pm 0.04$ \\
NGC4236 & HERACLES & $8.0$ & $9.08 \pm 0.15$ & $-0.69 \pm 0.04$ & $6.37 \pm 0.04$ & $6.38 \pm 0.01$ & $9.34 \pm 0.03$ & $9.46 \pm 0.03$ \\
NGC4254 & HERACLES & $5.2$ & $10.13 \pm 0.07$ & $0.71 \pm 0.04$ & $7.35 \pm 0.02$ & $9.64 \pm 0.03$ & $9.48 \pm 0.03$ & $9.54 \pm 0.01$ \\
NGC4258 & COMING & $4.0$ & $10.75 \pm 0.06$ & $0.01 \pm 0.12$ & $7.49 \pm 0.15$ & $9.27 \pm 0.02$ & $9.66 \pm 0.03$ & $9.83 \pm 0.03$ \\
NGC4299 & VertiCO & $8.4$ & $9.49 \pm 0.11$ & $0.02 \pm 0.03$ & $6.35 \pm 0.07$ & $7.65 \pm 0.09$ & $8.72 \pm 0.02$ & $9.25 \pm 0.02$ \\
NGC4302 & COMING & $5.3$ & $10.43 \pm 0.07$ & $-0.38 \pm 0.14$ & $7.17 \pm 0.03$ & $9.09 \pm 0.02$ & $9.20 \pm 0.01$ & $9.21 \pm 0.01$ \\
NGC4321 & HERACLES & $4.0$ & $10.69 \pm 0.10$ & $0.78 \pm 0.08$ & $7.57 \pm 0.05$ & $9.81 \pm 0.02$ & $9.46 \pm 0.02$ & $9.50 \pm 0.02$ \\
NGC4351 & VertiCO & $2.8$ & $9.16 \pm 0.10$ & $-0.97 \pm 0.07$ & $5.98 \pm 0.09$ & $7.52 \pm 0.04$ & $8.00 \pm 0.01$ & $8.31 \pm 0.01$ \\
NGC4380 & VertiCO & $2.4$ & $10.37 \pm 0.08$ & $-0.63 \pm 0.23$ & $6.99 \pm 0.03$ & $8.69 \pm 0.01$ & $8.40 \pm 0.01$ & $8.40 \pm 0.01$ \\
NGC4388 & HeViCS & $2.8$ & $10.33 \pm 0.07$ & $0.14 \pm 0.10$ & $6.79 \pm 0.04$ & $8.87 \pm 0.02$ & $8.73 \pm 0.01$ & $8.73 \pm 0.01$ \\
NGC4396 & VertiCO & $6.8$ & $9.35 \pm 0.11$ & $-0.56 \pm 0.07$ & $6.83 \pm 0.06$ & $7.94 \pm 0.04$ & $8.99 \pm 0.02$ & $8.99 \pm 0.02$ \\
NGC4402 & HeViCS & $3.2$ & $9.78 \pm 0.14$ & $-0.42 \pm 0.21$ & $6.81 \pm 0.02$ & $8.89 \pm 0.02$ & $8.45 \pm 0.01$ & $8.45 \pm 0.01$ \\
NGC4419 & Chung+17 & $1.2$ & $10.43 \pm 0.06$ & $-0.57 \pm 0.30$ & $6.75 \pm 0.07$ & $8.96 \pm 0.11$ & $8.00 \pm 0.03$ & $8.01 \pm 0.03$ \\
NGC4424 & VertiCO & $0.9$ & $8.61 \pm 0.09$ & $-2.06 \pm 0.24$ & $4.90 \pm 0.07$ & $7.04 \pm 0.01$ & $7.15 \pm 0.01$ & $7.15 \pm 0.01$ \\
NGC4438 & HeViCS & $0.6$ & $10.46 \pm 0.06$ & $-0.79 \pm 0.14$ & $6.45 \pm 0.02$ & $8.47 \pm 0.06$ & $8.38 \pm 0.01$ & $8.38 \pm 0.01$ \\
NGC4450 & VertiCO & $2.4$ & $10.75 \pm 0.05$ & $-0.63 \pm 0.11$ & $6.93 \pm 0.05$ & $8.59 \pm 0.01$ & $8.48 \pm 0.01$ & $8.47 \pm 0.01$ \\
NGC4501 & HeViCS & $3.3$ & $10.74 \pm 0.10$ & $0.27 \pm 0.08$ & $7.28 \pm 0.03$ & $9.39 \pm 0.02$ & $8.93 \pm 0.02$ & $8.95 \pm 0.02$ \\
NGC4522 & HeViCS & $5.9$ & $9.45 \pm 0.10$ & $-0.57 \pm 0.06$ & $6.34 \pm 0.05$ & $8.34 \pm 0.01$ & $8.56 \pm 0.01$ & $8.55 \pm 0.01$ \\
NGC4527 & COMING & $4.0$ & $10.65 \pm 0.08$ & $0.38 \pm 0.08$ & $7.43 \pm 0.06$ & $9.59 \pm 0.02$ & $9.66 \pm 0.04$ & $9.75 \pm 0.03$ \\
NGC4532 & VertiCO & $9.7$ & $9.24 \pm 0.12$ & $-0.12 \pm 0.02$ & $6.35 \pm 0.03$ & $7.83 \pm 0.02$ & $8.61 \pm 0.03$ & $9.08 \pm 0.02$ \\
NGC4535 & HeViCS & $5.0$ & $10.47 \pm 0.10$ & $0.39 \pm 0.05$ & $7.47 \pm 0.03$ & $9.39 \pm 0.02$ & $9.58 \pm 0.05$ & $9.63 \pm 0.04$ \\
NGC4536 & COMING & $4.3$ & $10.29 \pm 0.11$ & $0.50 \pm 0.05$ & $7.14 \pm 0.02$ & $9.11 \pm 0.01$ & $9.58 \pm 0.02$ & $9.65 \pm 0.03$ \\
NGC4559 & COMING & $6.0$ & $9.57 \pm 0.11$ & $-0.16 \pm 0.04$ & $6.75 \pm 0.08$ & $8.64 \pm 0.01$ & $9.51 \pm 0.04$ & $9.71 \pm 0.03$ \\
NGC4561 & VertiCO & $7.2$ & $9.61 \pm 0.12$ & $-0.06 \pm 0.05$ & $6.37 \pm 0.09$ & $7.55 \pm 0.24$ & $8.94 \pm 0.02$ & $9.56 \pm 0.01$ \\
NGC4569 & HERACLES & $2.4$ & $10.50 \pm 0.06$ & $-0.73 \pm 0.24$ & $6.89 \pm 0.02$ & $9.18 \pm 0.02$ & $8.57 \pm 0.06$ & $8.57 \pm 0.05$ \\
NGC4579 & COMING & $2.8$ & $11.11 \pm 0.05$ & $-0.04 \pm 0.22$ & $7.41 \pm 0.06$ & $9.50 \pm 0.01$ & $8.96 \pm 0.03$ & $8.96 \pm 0.03$ \\
NGC4580 & VertiCO & $1.6$ & $10.10 \pm 0.05$ & $-0.83 \pm 0.24$ & $6.79 \pm 0.03$ & $8.60 \pm 0.02$ & $7.63 \pm 0.05$ & $7.69 \pm 0.04$ \\
NGC4625 & HERACLES & $8.7$ & $8.90 \pm 0.09$ & $-1.02 \pm 0.06$ & $6.07 \pm 0.22$ & $7.20 \pm 0.02$ & $8.00 \pm 0.03$ & $8.68 \pm 0.03$ \\
NGC4631 & HERACLES & $6.5$ & $10.25 \pm 0.08$ & $0.56 \pm 0.03$ & $7.21 \pm 0.03$ & $9.06 \pm 0.02$ & $9.77 \pm 0.04$ & $10.09 \pm 0.03$ \\
NGC4651 & VertiCO & $5.1$ & $10.67 \pm 0.06$ & $0.44 \pm 0.04$ & $7.37 \pm 0.06$ & $8.98 \pm 0.01$ & $9.86 \pm 0.04$ & $9.90 \pm 0.04$ \\
NGC4654 & Chung+17 & $5.9$ & $10.18 \pm 0.10$ & $0.40 \pm 0.08$ & $7.16 \pm 0.07$ & $9.03 \pm 0.06$ & $9.36 \pm 0.03$ & $9.39 \pm 0.03$ \\
NGC4666 & COMING & $5.0$ & $10.59 \pm 0.08$ & $0.65 \pm 0.05$ & $7.32 \pm 0.06$ & $9.45 \pm 0.01$ & $9.61 \pm 0.08$ & $9.61 \pm 0.08$ \\
NGC4689 & Chung+17 & $4.7$ & $10.13 \pm 0.11$ & $-0.33 \pm 0.26$ & $6.97 \pm 0.03$ & $8.94 \pm 0.07$ & $8.63 \pm 0.02$ & $8.63 \pm 0.02$ \\
NGC4698 & VertiCO & $1.6$ & $10.86 \pm 0.06$ & $-1.05 \pm 0.27$ & $7.19 \pm 0.14$ & $8.11 \pm 0.02$ & $9.53 \pm 0.04$ & $9.55 \pm 0.03$ \\
NGC4713 & VertiCO & $6.8$ & $9.33 \pm 0.06$ & $-0.02 \pm 0.05$ & $6.77 \pm 0.07$ & $8.19 \pm 0.02$ & $8.73 \pm 0.02$ & $9.38 \pm 0.02$ \\
NGC4725 & HERACLES & $2.2$ & $10.87 \pm 0.07$ & $-0.03 \pm 0.05$ & $7.39 \pm 0.05$ & $8.72 \pm 0.02$ & $9.60 \pm 0.03$ & $9.70 \pm 0.03$ \\
NGC4736 & HERACLES & $2.3$ & $10.39 \pm 0.07$ & $-0.26 \pm 0.15$ & $6.39 \pm 0.03$ & $8.60 \pm 0.01$ & $8.54 \pm 0.02$ & $8.55 \pm 0.02$ \\
NGC4808 & VertiCO & $5.9$ & $9.71 \pm 0.12$ & $0.18 \pm 0.03$ & $6.96 \pm 0.03$ & $8.70 \pm 0.01$ & $9.15 \pm 0.02$ & $9.70 \pm 0.02$ \\
NGC4945 & Curran+01 & $6.1$ & $10.14 \pm 0.18$ & $0.81 \pm 0.24$ & $7.31 \pm 0.02$ & $9.52 \pm 0.08$ & $8.91 \pm 0.01$ & $8.92 \pm 0.01$ \\
NGC5033 & Curran+01 & $5.0$ & $10.60 \pm 0.10$ & $0.47 \pm 0.10$ & $7.56 \pm 0.14$ & $9.46 \pm 0.09$ & $9.84 \pm 0.03$ & $10.04 \pm 0.03$ \\
NGC5055 & COMING & $4.0$ & $10.77 \pm 0.08$ & $0.39 \pm 0.13$ & $7.65 \pm 0.08$ & $9.71 \pm 0.01$ & $9.86 \pm 0.02$ & $9.86 \pm 0.02$ \\
NGC5248 & COMING & $4.0$ & $10.18 \pm 0.11$ & $0.32 \pm 0.07$ & $7.19 \pm 0.05$ & $9.39 \pm 0.01$ & $9.40 \pm 0.04$ & $9.45 \pm 0.03$ \\
NGC5364 & COMING & $4.0$ & $10.13 \pm 0.10$ & $0.05 \pm 0.10$ & $7.33 \pm 0.07$ & $9.00 \pm 0.01$ & $9.48 \pm 0.06$ & $9.52 \pm 0.06$ \\
NGC5426 & COMING & $5.0$ & $10.40 \pm 0.11$ & $0.72 \pm 0.03$ & $7.64 \pm 0.16$ & $9.54 \pm 0.01$ & $10.27 \pm 0.08$ & $10.27 \pm 0.07$ \\
NGC5427 & COMING & $5.0$ & $10.38 \pm 0.09$ & $0.67 \pm 0.03$ & $7.40 \pm 0.13$ & $9.60 \pm 0.01$ & $9.81 \pm 0.02$ & $9.81 \pm 0.01$ \\
NGC5457 & HERACLES & $5.9$ & $10.15 \pm 0.06$ & $0.68 \pm 0.02$ & $7.67 \pm 0.05$ & $9.33 \pm 0.02$ & $10.11 \pm 0.02$ & $10.12 \pm 0.02$ \\
NGC5474 & HERACLES & $6.1$ & $9.00 \pm 0.05$ & $-0.80 \pm 0.09$ & $6.43 \pm 0.15$ & $6.50 \pm 0.02$ & $8.34 \pm 0.07$ & $8.99 \pm 0.06$ \\
NGC5480 & COMING & $4.9$ & $10.08 \pm 0.11$ & $0.30 \pm 0.10$ & $7.02 \pm 0.04$ & $9.31 \pm 0.01$ & $8.92 \pm 0.05$ & $9.25 \pm 0.05$ \\
NGC5713 & COMING & $4.0$ & $9.86 \pm 0.13$ & $0.30 \pm 0.04$ & $6.75 \pm 0.03$ & $9.03 \pm 0.01$ & $8.84 \pm 0.05$ & $9.34 \pm 0.05$ \\
NGC5907 & COMING & $5.2$ & $10.95 \pm 0.08$ & $0.48 \pm 0.05$ & $7.85 \pm 0.02$ & $9.58 \pm 0.01$ & $10.11 \pm 0.06$ & $10.30 \pm 0.06$ \\
NGC6814 & Curran+01 & $4.0$ & $10.91 \pm 0.11$ & $0.45 \pm 0.08$ & $7.54 \pm 0.09$ & $9.19 \pm 0.08$ & $9.36 \pm 0.04$ & $9.71 \pm 0.04$ \\
NGC6946 & HERACLES & $5.9$ & $10.43 \pm 0.14$ & $0.85 \pm 0.03$ & $7.48 \pm 0.08$ & $9.68 \pm 0.02$ & $9.73 \pm 0.02$ & $9.73 \pm 0.02$ \\
NGC7331 & COMING & $3.9$ & $10.97 \pm 0.09$ & $0.63 \pm 0.13$ & $7.79 \pm 0.02$ & $9.79 \pm 0.01$ & $9.93 \pm 0.02$ & $10.08 \pm 0.02$ \\
NGC7625 & COMING & $1.0$ & $10.08 \pm 0.11$ & $0.27 \pm 0.13$ & $6.96 \pm 0.09$ & $9.20 \pm 0.01$ & $8.87 \pm 0.01$ & $9.38 \pm 0.03$ \\
NGC7714 & COMING & $3.1$ & $9.94 \pm 0.10$ & $0.78 \pm 0.02$ & $6.36 \pm 0.17$ & $9.18 \pm 0.01$ & $9.65 \pm 0.09$ & $9.65 \pm 0.09$ \\
NGC7715 & COMING & $9.6$ & $9.07 \pm 0.02$ & $-0.82 \pm 0.02$ & $5.75 \pm 0.50$ & $8.92 \pm 0.02$ & $9.83 \pm 0.07$ & $9.83 \pm 0.07$ \\
NGC7721 & COMING & $4.9$ & $10.49 \pm 0.09$ & $0.53 \pm 0.08$ & $7.74 \pm 0.09$ & $9.13 \pm 0.01$ & $9.72 \pm 0.04$ & $10.08 \pm 0.04$ \\
UGC05720 & COMING & $9.8$ & $9.49 \pm 0.14$ & $0.19 \pm 0.03$ & $6.20 \pm 0.05$ & $7.91 \pm 0.05$ & $8.51 \pm 0.02$ & $8.89 \pm 0.02$ \\
\hline
\end{tabular}
\end{adjustbox}
\footnotesize {{\bf Notes.} ID of the target; Survey as in Table~\ref{table:surveys}; $HT$; logarithm of \mstar, $SFR$, $M_{dust}$, \mh, \mhiop, \mhitot.
The uncertainty on \mh\ does not include the systematic uncertainty of $\sim30\%$ due to the adoption of an \aco~\cite{Bolatto13}.}
\end{table*}
\FloatBarrier 
%

\section{Comparison with \mh\ from C20}
\label{sec:motivation}
C20 presented \mh\ estimates for 255 late-type DustPedia galaxies. Due to the unavailability of maps for most galaxies, the total \lco (and thus the total \mh) is estimated from single-pointing observation at the galactic center, assuming an exponential model for the disk.
The empirical model is that of  \cite{Boselli14}, which is a 3D extension of the model by \cite{Lisenfeld11}; it assumes that the molecular gas density follows an exponential distribution, both along the radius (with scale-radius $R_{CO}$) and above the plane (with scale-height $z_{CO}$), and that both scale-lengths are directly proportional to the optical scale-radius ($R_{D25}$).
In particular, C20 assumed $R_{CO}/R_{D25}=0.17$, as found by \citet{Casasola17} on the largest DustPedia galaxies, and $z_{CO}/R_{D25}=0.01$ as derived by \citet{Boselli14} from observations and models of edge-on galaxies.

\begin{figure}[t]
	\includegraphics[width = \columnwidth, keepaspectratio=True]{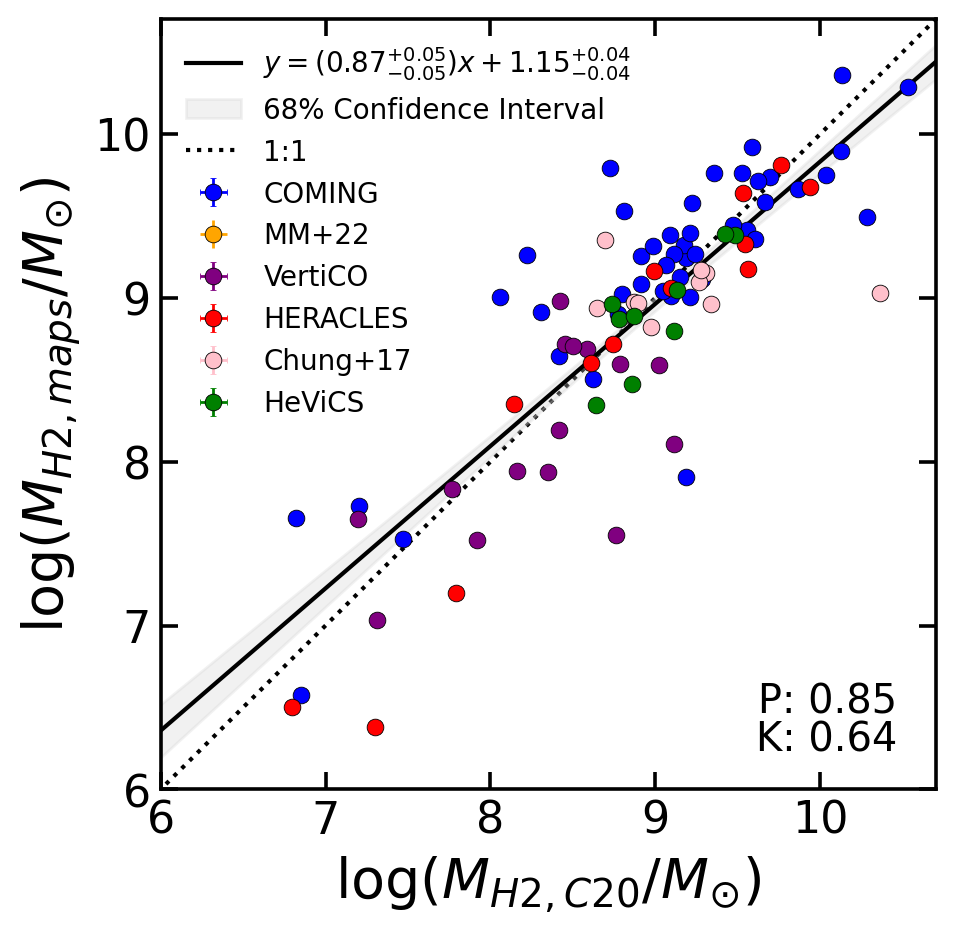}
	\caption{Comparison of the best-estimates of \mh\ presented in this work (y-axes) and retrieved by C20 (x-axes).
	Objects are color-coded based on the survey from which data were taken.
    Correlation coefficients (namely, $P$ and $K$) are reported in the bottom right corner.
    \label{fig:appC20Za}
    }
\end{figure}

There are 103 galaxies in common between C20 and this work (among which 8 have only upper limits of \mh\ from the former and one from the latter).
The comparison between \mh\ estimates for the galaxies in common is shown in  Fig.~\ref{fig:appC20Za}. There is an overall agreement between the two sets of estimates, but a significant scatter ($\sim0.35\pm0.04$ dex) is visible.
We investigated the objects showing a large variation in the estimate of \mh\ from C20 to this work for any selection effect.
In particular, we focused on the presence of AGN activity, and any trend with the Hubble stage.
Regarding the former, AGN are usually associated with an increased molecular gas emission in the central region (e.g., \citealt{SolomonVandenBout05, Ellison21}), which may lead to an overestimation of the total \mh\ assuming the model by \cite{Boselli14}.
Using the classification of active galaxies provided in C20, we did not find any relation between nuclear activity and the object deviating more in \mh\ estimate.
Indeed, previous studies on local AGN did not find any conclusive results on the systematic difference in terms of \mh\ content between AGN and non-AGN galaxies at galactic scales, even if increased global SFR (e.g., \citealt{Salvestrini20, Salvestrini22}) and SFR surface densities have been observed in the local Universe (e.g., \citealt{Casasola15}).
Regarding the latter, even if all Hubble stages are not evenly represented in our sample, we did not observe any correlation between the Hubble stage and the offset in \mh.
We found it more likely that deviation from the average exponential model is due to the presence of unresolved structures (at the scales of the mm observations available) like bars, which may lead to an increased concentration of molecular gas in the central part of the galaxy.

The differences shown in Fig.~\ref{fig:appC20Za} are likely driven by the model assumption in extrapolating the global \mh: although the exponential decreasing disk model reproduces well the average radial distribution of the molecular gas, as validated by several literature works (e.g., \citealt{Lisenfeld11, Casasola17}), it remains an approximation that may tamper the global \mh\ determination of single galaxies.
This issue is discussed in section \ref{sec:rad_profile}.

\begin{figure*}[ht]
        \centering
	\includegraphics[width = 0.85\textwidth, keepaspectratio=True]{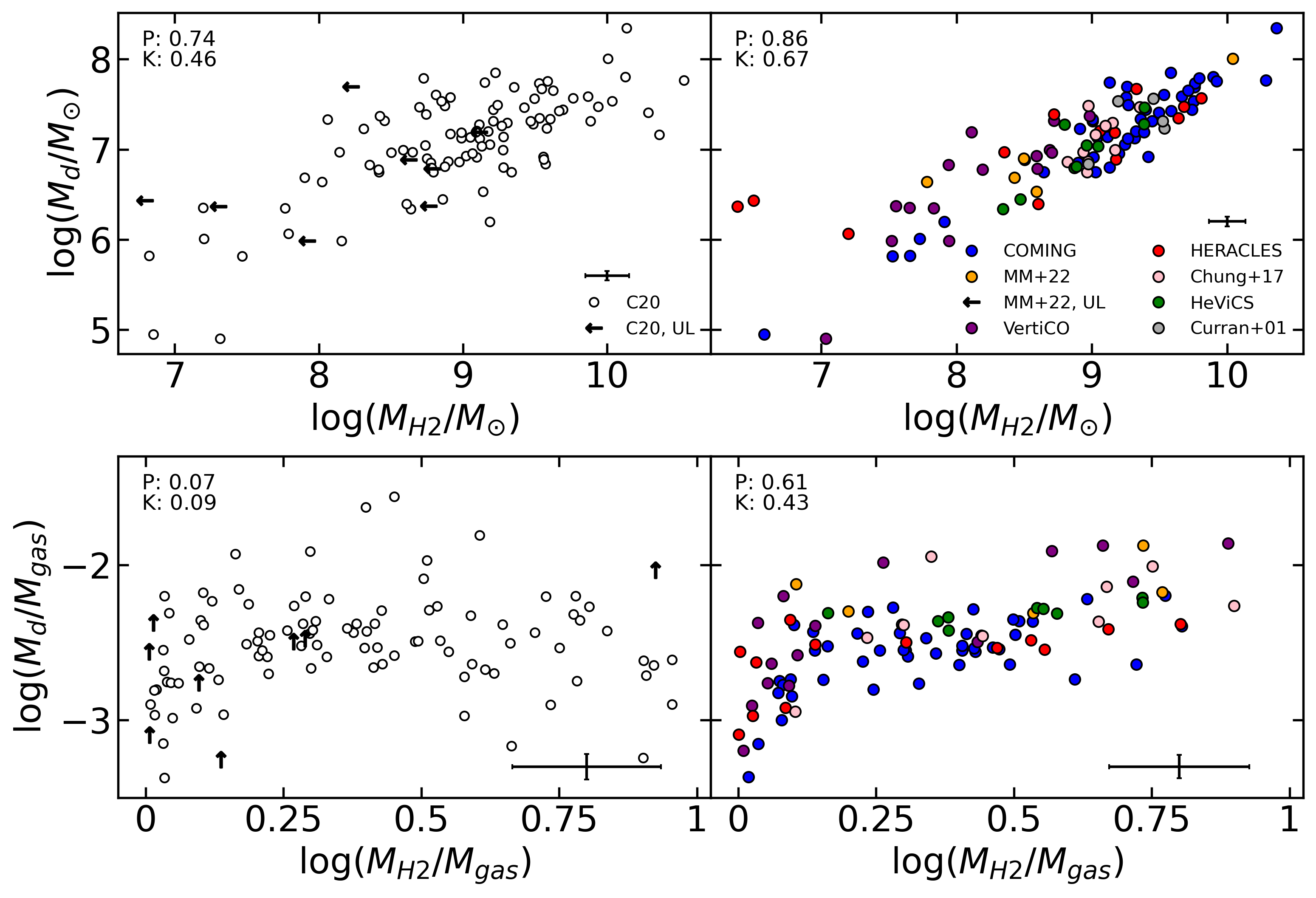}
	\caption{Effect of the adoption of \mh\ from C20 or from this work. 
        Top row: Dust vs molecular gas mass, adopting the \mh\ estimate from C20 and Table \ref{table:sample} (left and right panel, respectively).
	The correlation coefficients (namely, $P$ and $K$) are reported in the bottom left corner of both panels.
        In the right panel, objects are color-coded on the base of the survey from which data were taken.
        \mh\ upper limits are marked as shown in the legends.
        Bottom row: DTGR vs molecular fraction, with \mh\ as in the top row.
        Upper limits and galaxies are shown as described in the corresponding upper panels.}
	\label{fig:motivation}
\end{figure*}

In Fig.~\ref{fig:motivation} we showed the impact of using the new map-derived \mh\ estimates on the analysis of scaling relations. The top row
(analogous to Fig.~3 in C20) presents the dust mass as a function of the molecular gas mass,
for the C20 \mh estimates (left) and for the estimates of this work (right). We remind that
the the dust mass \mdust\ is retrieved from the same DustPedia archive both in this work and C20.
In the figure, we also indicated the quality of the correlation by using the $P$ (this was used by C20) and the non-parametric CC of Kendall.
A major improvement provided by the adoption of \mh\ obtained from global mapping of the CO emission results is a tighter relation, with a significant increase in all CCs. 
When using \mh\ detections from C20 (left panel), the correlation has, e.g., a $P=0.74$ (0.72 for the full sample of 252 galaxies analyzed by C20). Instead, it is R=0.86 when using the \mh\ detections from the mapping observations we collected in this work. 
It becomes $P=0.84$ when using the 115 galaxies with CO detection of Table~\ref{table:sample}.
$P$ increases when moving from the \mdust-\mh\ relation shown in the left panel to that in the right panel.
This indicates that the scaling relation that uses the \mh\ of this work shows a lower dispersion, due to a stronger monotonic trend.

The use of full map-based \mh\ estimates might
also prevents the insurgence of spurious trends. 
An example is shown in the bottom row of Fig.~\ref{fig:motivation}
(analogous to Fig.~14 in C20), where the 
dust-to-total gas (DTGR) is plotted versus 
the molecular gas fraction (the C20 $M_{HI, R25}$ values are used for atomic gas). On their full sample, C20 found a puzzling trend of DTGR increasing with the molecular gas fraction, reaching a maximum, and then decreasing for larger values of the fraction (several possible explanation for the trend are proposed in their section 5.5). The same ``hill-like'' shape can be seen for the subset of
C20 estimates are considered here (left panel).
The trend, however, is not present when we used the \mh\ from Table~\ref{table:sample} (right panel): the DTGR shows a shallow increase ($P=0.61$ and $K=0.43$, respectively) at large molecular-to-total gas ratios ($>0.1$).
The ``hill-like'' shape found by C20 is likely a consequence of the fact that some of the \mh\ provided by C20 for \hh-dominated galaxies are overestimated, thus making the DTGR smaller at increasing dominance of the molecular phase over the atomic one.
\FloatBarrier 

\section{Molecular gas masses and metallicity}
\label{app:gas_mass_Z}
\begin{figure}[ht]
	\includegraphics[width =\columnwidth, keepaspectratio=True]{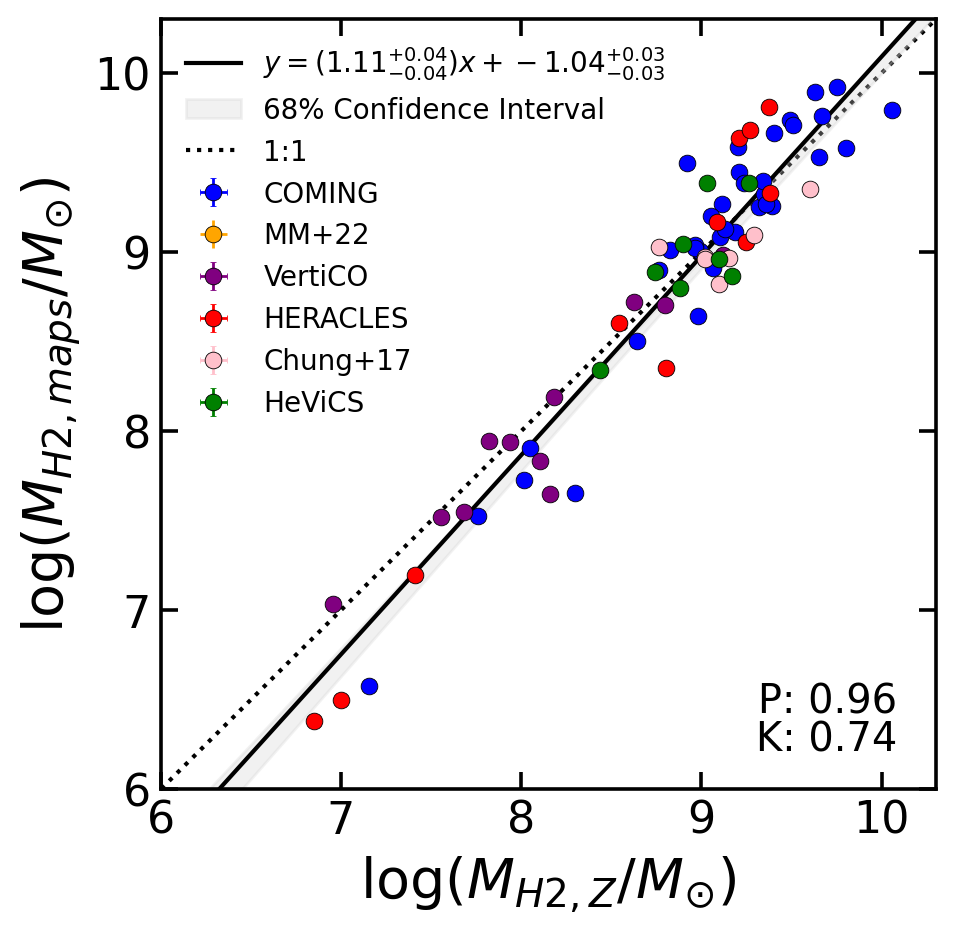}
	\caption{
    Comparison of \mh\ derived with a constant Milky-Way like \aco\ (y-axes) and derived with a \aco~based on the prescription by \cite{Amorin16} (x-axes).
    Correlation coefficients (namely, $P$ and $K$) are reported in the bottom right corner.
    \label{fig:appC20Z}
    }
\end{figure}

The CO-to-H$_{\rm 2}$ conversion factor, $\alpha_{CO}$, strongly depends on the physical conditions (e.g., the presence of highly ionizing radiation fields) and composition (e.g., gas-phase metallicity) of the ISM (see \citealt{Bolatto13} for a detailed review).
In section \ref{sec:results}, we present the results of our analysis which are based on the assumption of a constant $\alpha_{CO}=3.26$ \uaco\ to derive \mh.
However, several literature works suggested a strong dependence of the \aco\ on the gas-phase metallicity (e.g., \citealt{Bolatto13, Accurso17}).
In the right panel of Fig.~\ref{fig:appC20Z}, we compared the \mh\ presented in Sect.~\ref{sec:co_maps} with those derived with a \aco\ dependent on the gas-phase metallicity.
Among the different recipes for the gas-phase metallicity for DustPedia galaxies provided by \cite{DeVis19}, we opted for that by \cite{PP04}, which was available for 95 (83) out of 121 (108) galaxies in the present sample (in the sample with Hubble type $\leq7$; \lco upper limits are not included).
We then adopted the prescription by \cite{Amorin16}, as in C20.
In Fig.~\ref{fig:appC20Z}, the distribution of \mh\ obtained with a metallicity-dependent \aco (y-axes) are almost systematically larger ($\sim 1.5-3 \times$) than those obtained with a constant \aco.
We tested the impact of the metallicity-dependent \aco assumption on the scaling relations presented in section \ref{sec:results}, but we did not find any significant deviation from the trend obtained with a constant \aco, apart from relatively lower $\delta_{intr}$, of the order of few 0.01~dex.
This lower intrinsic dispersion is likely driven by the reduction of the sample size ($\sim86\%$ of the present one) rather than more accurate \mh\ measurements.

%
\section{The gKS dependence on the SFR recipe}
\label{app:gKS_multiSFR}
\begin{figure*}[ht]
        \centering
	\includegraphics[width = \textwidth, keepaspectratio=True]{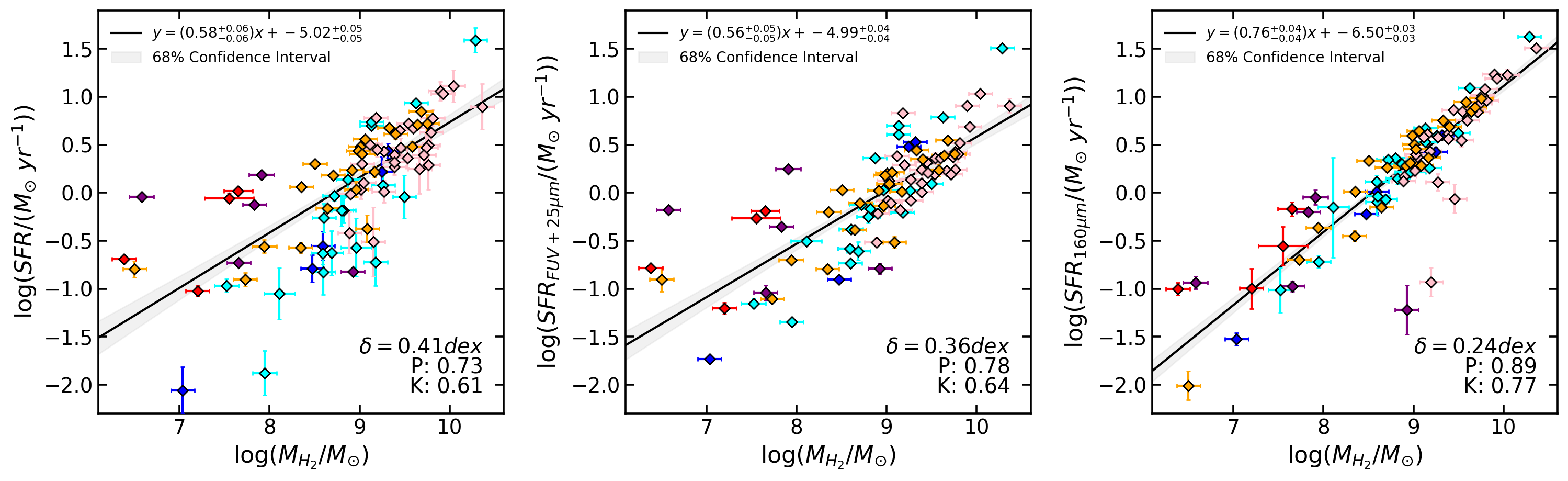}
	\caption{SFR vs molecular gas mass.
    Different SFR estimates vs \mh: SFR derived from the SED fitting with CIGALE (\citealt{Nersesian19}; left panel), and from the hybrid FUV$+$24$\mu$m recipe (\citealt{Leroy19}; right panel]).
    The 85 galaxies are color-coded with the morphological type.
    The best-fit parameters for each scaling relation are reported in the upper-left corner, while the best-fit line is represented with the black solid line.
    The intrinsic dispersion ($\delta_{intr}$) is reported in the bottom-right corner, followed by the correlation coefficients (namely, $P$ and $K$).}
    \label{fig:multiSFR}
\end{figure*}

In section \ref{sec:MGMS_KS_MGD}, we presented the gKs relation for the entire sample of galaxies.
The relatively tight correlation represents the physical process (i.e., the star formation activity) that turns the molecular gas into new stars.
However, different approaches allow us to estimate the SFR based on different tracers and calibrated in diverse environments (both Galactic and extragalactic; e.g., see \citealt{KennicuttEvans12, Figueira22}).
Here, based on the wealth of multi-wavelength data of the DustPedia archive, we tested the consistency of the SFR estimates by \cite{Nersesian19} from the global SED fitting as a good proxy of the ongoing star formation activity to study the gKS.
To this end, we considered two independent SFR estimates by using the prescriptions by \cite{KennicuttEvans12}: \emph{i)} the far-UV luminosity corrected for obscuration by using the mid-IR luminosity at 24\micron; the 160\micron~luminosity as a proxy of the peak of the dust reprocessed emission in the far-IR.
All three prescriptions allowed us to determine the SFR averaged over a timescale of $\sim100$ Myr. 
We referred to \cite{KennicuttEvans12} for a detailed discussion on the timescales probed by different SFR diagnostics. 
For homogeneity reasons, the analysis was limited to 85 sources out of the 121 objects from the sample presented in section \ref{sec:MGMS_KS_MGD}, since we required each galaxy to be observed in all three bands (namely, far-UV, at 24\micron, and at 160\micron).
No upper limits on \mh\ were present in this subsample.
In Fig.~\ref{fig:multiSFR}, we presented the correlation between \mh\ (y-axes) and the SFR estimates (x-axes) derived from the SED fitting (SFR$_{SED}$, left panel), with hybrid UV+IR proxy (SFR$_{FUV+24 \mu m}$, central panel), and from the far-IR luminosity at 160\micron~(SFR$_{160 \mu m}$, right panel).
The slopes of the relations shown in the first two panels from the left of Fig.~\ref{fig:multiSFR} are fully consistent within uncertainties ($0.58\pm0.06$ and $0.56\pm0.06$, respectively), while the SFR$_{FUV+24 \mu m}$-\mh\ relation has a steeper slope ($0.76\pm0.04$; see Appendix~\ref{app:corr_coeff} and Table~\ref{table:corr} for best-fit parameters).
Furthermore, the relation between SFR$_{160 \mu m}$ and \mh\ is the strongest in terms of CCs, and with the lowest intrinsic scatter ($\delta_{intr}=0.24$) among the three relations presented in Fig.~\ref{fig:multiSFR}.
While the SFRs estimated from the SED fitting and with the hybrid UV-IR proxy show relatively strong CCs with the molecular gas mass (e.g., $P=0.73$ and 0.78, and $K=0.61$ and $0.64$, respectively), the \mh-SFR$_{160 \mu m}$ is characterized by a strong correlation (e.g., $P=0.89$, $K= 0.7$, respectively).
This tight trend results from the physical connection between the molecular gas reservoir and the dust heated by the emission from young massive stars around star-forming regions (e.g., \citealt{Calzetti10}, and reference therein).
Alternatively, the emission at 160 $\mu$m is proportional to the dust content, since the emission close to the peak of the far-IR bump is associated with the dust heated by the interstellar radiation field (e.g., \citealt{Bianchi22b}).
The tight correlation shown in the right panel of Fig.~\ref{fig:multiSFR} can represent the connection between \mh\ and \mdust.
This suggests caution in using SFR$_{160 \mu m}$ as a proxy for SFR. 
On the other hand, the SFR$_{FUV+24 \mu m}$ leads to results generally consistent with those from the global SED fit of a galaxy.
We further discussed the relation between the far-IR emission and the molecular gas content in section~\ref{sec:photometry}.
Regarding the SFR estimate derived from the SED fitting, the dispersion of the relation with \mh\ is likely driven in the first case by potential degeneracy between different models (e.g., dusty star formation and older stellar population), which can be difficult to disentangle especially in the case of low luminosity regimes.
Indeed, this is likely the case of the sources with $HT>7$ that lie in the central lower region of the left panel of Fig.~\ref{fig:multiSFR}, which likely have their SFR overestimates from the SED fitting, compared to that derived from the 160$\mu$m emission.
Given their morphological classification, we expected them to have relatively low SFR and a larger population of old stars.
The same issue is likely affecting the correlation in the central panel of Fig.~\ref{fig:multiSFR}, where the UV-corrected emission is likely overestimated. 
Here, we showed that the dispersion of the gKS can be reduced by adopting an accurate prescription for the SFR, which can be particularly important in the case of galaxies spanning a wide range of morphological types.

\FloatBarrier 
\onecolumn
\section{Results of the analysis of scaling relations}
\label{app:corr_coeff}

\begin{table*}[ht]
\caption{2D Scaling relations}
\label{table:corr}      
\centering          
\begin{tabular}[\textwidth]{l l l l l l l l l}     
\hline     
    Relation                &   Sample  & UL   & $\alpha$ & $\beta$ & $\delta_{intr}$ & P &   K  &   Fig. \\ 
    (1)                     &   (2)     & (3) &  (4)     &   (5)   &     (6)         &   (7) &   (9)  &  (10) \\ 
\hline
$M_{HI, R25}$--$M_{dust}$  &  120  &  1  &  $ 0.80^{+0.06}_{-0.06 }$  &  $ 3.52^{+0.04}_{-0.04 }$  &  $ 0.37^{+0.03}_{-0.03 }$  &  0.72  &  0.60  &  \ref{fig:mdust_mgas} \\
$M_{HI, tot}$--$M_{dust}$  &  120  &  1  &  $ 0.77^{+0.07}_{-0.07 }$  &  $ 3.88^{+0.04}_{-0.04 }$  &  $ 0.43^{+0.04}_{-0.03 }$  &  0.71  &  0.55  &  \ref{fig:mdust_mgas} \\
$M_{HI, R25}$--$M_{\star}$  &  120  &  0  &  $ 0.93^{+0.05}_{-0.05 }$  &  $ 3.63^{+0.03}_{-0.03 }$  &  $ 0.29^{+0.03}_{-0.02 }$  &  0.89  &  0.65  &  \ref{fig:mdust_mgas} \\
$M_{H2}$--$M_{dust}$  &  120  &  5  &  $ 1.25^{+0.06}_{-0.06 }$  &  $ 0.08^{+0.04}_{-0.04 }$  &  $ 0.38^{+0.03}_{-0.03 }$  &  0.86  &  0.66  &  \ref{fig:mdust_mgas} \\
$M_{gas}$--$M_{dust}$  &  120  &  6  &  $ 0.93^{+0.04}_{-0.04 }$  &  $ 2.87^{+0.02}_{-0.02 }$  &  $ 0.21^{+0.02}_{-0.02 }$  &  0.86  &  0.70  &  \ref{fig:mdust_mgas} \\
$M_{gas, tot}$--$M_{dust}$  &  120  &  6  &  $ 0.89^{+0.05}_{-0.05 }$  &  $ 3.25^{+0.03}_{-0.03 }$  &  $ 0.25^{+0.03}_{-0.02 }$  &  0.85  &  0.66  &  \ref{fig:mdust_mgas} \\
$SFR$--$M_{\star}$  &  121  &  0  &  $ 0.69^{+0.07}_{-0.07 }$  &  $ -6.98^{+0.05}_{-0.05 }$  &  $ 0.52^{+0.04}_{-0.04 }$  &  0.64  &  0.38  &  \\
$M_{H_2}$--$M_{\star}$  &  121  &  6  &  $ 1.17^{+0.07}_{-0.07 }$  &  $ -3.03^{+0.05}_{-0.04 }$  &  $ 0.45^{+0.04}_{-0.03 }$  &  0.84  &  0.56  &  \ref{fig:MGMS_KS_MGD} \\
$M_{H_2}$--$SFR$  &  121  &  6  &  $ 1.09^{+0.08}_{-0.08 }$  &  $ 8.74^{+0.05}_{-0.05 }$  &  $ 0.52^{+0.05}_{-0.04 }$  &  0.74  &  0.61  &  \ref{fig:MGMS_KS_MGD} \\
$M_{H_2}$--$M_{dust}$ &  121  &  6  &  $ 1.26^{+0.06}_{-0.06 }$  &  $ 0.04^{+0.04}_{-0.04 }$  &  $ 0.38^{+0.04}_{-0.03 }$  &  0.86  &  0.66  &  \ref{fig:MGMS_KS_MGD} \\
$M_{H_2}$--$M_{HI, R25}$  &   120  &  0  &  $ 0.94 ^{ 0.10 }_{ 0.10 }$  &  $ 0.28 ^{ 0.06 }_{ 0.06 }$  &  $ 0.68 ^{ 0.05 }_{ 0.05 }$  &  0.63  &  0.47  & \ref{fig:gas_phase_trans} \\
$M_{H_2/HI, R25}$--$M_{\star}$ &  120  &  0  &  $ 0.60 ^{ 0.09 }_{ 0.08 }$  &  $ -6.35 ^{ 0.05 }_{ 0.05 }$  &  $ 0.56 ^{ 0.04 }_{ 0.04 }$  &  0.51  &  0.25  & \ref{fig:gas_phase_trans} \\
$M_{H_2/HI, R25}$--$M_{dust/\star}$ &  120  &  0  &  $ -0.67 ^{ 0.22 }_{ 0.22 }$  &  $ -2.41 ^{ 0.06 }_{ 0.06 }$  &  $ 0.64 ^{ 0.05 }_{ 0.04 }$  &  -0.26  &  -0.17  & \ref{fig:gas_phase_trans} \\
$M_{H_2}/M_{\star}$--$sSFR$  &  121  &  6  &  $ 0.44^{+0.07}_{-0.07 }$  &  $ 3.15^{+0.04}_{-0.04 }$  &  $ 0.37^{+0.03}_{-0.03 }$  &  0.47  &  0.37  &  \ref{fig:fgas_sfe} \\
SFE--$M_{\star}$  &  121  &  6  &  $ -0.53^{+0.07}_{-0.07 }$  &  $ -3.37^{+0.04}_{-0.04 }$  &  $ 0.41^{+0.04}_{-0.03 }$  &  -0.57  &  -0.38  &  \ref{fig:fgas_sfe} \\
\hline
$SFR_{SED}$--$M_{H_2}$  &  85  &  0  &  $ 0.58^{+0.06}_{-0.06 }$  &  $ -5.02^{+0.05}_{-0.05 }$  &  $ 0.41^{+0.04}_{-0.03 }$  &  0.73  &  0.61  &  \ref{fig:multiSFR} \\
$SFR_{\rm FUV+24 \mu m}$--$M_{H_2}$ &  85  &  0  &  $ 0.56^{+0.05}_{-0.05 }$  &  $ -4.99^{+0.04}_{-0.04 }$  &  $ 0.37^{+0.03}_{-0.03 }$  &  0.78  &  0.64  &  \ref{fig:multiSFR} \\
$SFR_{\rm 160 \mu m}$--$M_{H_2}$  & 85  &  0  &  $ 0.76^{+0.04}_{-0.04 }$  &  $ -6.50^{+0.03}_{-0.03 }$  &  $ 0.24^{+0.03}_{-0.02 }$  &  0.89  &  0.77 &  \ref{fig:multiSFR} \\
\hline
\end{tabular}
\flushleft 
\footnotesize {{\bf Notes.} Scaling relations: best-fit parameters and correlation coefficients. Columns: (1) Scaling relation; (2) Total number of objects with available atomic and/or molecular gas measurements; (3) Number of detection of the corresponding gas measurements; (4), (5) and (6) Best-fit parameters, namely slope $\alpha$, normalization $\beta$, and intrinsic dispersion $\delta_{intr}$; (7), (8) and (9) Pearson CC and Kendall rank CCs; (10) Reference to figures in the main text (when available).
CCs are calculated considering only CO and HI detections.}
\end{table*}

\FloatBarrier 

\begin{table*}[ht]
\caption{3D Scaling relations}
\label{table:corr_3d}      
\centering          
\begin{tabular}[\textwidth]{l l l l l l l l}     
\hline     
    Relation                &   Sample  & UL   & $\alpha$ & $\beta$ & $\gamma$ & $\delta_{intr}$ &    Fig. \\ 
    (1)                     &   (2)     & (3) &  (4)     &   (5)   &     (6)         &   (7) &   \\ 
\hline
x: \mstar, y: SFR, z:\mh\ &   115  & 0  & $0.79^{+0.07}_{-0.07}$ & $0.51^{+0.07}_{-0.07}$ & $8.86^{+0.03}_{-0.03}$ & $0.24^{+0.05}_{-0.05}$ &    Fig.~\ref{fig:3d}  \\ 
x: \mdust, y: \mhiop, z:\mh\ &   115  & 0   & $1.38^{+0.10}_{-0.10}$ & $-0.24^{+0.10}_{-0.10}$ & $1.38^{+0.54}_{-0.53}$ & $0.34^{+0.03}_{-0.03}$ &    Fig.~\ref{fig:3d}  \\ 
\hline

\end{tabular}
\flushleft 
\footnotesize {{\bf Notes.} Three-dimensional scaling relations: best-fit parameters and correlation coefficients. Columns: (1) Scaling relation; (2) Total number of objects with available atomic and/or molecular gas measurements; (3) Number of detection of the corresponding gas measurements; (4), (5) and (6) Best-fit parameters, namely slopes $\alpha$ and $\beta$, normalization $\gamma$, and intrinsic dispersion $\delta_{intr}$; (7) Reference to figure.}
\end{table*}

\FloatBarrier 

\section{Figures and tables of section~\ref{sec:photometry}}
\label{app:photometry}

\begin{figure*}[htp]
        \centering
	\includegraphics[width =\textwidth, keepaspectratio=True]{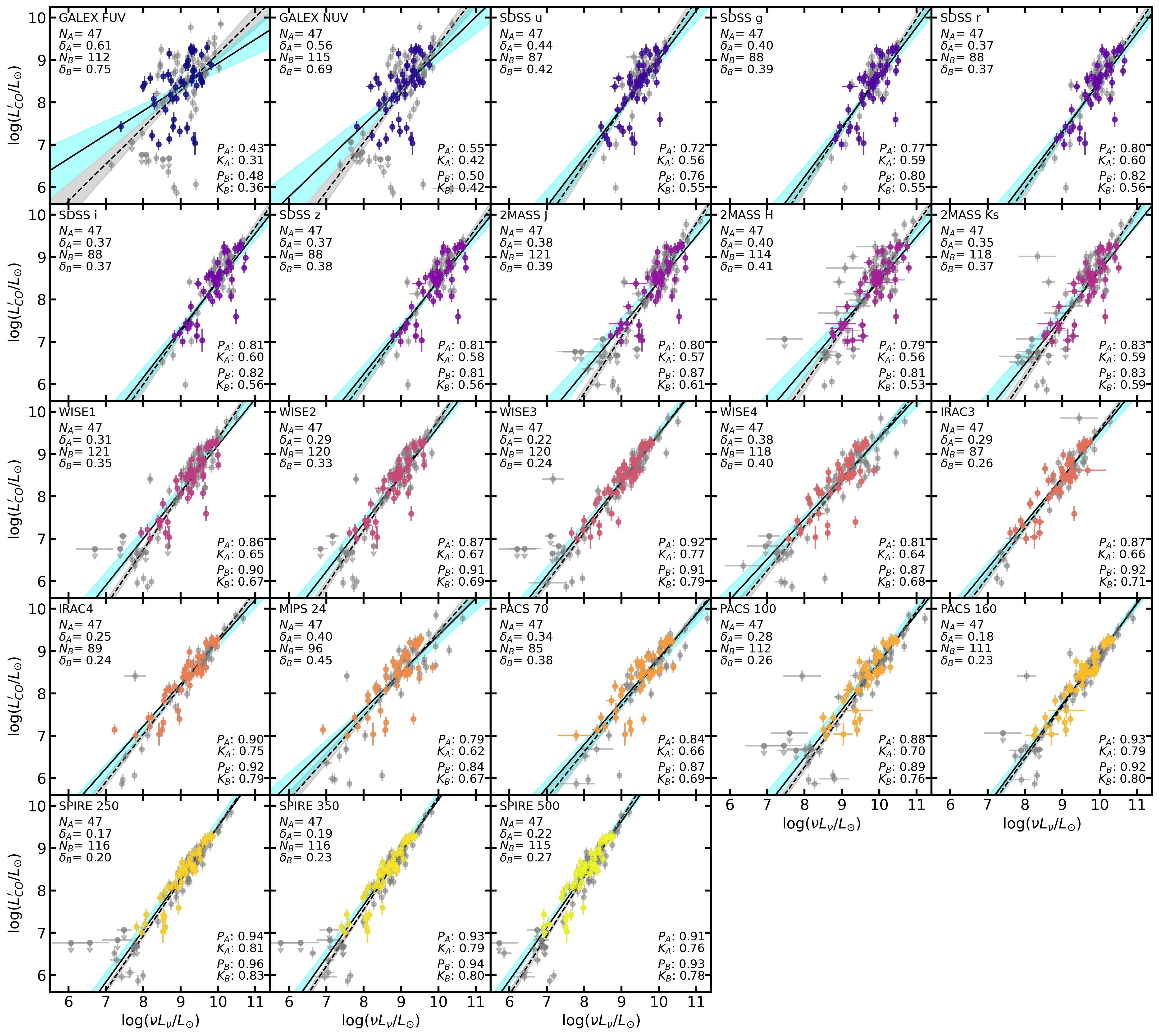}
	\caption{\lco\ versus  monochromatic luminosities.
 The names of telescopes and instruments used to derive the luminosities on the x-axis are reported in the upper-left corner of each panel.
 The galaxies from sample A (see text) are colored in each panel, while the remaining sources from sample B are in gray in each panel.
 The best-fit relation on data from sample A (B) is the solid (dashed) black line, and the 68\% confidence levels are shown as a cyan (light gray) region.
 Beneath, the number of galaxies used in the fitting procedure and the measured intrinsic dispersion for samples A and B, respectively.
 Correlation coefficients (namely, $P$ and $K$) for samples A and B are shown in the lower-right corner of each panel.\\
 }
    \label{fig:photometry}
\end{figure*}
\FloatBarrier 
\begin{table*}[ht]
\caption{\lco\ vs. monochromatic luminosities}
\label{table:corr_CO}      
\centering          
\begin{tabular}[\textwidth]{l l l l l l l l l}     
\hline     
    Relation                &   Sample  & UL   & $\alpha$ & $\beta$ & $\delta_{intr}$ & P &   K  &   Fig. \\ 
    (1)                     &   (2)     & (3) &  (4)     &   (5)   &     (6)         &   (7) &   (8)  &  (9) \\ 
\hline
$L^{\prime}_{CO}$--$\nu L_{\nu, FUV, A}$	 &  47  &  0  &  $ 0.56^{+0.18}_{-0.17 }$  &  $ 3.28^{+0.09}_{-0.09 }$  &  $ 0.61^{+0.08}_{-0.06 }$  &  0.43  &  0.31  &  \ref{fig:photometry} \\
$L^{\prime}_{CO}$--$\nu L_{\nu, FUV, B}$	 &  112  &  5  &  $ 0.87^{+0.13}_{-0.13 }$  &  $ 0.47^{+0.07}_{-0.07 }$  &  $ 0.75^{+0.06}_{-0.05 }$  &  0.48  &  0.36  &  \ref{fig:photometry} \\
$L^{\prime}_{CO}$--$\nu L_{\nu, NUV, A}$	 &  47  &  0  &  $ 0.85^{+0.20}_{-0.19 }$  &  $ 0.67^{+0.08}_{-0.08 }$  &  $ 0.56^{+0.08}_{-0.06 }$  &  0.55  &  0.42  &  \ref{fig:photometry} \\
$L^{\prime}_{CO}$--$\nu L_{\nu, NUV, B}$	 &  115  &  6  &  $ 1.16^{+0.12}_{-0.12 }$  &  $ -2.21^{+0.07}_{-0.07 }$  &  $ 0.69^{+0.06}_{-0.05 }$  &  0.50  &  0.42  &  \ref{fig:photometry} \\
$L^{\prime}_{CO}$--$\nu L_{\nu, u, A}$  	 &  47  &  0  &  $ 1.15^{+0.16}_{-0.16 }$  &  $ -2.53^{+0.07}_{-0.07 }$  &  $ 0.44^{+0.06}_{-0.05 }$  &  0.72  &  0.56  &  \ref{fig:photometry} \\
$L^{\prime}_{CO}$--$\nu L_{\nu, u, B}$  	 &  87  &  0  &  $ 1.29^{+0.12}_{-0.12 }$  &  $ -3.76^{+0.05}_{-0.05 }$  &  $ 0.42^{+0.04}_{-0.03 }$  &  0.76  &  0.55  &  \ref{fig:photometry} \\
$L^{\prime}_{CO}$--$\nu L_{\nu, g, A}$  	 &  47  &  0  &  $ 1.18^{+0.14}_{-0.14 }$  &  $ -3.19^{+0.06}_{-0.06 }$  &  $ 0.40^{+0.06}_{-0.04 }$  &  0.77  &  0.59  &  \ref{fig:photometry} \\
$L^{\prime}_{CO}$--$\nu L_{\nu, g, B}$  	 &  88  &  0  &  $ 1.30^{+0.10}_{-0.11 }$  &  $ -4.37^{+0.04}_{-0.04 }$  &  $ 0.39^{+0.04}_{-0.03 }$  &  0.80  &  0.55  &  \ref{fig:photometry} \\
$L^{\prime}_{CO}$--$\nu L_{\nu, r, A}$  	 &  47  &  0  &  $ 1.16^{+0.13}_{-0.12 }$  &  $ -3.15^{+0.06}_{-0.06 }$  &  $ 0.37^{+0.06}_{-0.04 }$  &  0.80  &  0.60  &  \ref{fig:photometry} \\
$L^{\prime}_{CO}$--$\nu L_{\nu, r, B}$  	 &  88  &  0  &  $ 1.28^{+0.09}_{-0.10 }$  &  $ -4.27^{+0.04}_{-0.04 }$  &  $ 0.37^{+0.04}_{-0.03 }$  &  0.82  &  0.56  &  \ref{fig:photometry} \\
$L^{\prime}_{CO}$--$\nu L_{\nu, I, A}$  	 &  47  &  0  &  $ 1.13^{+0.12}_{-0.12 }$  &  $ -2.91^{+0.06}_{-0.06 }$  &  $ 0.37^{+0.06}_{-0.04 }$  &  0.81  &  0.60  &  \ref{fig:photometry} \\
$L^{\prime}_{CO}$--$\nu L_{\nu, I, B}$  	 &  88  &  0  &  $ 1.24^{+0.09}_{-0.09 }$  &  $ -3.90^{+0.04}_{-0.04 }$  &  $ 0.37^{+0.04}_{-0.03 }$  &  0.82  &  0.56  &  \ref{fig:photometry} \\
$L^{\prime}_{CO}$--$\nu L_{\nu, z, A}$  	 &  47  &  0  &  $ 1.10^{+0.12}_{-0.11 }$  &  $ -2.57^{+0.06}_{-0.06 }$  &  $ 0.37^{+0.05}_{-0.04 }$  &  0.81  &  0.58  &  \ref{fig:photometry} \\
$L^{\prime}_{CO}$--$\nu L_{\nu, z, B}$  	 &  88  &  0  &  $ 1.20^{+0.09}_{-0.09 }$  &  $ -3.53^{+0.04}_{-0.04 }$  &  $ 0.38^{+0.04}_{-0.03 }$  &  0.81  &  0.56  &  \ref{fig:photometry} \\
$L^{\prime}_{CO}$--$\nu L_{\nu, J, A}$  	 &  47  &  0  &  $ 1.07^{+0.12}_{-0.12 }$  &  $ -2.15^{+0.06}_{-0.06 }$  &  $ 0.38^{+0.06}_{-0.04 }$  &  0.80  &  0.57  &  \ref{fig:photometry} \\
$L^{\prime}_{CO}$--$\nu L_{\nu, J, B}$  	 &  121  &  6  &  $ 1.34^{+0.07}_{-0.07 }$  &  $ -4.90^{+0.04}_{-0.04 }$  &  $ 0.39^{+0.03}_{-0.03 }$  &  0.87  &  0.61  &  \ref{fig:photometry} \\
$L^{\prime}_{CO}$--$\nu L_{\nu, H, A}$  	 &  47  &  0  &  $ 1.04^{+0.13}_{-0.13 }$  &  $ -1.86^{+0.06}_{-0.06 }$  &  $ 0.40^{+0.06}_{-0.05 }$  &  0.79  &  0.56  &  \ref{fig:photometry} \\
$L^{\prime}_{CO}$--$\nu L_{\nu, H, B}$  	 &  114  &  4  &  $ 1.25^{+0.08}_{-0.08 }$  &  $ -4.00^{+0.04}_{-0.04 }$  &  $ 0.41^{+0.04}_{-0.03 }$  &  0.81  &  0.53  &  \ref{fig:photometry} \\
$L^{\prime}_{CO}$--$\nu L_{\nu, Ks, A}$ 	 &  47  &  0  &  $ 1.10^{+0.11}_{-0.11 }$  &  $ -2.20^{+0.06}_{-0.06 }$  &  $ 0.35^{+0.05}_{-0.04 }$  &  0.83  &  0.59  &  \ref{fig:photometry} \\
$L^{\prime}_{CO}$--$\nu L_{\nu, Ks, B}$ 	 &  118  &  6  &  $ 1.31^{+0.07}_{-0.07 }$  &  $ -4.31^{+0.04}_{-0.04 }$  &  $ 0.37^{+0.03}_{-0.03 }$  &  0.83  &  0.59  &  \ref{fig:photometry} \\
$L^{\prime}_{CO}$--$\nu L_{\nu, W1, A}$ 	 &  47  &  0  &  $ 1.10^{+0.10}_{-0.10 }$  &  $ -1.70^{+0.05}_{-0.05 }$  &  $ 0.31^{+0.05}_{-0.04 }$  &  0.86  &  0.65  &  \ref{fig:photometry} \\
$L^{\prime}_{CO}$--$\nu L_{\nu, W1, B}$ 	 &  121  &  6  &  $ 1.32^{+0.06}_{-0.06 }$  &  $ -3.78^{+0.03}_{-0.03 }$  &  $ 0.35^{+0.03}_{-0.02 }$  &  0.90  &  0.67  &  \ref{fig:photometry} \\
$L^{\prime}_{CO}$--$\nu L_{\nu, W2, A}$ 	 &  47  &  0  &  $ 1.14^{+0.09}_{-0.09 }$  &  $ -1.68^{+0.05}_{-0.05 }$  &  $ 0.29^{+0.05}_{-0.04 }$  &  0.87  &  0.67  &  \ref{fig:photometry} \\
$L^{\prime}_{CO}$--$\nu L_{\nu, W2, B}$ 	 &  120  &  5  &  $ 1.30^{+0.06}_{-0.06 }$  &  $ -3.19^{+0.03}_{-0.03 }$  &  $ 0.33^{+0.03}_{-0.02 }$  &  0.91  &  0.69  &  \ref{fig:photometry} \\
$L^{\prime}_{CO}$--$\nu L_{\nu, W3, A}$ 	 &  47  &  0  &  $ 1.11^{+0.07}_{-0.07 }$  &  $ -1.60^{+0.04}_{-0.04 }$  &  $ 0.22^{+0.04}_{-0.03 }$  &  0.92  &  0.77  &  \ref{fig:photometry} \\
$L^{\prime}_{CO}$--$\nu L_{\nu, W3, B}$ 	 &  120  &  6  &  $ 1.17^{+0.04}_{-0.04 }$  &  $ -2.20^{+0.03}_{-0.03 }$  &  $ 0.24^{+0.03}_{-0.02 }$  &  0.91  &  0.79  &  \ref{fig:photometry} \\
$L^{\prime}_{CO}$--$\nu L_{\nu, W4, A}$ 	 &  47  &  0  &  $ 0.95^{+0.11}_{-0.10 }$  &  $ -0.16^{+0.06}_{-0.06 }$  &  $ 0.38^{+0.06}_{-0.04 }$  &  0.81  &  0.64  &  \ref{fig:photometry} \\
$L^{\prime}_{CO}$--$\nu L_{\nu, W4, B}$ 	 &  118  &  4  &  $ 1.05^{+0.05}_{-0.05 }$  &  $ -1.13^{+0.04}_{-0.04 }$  &  $ 0.40^{+0.03}_{-0.03 }$  &  0.87  &  0.68  &  \ref{fig:photometry} \\
$L^{\prime}_{CO}$--$\nu L_{\nu, I3, A}$ 	 &  47  &  0  &  $ 1.07^{+0.09}_{-0.09 }$  &  $ -1.26^{+0.05}_{-0.05 }$  &  $ 0.29^{+0.05}_{-0.03 }$  &  0.87  &  0.66  &  \ref{fig:photometry} \\
$L^{\prime}_{CO}$--$\nu L_{\nu, I3, B}$ 	 &  87  &  1  &  $ 1.16^{+0.05}_{-0.05 }$  &  $ -2.05^{+0.03}_{-0.03 }$  &  $ 0.26^{+0.03}_{-0.03 }$  &  0.92  &  0.71  &  \ref{fig:photometry} \\
$L^{\prime}_{CO}$--$\nu L_{\nu, I4, A}$ 	 &  47  &  0  &  $ 1.00^{+0.07}_{-0.07 }$  &  $ -0.80^{+0.04}_{-0.04 }$  &  $ 0.25^{+0.04}_{-0.03 }$  &  0.90  &  0.75  &  \ref{fig:photometry} \\
$L^{\prime}_{CO}$--$\nu L_{\nu, I4, B}$ 	 &  89  &  0  &  $ 1.12^{+0.04}_{-0.04 }$  &  $ -1.94^{+0.03}_{-0.03 }$  &  $ 0.24^{+0.03}_{-0.02 }$  &  0.92  &  0.79  &  \ref{fig:photometry} \\
$L^{\prime}_{CO}$--$\nu L_{\nu, 24, A}$ 	 &  47  &  0  &  $ 0.87^{+0.10}_{-0.11 }$  &  $ 0.56^{+0.06}_{-0.06 }$  &  $ 0.40^{+0.06}_{-0.04 }$  &  0.79  &  0.62  &  \ref{fig:photometry} \\
$L^{\prime}_{CO}$--$\nu L_{\nu, 24, B}$ 	 &  96  &  0  &  $ 0.99^{+0.07}_{-0.06 }$  &  $ -0.45^{+0.05}_{-0.05 }$  &  $ 0.45^{+0.04}_{-0.03 }$  &  0.84  &  0.67  &  \ref{fig:photometry} \\
$L^{\prime}_{CO}$--$\nu L_{\nu, 70, A}$ 	 &  47  &  0  &  $ 1.02^{+0.10}_{-0.10 }$  &  $ -1.46^{+0.05}_{-0.05 }$  &  $ 0.34^{+0.05}_{-0.04 }$  &  0.84  &  0.66  &  \ref{fig:photometry} \\
$L^{\prime}_{CO}$--$\nu L_{\nu, 70, B}$ 	 &  85  &  0  &  $ 1.09^{+0.07}_{-0.07 }$  &  $ -2.07^{+0.04}_{-0.04 }$  &  $ 0.38^{+0.04}_{-0.03 }$  &  0.87  &  0.69  &  \ref{fig:photometry} \\
$L^{\prime}_{CO}$--$\nu L_{\nu, 100, B}$	 &  47  &  0  &  $ 1.13^{+0.09}_{-0.09 }$  &  $ -2.46^{+0.05}_{-0.04 }$  &  $ 0.28^{+0.05}_{-0.03 }$  &  0.88  &  0.70  &  \ref{fig:photometry} \\
$L^{\prime}_{CO}$--$\nu L_{\nu, 100, B}$	 &  112  &  5  &  $ 1.23^{+0.05}_{-0.05 }$  &  $ -3.55^{+0.03}_{-0.03 }$  &  $ 0.26^{+0.03}_{-0.02 }$  &  0.89  &  0.76  &  \ref{fig:photometry} \\
$L^{\prime}_{CO}$--$\nu L_{\nu, 600, B}$	 &  47  &  0  &  $ 1.19^{+0.07}_{-0.07 }$  &  $ -2.94^{+0.03}_{-0.03 }$  &  $ 0.18^{+0.04}_{-0.03 }$  &  0.93  &  0.79  &  \ref{fig:photometry} \\
$L^{\prime}_{CO}$--$\nu L_{\nu, 160, B}$	 &  111  &  3  &  $ 1.22^{+0.04}_{-0.04 }$  &  $ -3.33^{+0.03}_{-0.03 }$  &  $ 0.23^{+0.03}_{-0.02 }$  &  0.92  &  0.80  &  \ref{fig:photometry} \\
$L^{\prime}_{CO}$--$\nu L_{\nu, 250, A}$	 &  47  &  0  &  $ 1.22^{+0.06}_{-0.06 }$  &  $ -2.61^{+0.03}_{-0.03 }$  &  $ 0.17^{+0.04}_{-0.03 }$  &  0.94  &  0.81  &  \ref{fig:photometry} \\
$L^{\prime}_{CO}$--$\nu L_{\nu, 250, B}$	 &  116  &  6  &  $ 1.29^{+0.04}_{-0.04 }$  &  $ -3.41^{+0.02}_{-0.02 }$  &  $ 0.20^{+0.02}_{-0.02 }$  &  0.96  &  0.83  &  \ref{fig:photometry} \\
$L^{\prime}_{CO}$--$\nu L_{\nu, 350, A}$	 &  47  &  0  &  $ 1.24^{+0.07}_{-0.07 }$  &  $ -2.18^{+0.03}_{-0.03 }$  &  $ 0.19^{+0.04}_{-0.03 }$  &  0.93  &  0.79  &  \ref{fig:photometry} \\
$L^{\prime}_{CO}$--$\nu L_{\nu, 350, B}$	 &  116  &  6  &  $ 1.33^{+0.04}_{-0.04 }$  &  $ -3.12^{+0.03}_{-0.03 }$  &  $ 0.23^{+0.03}_{-0.02 }$  &  0.94  &  0.80  &  \ref{fig:photometry} \\
$L^{\prime}_{CO}$--$\nu L_{\nu, 500, A}$	 &  47  &  0  &  $ 1.25^{+0.08}_{-0.08 }$  &  $ -1.57^{+0.04}_{-0.04 }$  &  $ 0.22^{+0.04}_{-0.03 }$  &  0.91  &  0.76  &  \ref{fig:photometry} \\
$L^{\prime}_{CO}$--$\nu L_{\nu, 500, B}$	 &  115  &  5  &  $ 1.39^{+0.05}_{-0.05 }$  &  $ -2.78^{+0.03}_{-0.03 }$  &  $ 0.27^{+0.03}_{-0.02 }$  &  0.93  &  0.78  &  \ref{fig:photometry} \\
\hline
\end{tabular}
\flushleft \footnotesize {{\bf Notes.} Best-fit parameters and correlation coefficients. Columns: (1) Scaling relation; (2) Total number of objects with available \lco\ and monochromatic luminosity; (3) Number of \lco\ detection; (4), (5) and (6) Best-fit parameters, namely slope $\alpha$, normalization $\gamma$, and intrinsic dispersion $\delta_{intr}$; (7) and (8) are $P$ and $K$; (9) Reference to figure.}
\end{table*}
\FloatBarrier 
%
\begin{figure*}
	\includegraphics[width =\textwidth, keepaspectratio=True]{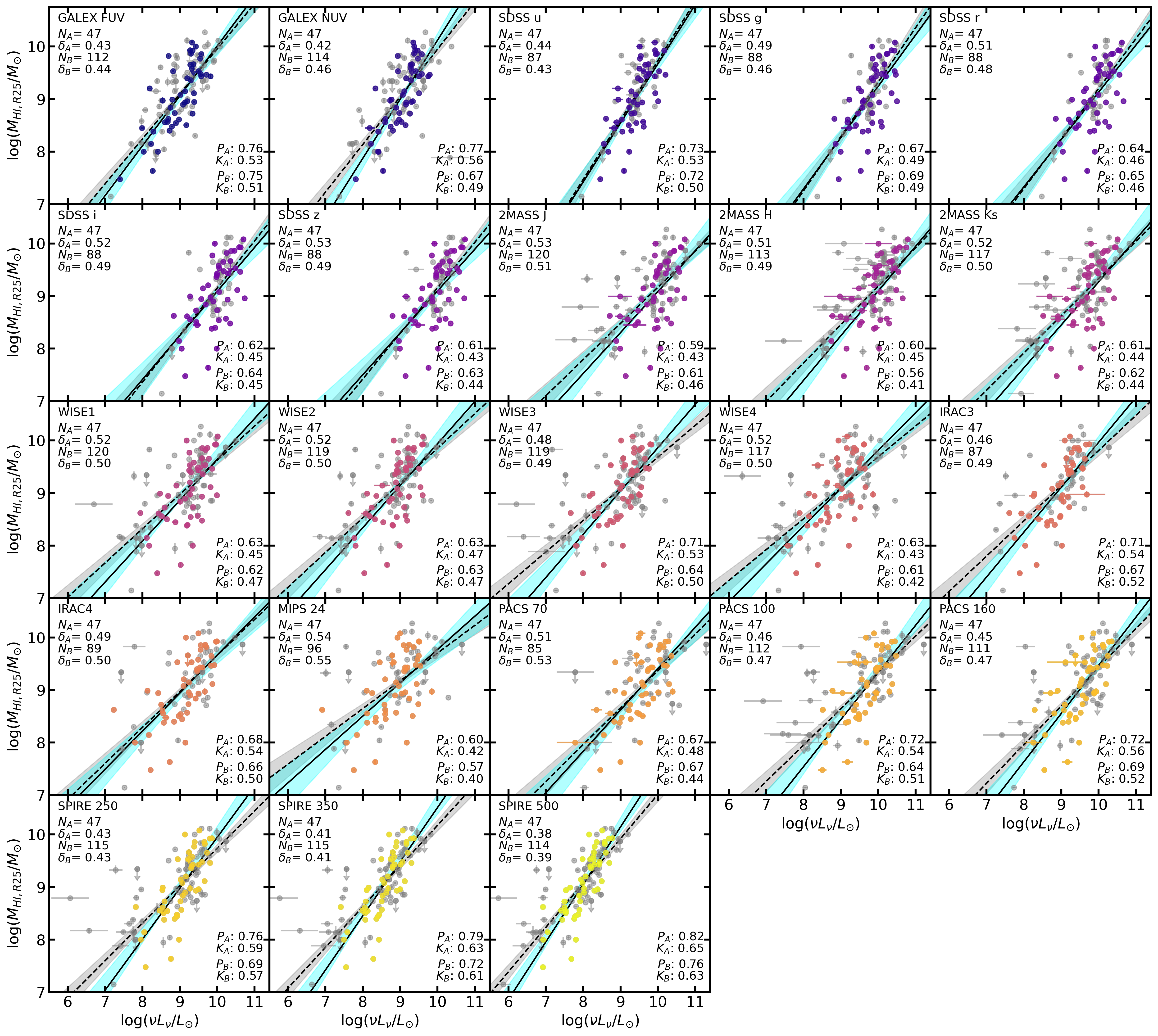}
	\caption{\mhiop~versus monochromatic luminosities.
See the caption of Fig.~\ref{fig:photometry}.}
    \label{fig:hi_photometry}
\end{figure*}
\FloatBarrier 
\begin{table*}[ht]
\caption{$M_{HI, R25}$ vs. monochromatic luminosities}
\label{table:corr_HIr25}      
\centering          
\begin{tabular}[\textwidth]{l l l l l l l l l}     
\hline     
    Relation                &   Sample  & UL   & $\alpha$ & $\beta$ & $\delta_{intr}$ & P &   K  &   Fig. \\ 
    (1)                     &   (2)     & (3) &  (4)     &   (5)   &     (6)         &   (7) &   (8)  &  (9) \\ 
\hline
$M_{HI, R25}$--$\nu L_{\nu, FUV, A}$	&  47  &  1  &  $ 0.95 ^{+ 0.12 }_{- 0.13 }$  &  $ 0.49 ^{+ 0.06 }_{- 0.06 }$  &  $ 0.43 ^{+ 0.05 }_{- 0.04 }$  &  0.76  &  0.53  &  \ref{fig:hi_photometry} \\
$M_{HI, R25}$--$\nu L_{\nu, FUV, B}$	 &  112  &  6  &  $ 0.86 ^{+ 0.07 }_{- 0.08 }$  &  $ 1.39 ^{+ 0.04 }_{- 0.04 }$  &  $ 0.44 ^{+ 0.03 }_{- 0.03 }$  &  0.75  &  0.51  &  \ref{fig:hi_photometry} \\
$M_{HI, R25}$--$\nu L_{\nu, NUV, A}$	 &  47  &  1  &  $ 1.12 ^{+ 0.14 }_{- 0.14 }$  &  $ -1.13 ^{+ 0.06 }_{- 0.06 }$  &  $ 0.42 ^{+ 0.05 }_{- 0.04 }$  &  0.77  &  0.56  &  \ref{fig:hi_photometry} \\
$M_{HI, R25}$--$\nu L_{\nu, NUV, B}$	 &  115  &  8  &  $ 0.91 ^{+ 0.08 }_{- 0.08 }$  &  $ 0.89 ^{+ 0.04 }_{- 0.04 }$  &  $ 0.46 ^{+ 0.04 }_{- 0.03 }$  &  0.67  &  0.49  &  \ref{fig:hi_photometry} \\
$M_{HI, R25}$--$\nu L_{\nu, u, A}$  	 &  47  &  1  &  $ 1.13 ^{+ 0.15 }_{- 0.16 }$  &  $ -1.58 ^{+ 0.07 }_{- 0.06 }$  &  $ 0.44 ^{+ 0.06 }_{- 0.04 }$  &  0.73  &  0.53  &  \ref{fig:hi_photometry} \\
$M_{HI, R25}$--$\nu L_{\nu, u, B}$  	 &  87  &  3  &  $ 1.13 ^{+ 0.12 }_{- 0.12 }$  &  $ -1.58 ^{+ 0.05 }_{- 0.05 }$  &  $ 0.43 ^{+ 0.04 }_{- 0.03 }$  &  0.72  &  0.50  &  \ref{fig:hi_photometry} \\
$M_{HI, R25}$--$\nu L_{\nu, g, A}$  	 &  47  &  1  &  $ 0.98 ^{+ 0.16 }_{- 0.16 }$  &  $ -0.52 ^{+ 0.07 }_{- 0.07 }$  &  $ 0.49 ^{+ 0.06 }_{- 0.05 }$  &  0.67  &  0.49  &  \ref{fig:hi_photometry} \\
$M_{HI, R25}$--$\nu L_{\nu, g, B}$  	 &  88  &  3  &  $ 1.03 ^{+ 0.12 }_{- 0.11 }$  &  $ -1.02 ^{+ 0.05 }_{- 0.05 }$  &  $ 0.46 ^{+ 0.04 }_{- 0.03 }$  &  0.69  &  0.49  &  \ref{fig:hi_photometry} \\
$M_{HI, R25}$--$\nu L_{\nu, r, A}$  	 &  47  &  1  &  $ 0.90 ^{+ 0.16 }_{- 0.16 }$  &  $ 0.13 ^{+ 0.07 }_{- 0.08 }$  &  $ 0.51 ^{+ 0.06 }_{- 0.05 }$  &  0.64  &  0.46  &  \ref{fig:hi_photometry} \\
$M_{HI, R25}$--$\nu L_{\nu, r, B}$  	 &  88  &  3  &  $ 0.95 ^{+ 0.12 }_{- 0.12 }$  &  $ -0.32 ^{+ 0.05 }_{- 0.05 }$  &  $ 0.48 ^{+ 0.04 }_{- 0.03 }$  &  0.65  &  0.46  &  \ref{fig:hi_photometry} \\
$M_{HI, R25}$--$\nu L_{\nu, I, A}$  	 &  47  &  1  &  $ 0.84 ^{+ 0.16 }_{- 0.17 }$  &  $ 0.71 ^{+ 0.08 }_{- 0.08 }$  &  $ 0.52 ^{+ 0.07 }_{- 0.05 }$  &  0.62  &  0.45  &  \ref{fig:hi_photometry} \\
$M_{HI, R25}$--$\nu L_{\nu, I, B}$  	 &  88  &  3  &  $ 0.90 ^{+ 0.12 }_{- 0.11 }$  &  $ 0.19 ^{+ 0.05 }_{- 0.05 }$  &  $ 0.49 ^{+ 0.04 }_{- 0.03 }$  &  0.64  &  0.45  &  \ref{fig:hi_photometry} \\
$M_{HI, R25}$--$\nu L_{\nu, z, A}$  	 &  47  &  1  &  $ 0.80 ^{+ 0.16 }_{- 0.16 }$  &  $ 1.12 ^{+ 0.08 }_{- 0.08 }$  &  $ 0.53 ^{+ 0.07 }_{- 0.05 }$  &  0.61  &  0.43  &  \ref{fig:hi_photometry} \\
$M_{HI, R25}$--$\nu L_{\nu, z, B}$  	 &  88  &  3  &  $ 0.87 ^{+ 0.12 }_{- 0.11 }$  &  $ 0.49 ^{+ 0.05 }_{- 0.05 }$  &  $ 0.49 ^{+ 0.05 }_{- 0.03 }$  &  0.63  &  0.44  &  \ref{fig:hi_photometry} \\
$M_{HI, R25}$--$\nu L_{\nu, J, A}$  	 &  47  &  1  &  $ 0.78 ^{+ 0.16 }_{- 0.16 }$  &  $ 1.36 ^{+ 0.08 }_{- 0.08 }$  &  $ 0.53 ^{+ 0.07 }_{- 0.05 }$  &  0.59  &  0.43  &  \ref{fig:hi_photometry} \\
$M_{HI, R25}$--$\nu L_{\nu, J, B}$  	 &  121  &  8  &  $ 0.69 ^{+ 0.08 }_{- 0.08 }$  &  $ 2.28 ^{+ 0.05 }_{- 0.05 }$  &  $ 0.51 ^{+ 0.04 }_{- 0.03 }$  &  0.61  &  0.46  &  \ref{fig:hi_photometry} \\
$M_{HI, R25}$--$\nu L_{\nu, H, A}$  	 &  47  &  1  &  $ 0.85 ^{+ 0.16 }_{- 0.17 }$  &  $ 0.66 ^{+ 0.08 }_{- 0.08 }$  &  $ 0.51 ^{+ 0.07 }_{- 0.05 }$  &  0.60  &  0.45  &  \ref{fig:hi_photometry} \\
$M_{HI, R25}$--$\nu L_{\nu, H, B}$  	 &  114  &  7  &  $ 0.73 ^{+ 0.09 }_{- 0.09 }$  &  $ 1.88 ^{+ 0.05 }_{- 0.05 }$  &  $ 0.49 ^{+ 0.04 }_{- 0.03 }$  &  0.56  &  0.41  &  \ref{fig:hi_photometry} \\
$M_{HI, R25}$--$\nu L_{\nu, Ks, A}$ 	 &  47  &  1  &  $ 0.82 ^{+ 0.16 }_{- 0.15 }$  &  $ 1.19 ^{+ 0.08 }_{- 0.08 }$  &  $ 0.52 ^{+ 0.06 }_{- 0.05 }$  &  0.61  &  0.44  &  \ref{fig:hi_photometry} \\
$M_{HI, R25}$--$\nu L_{\nu, Ks, B}$ 	 &  118  &  8  &  $ 0.71 ^{+ 0.08 }_{- 0.08 }$  &  $ 2.24 ^{+ 0.05 }_{- 0.05 }$  &  $ 0.50 ^{+ 0.04 }_{- 0.03 }$  &  0.62  &  0.44  &  \ref{fig:hi_photometry} \\
$M_{HI, R25}$--$\nu L_{\nu, W1, A}$ 	 &  47  &  1  &  $ 0.78 ^{+ 0.14 }_{- 0.14 }$  &  $ 1.93 ^{+ 0.08 }_{- 0.08 }$  &  $ 0.52 ^{+ 0.06 }_{- 0.05 }$  &  0.63  &  0.45  &  \ref{fig:hi_photometry} \\
$M_{HI, R25}$--$\nu L_{\nu, W1, B}$ 	 &  121  &  8  &  $ 0.65 ^{+ 0.07 }_{- 0.07 }$  &  $ 3.11 ^{+ 0.05 }_{- 0.05 }$  &  $ 0.50 ^{+ 0.04 }_{- 0.03 }$  &  0.62  &  0.47  &  \ref{fig:hi_photometry} \\
$M_{HI, R25}$--$\nu L_{\nu, W2, A}$ 	 &  47  &  1  &  $ 0.80 ^{+ 0.15 }_{- 0.15 }$  &  $ 1.99 ^{+ 0.08 }_{- 0.08 }$  &  $ 0.52 ^{+ 0.07 }_{- 0.05 }$  &  0.63  &  0.47  &  \ref{fig:hi_photometry} \\
$M_{HI, R25}$--$\nu L_{\nu, W2, B}$ 	 &  120  &  8  &  $ 0.66 ^{+ 0.07 }_{- 0.07 }$  &  $ 3.25 ^{+ 0.05 }_{- 0.05 }$  &  $ 0.50 ^{+ 0.04 }_{- 0.03 }$  &  0.63  &  0.47  &  \ref{fig:hi_photometry} \\
$M_{HI, R25}$--$\nu L_{\nu, W3, A}$ 	 &  47  &  1  &  $ 0.82 ^{+ 0.13 }_{- 0.13 }$  &  $ 1.68 ^{+ 0.07 }_{- 0.07 }$  &  $ 0.48 ^{+ 0.06 }_{- 0.05 }$  &  0.71  &  0.53  &  \ref{fig:hi_photometry} \\
$M_{HI, R25}$--$\nu L_{\nu, W3, B}$ 	 &  120  &  8  &  $ 0.60 ^{+ 0.07 }_{- 0.07 }$  &  $ 3.67 ^{+ 0.05 }_{- 0.04 }$  &  $ 0.49 ^{+ 0.04 }_{- 0.03 }$  &  0.64  &  0.50  &  \ref{fig:hi_photometry} \\
$M_{HI, R25}$--$\nu L_{\nu, W4, A}$ 	 &  47  &  1  &  $ 0.72 ^{+ 0.14 }_{- 0.14 }$  &  $ 2.62 ^{+ 0.08 }_{- 0.07 }$  &  $ 0.52 ^{+ 0.06 }_{- 0.05 }$  &  0.63  &  0.43  &  \ref{fig:hi_photometry} \\
$M_{HI, R25}$--$\nu L_{\nu, W4, B}$ 	 &  118  &  8  &  $ 0.59 ^{+ 0.07 }_{- 0.07 }$  &  $ 3.81 ^{+ 0.05 }_{- 0.05 }$  &  $ 0.50 ^{+ 0.04 }_{- 0.03 }$  &  0.61  &  0.42  &  \ref{fig:hi_photometry} \\
$M_{HI, R25}$--$\nu L_{\nu, I3, A}$ 	 &  47  &  1  &  $ 0.85 ^{+ 0.13 }_{- 0.13 }$  &  $ 1.39 ^{+ 0.07 }_{- 0.07 }$  &  $ 0.46 ^{+ 0.06 }_{- 0.05 }$  &  0.71  &  0.54  &  \ref{fig:hi_photometry} \\
$M_{HI, R25}$--$\nu L_{\nu, I3, B}$ 	 &  87  &  3  &  $ 0.67 ^{+ 0.08 }_{- 0.08 }$  &  $ 3.11 ^{+ 0.05 }_{- 0.05 }$  &  $ 0.49 ^{+ 0.04 }_{- 0.03 }$  &  0.67  &  0.52  &  \ref{fig:hi_photometry} \\
$M_{HI, R25}$--$\nu L_{\nu, I4, A}$ 	 &  47  &  1  &  $ 0.72 ^{+ 0.12 }_{- 0.12 }$  &  $ 2.39 ^{+ 0.07 }_{- 0.07 }$  &  $ 0.49 ^{+ 0.06 }_{- 0.05 }$  &  0.68  &  0.54  &  \ref{fig:hi_photometry} \\
$M_{HI, R25}$--$\nu L_{\nu, I4, B}$ 	 &  89  &  5  &  $ 0.68 ^{+ 0.08 }_{- 0.09 }$  &  $ 2.81 ^{+ 0.05 }_{- 0.05 }$  &  $ 0.50 ^{+ 0.04 }_{- 0.04 }$  &  0.66  &  0.50  &  \ref{fig:hi_photometry} \\
$M_{HI, R25}$--$\nu L_{\nu, 24, A}$ 	 &  47  &  1  &  $ 0.63 ^{+ 0.14 }_{- 0.13 }$  &  $ 3.39 ^{+ 0.08 }_{- 0.08 }$  &  $ 0.54 ^{+ 0.07 }_{- 0.05 }$  &  0.60  &  0.42  &  \ref{fig:hi_photometry} \\
$M_{HI, R25}$--$\nu L_{\nu, 24, B}$ 	 &  96  &  5  &  $ 0.53 ^{+ 0.08 }_{- 0.08 }$  &  $ 4.42 ^{+ 0.06 }_{- 0.06 }$  &  $ 0.55 ^{+ 0.05 }_{- 0.04 }$  &  0.57  &  0.40  &  \ref{fig:hi_photometry} \\
$M_{HI, R25}$--$\nu L_{\nu, 70, A}$ 	 &  47  &  1  &  $ 0.79 ^{+ 0.14 }_{- 0.14 }$  &  $ 1.48 ^{+ 0.08 }_{- 0.07 }$  &  $ 0.51 ^{+ 0.07 }_{- 0.05 }$  &  0.67  &  0.48  &  \ref{fig:hi_photometry} \\
$M_{HI, R25}$--$\nu L_{\nu, 70, B}$ 	 &  85  &  6  &  $ 0.70 ^{+ 0.10 }_{- 0.09 }$  &  $ 2.35 ^{+ 0.06 }_{- 0.06 }$  &  $ 0.53 ^{+ 0.05 }_{- 0.04 }$  &  0.67  &  0.44  &  \ref{fig:hi_photometry} \\
$M_{HI, R25}$--$\nu L_{\nu, 100, B}$	 &  47  &  1  &  $ 0.90 ^{+ 0.13 }_{- 0.13 }$  &  $ 0.46 ^{+ 0.07 }_{- 0.07 }$  &  $ 0.46 ^{+ 0.06 }_{- 0.04 }$  &  0.72  &  0.54  &  \ref{fig:hi_photometry} \\
$M_{HI, R25}$--$\nu L_{\nu, 100, B}$	 &  112  &  6  &  $ 0.69 ^{+ 0.07 }_{- 0.07 }$  &  $ 2.42 ^{+ 0.05 }_{- 0.05 }$  &  $ 0.47 ^{+ 0.04 }_{- 0.03 }$  &  0.64  &  0.51  &  \ref{fig:hi_photometry} \\
$M_{HI, R25}$--$\nu L_{\nu, 600, B}$	 &  47  &  1  &  $ 0.92 ^{+ 0.13 }_{- 0.13 }$  &  $ 0.31 ^{+ 0.07 }_{- 0.07 }$  &  $ 0.45 ^{+ 0.06 }_{- 0.04 }$  &  0.72  &  0.56  &  \ref{fig:hi_photometry} \\
$M_{HI, R25}$--$\nu L_{\nu, 160, B}$	 &  111  &  7  &  $ 0.74 ^{+ 0.07 }_{- 0.07 }$  &  $ 2.07 ^{+ 0.05 }_{- 0.04 }$  &  $ 0.47 ^{+ 0.04 }_{- 0.03 }$  &  0.69  &  0.52  &  \ref{fig:hi_photometry} \\
$M_{HI, R25}$--$\nu L_{\nu, 250, A}$	 &  47  &  1  &  $ 0.97 ^{+ 0.12 }_{- 0.13 }$  &  $ 0.34 ^{+ 0.06 }_{- 0.06 }$  &  $ 0.43 ^{+ 0.06 }_{- 0.04 }$  &  0.76  &  0.59  &  \ref{fig:hi_photometry} \\
$M_{HI, R25}$--$\nu L_{\nu, 250, B}$	 &  116  &  8  &  $ 0.71 ^{+ 0.07 }_{- 0.07 }$  &  $ 2.61 ^{+ 0.04 }_{- 0.04 }$  &  $ 0.43 ^{+ 0.03 }_{- 0.03 }$  &  0.69  &  0.57  &  \ref{fig:hi_photometry} \\
$M_{HI, R25}$--$\nu L_{\nu, 350, A}$	 &  47  &  1  &  $ 1.03 ^{+ 0.12 }_{- 0.12 }$  &  $ 0.29 ^{+ 0.06 }_{- 0.06 }$  &  $ 0.41 ^{+ 0.05 }_{- 0.04 }$  &  0.79  &  0.63  &  \ref{fig:hi_photometry} \\
$M_{HI, R25}$--$\nu L_{\nu, 350, B}$	 &  116  &  8  &  $ 0.77 ^{+ 0.07 }_{- 0.06 }$  &  $ 2.44 ^{+ 0.04 }_{- 0.04 }$  &  $ 0.41 ^{+ 0.03 }_{- 0.03 }$  &  0.72  &  0.61  &  \ref{fig:hi_photometry} \\
$M_{HI, R25}$--$\nu L_{\nu, 500, A}$	 &  47  &  1  &  $ 1.09 ^{+ 0.11 }_{- 0.12 }$  &  $ 0.35 ^{+ 0.06 }_{- 0.06 }$  &  $ 0.38 ^{+ 0.05 }_{- 0.04 }$  &  0.82  &  0.65  &  \ref{fig:hi_photometry} \\
$M_{HI, R25}$--$\nu L_{\nu, 500, B}$	 &  115  &  8  &  $ 0.84 ^{+ 0.07 }_{- 0.07 }$  &  $ 2.33 ^{+ 0.04 }_{- 0.04 }$  &  $ 0.39 ^{+ 0.03 }_{- 0.02 }$  &  0.76  &  0.63  &  \ref{fig:hi_photometry} \\
\hline 
\end{tabular}
\\
\footnotesize {{\bf Notes.} Best-fit parameters and correlation coefficients. Columns: (1) Scaling relation; (2) Total number of objects with available $M_{HI, R25}$ and monochromatic luminosity; (3) Number of $M_{HI, R25}$ detections; (4), (5) and (6) Best-fit parameters, namely slope $\alpha$, normalization $\gamma$, and intrinsic dispersion $\delta_{intr}$; (7) and (8) are $P$ and $K$; (9) Reference to figure.}
\end{table*}
\FloatBarrier 
%
\begin{figure*}
	\includegraphics[width =\textwidth, keepaspectratio=True]{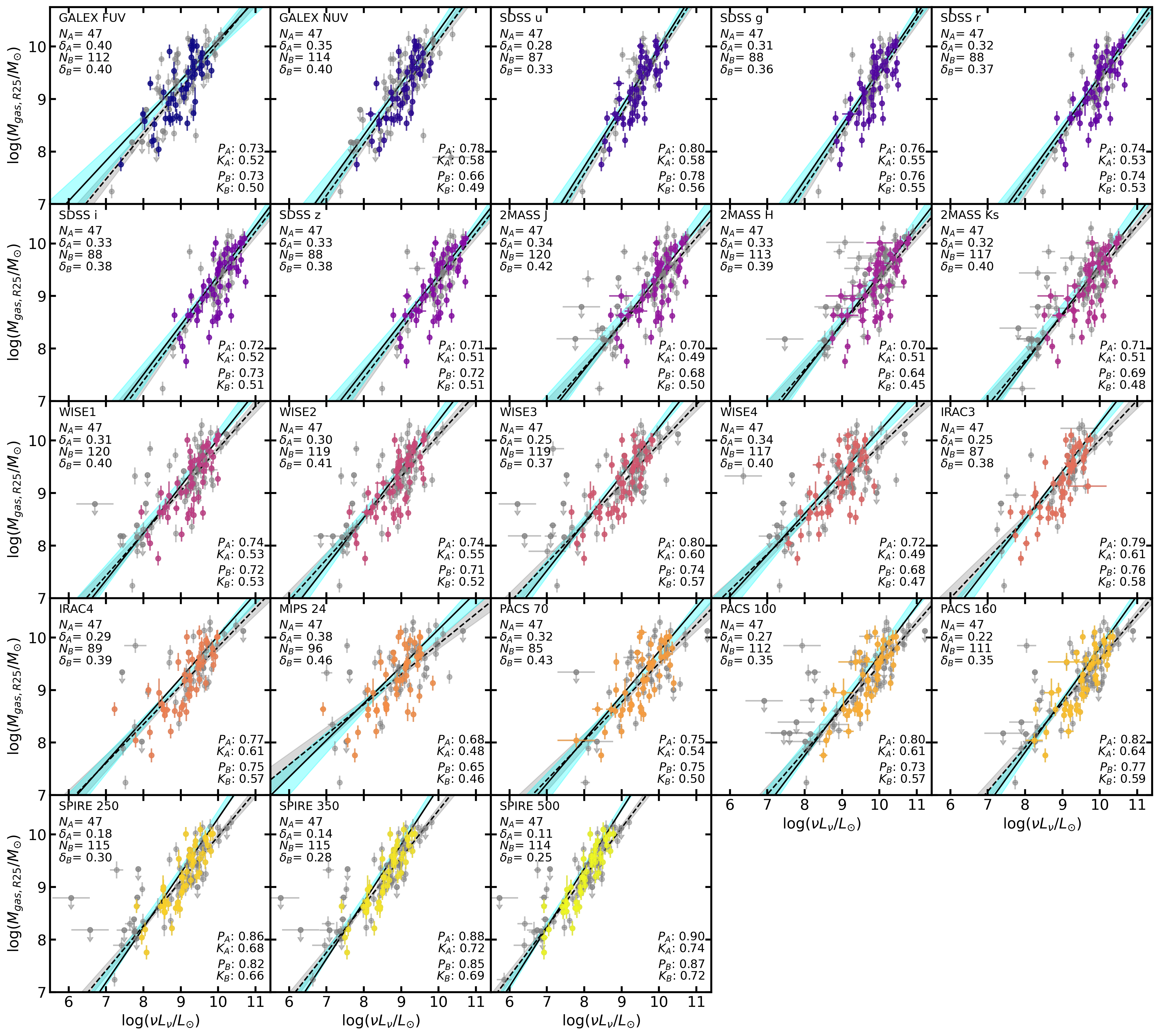}
	\caption{$M_{gas, R25}$ versus monochromatic luminosities.
See the caption of Fig.~\ref{fig:photometry}.}
    \label{fig:gas_photometry}
\end{figure*}
\FloatBarrier 
\begin{table*}[ht]
\caption{$M_{gas, R25}$ vs. monochromatic luminosities}
\label{table:corr_gasr25}      
\centering          
\begin{tabular}[\textwidth]{l l l l l l l l l}     
\hline     
    Relation                &   Sample  & UL   & $\alpha$ & $\beta$ & $\delta_{intr}$ & P &   K  &   Fig. \\ 
    (1)                     &   (2)     & (3) &  (4)     &   (5)   &     (6)         &   (7) &   (8)  &  (9) \\ 
\hline
$M_{gas, R25}$--$\nu L_{\nu, FUV, A}$	&  47  &  1  &  $ 0.77 ^{+ 0.12 }_{- 0.12 }$  &  $ 2.44 ^{+ 0.06 }_{- 0.06 }$  &  $ 0.40 ^{+ 0.06 }_{- 0.04 }$  &  0.73  &  0.52  &  \ref{fig:gas_photometry} \\
$M_{gas, R25}$--$\nu L_{\nu, FUV, B}$	 &  112  &  11  &  $ 0.86 ^{+ 0.07 }_{- 0.07 }$  &  $ 1.45 ^{+ 0.04 }_{- 0.04 }$  &  $ 0.40 ^{+ 0.04 }_{- 0.03 }$  &  0.73  &  0.50  &  \ref{fig:gas_photometry} \\
$M_{gas, R25}$--$\nu L_{\nu, NUV, A}$	 &  47  &  1  &  $ 0.99 ^{+ 0.12 }_{- 0.12 }$  &  $ 0.36 ^{+ 0.05 }_{- 0.06 }$  &  $ 0.35 ^{+ 0.05 }_{- 0.04 }$  &  0.78  &  0.58  &  \ref{fig:gas_photometry} \\
$M_{gas, R25}$--$\nu L_{\nu, NUV, B}$	 &  115  &  13  &  $ 0.98 ^{+ 0.08 }_{- 0.08 }$  &  $ 0.29 ^{+ 0.04 }_{- 0.04 }$  &  $ 0.40 ^{+ 0.04 }_{- 0.03 }$  &  0.66  &  0.49  &  \ref{fig:gas_photometry} \\
$M_{gas, R25}$--$\nu L_{\nu, u, A}$  	 &  47  &  1  &  $ 1.13 ^{+ 0.11 }_{- 0.11 }$  &  $ -1.33 ^{+ 0.05 }_{- 0.05 }$  &  $ 0.28 ^{+ 0.04 }_{- 0.03 }$  &  0.80  &  0.58  &  \ref{fig:gas_photometry} \\
$M_{gas, R25}$--$\nu L_{\nu, u, B}$  	 &  87  &  3  &  $ 1.14 ^{+ 0.09 }_{- 0.09 }$  &  $ -1.59 ^{+ 0.04 }_{- 0.04 }$  &  $ 0.33 ^{+ 0.04 }_{- 0.03 }$  &  0.78  &  0.56  &  \ref{fig:gas_photometry} \\
$M_{gas, R25}$--$\nu L_{\nu, g, A}$  	 &  47  &  1  &  $ 1.05 ^{+ 0.11 }_{- 0.11 }$  &  $ -0.95 ^{+ 0.05 }_{- 0.05 }$  &  $ 0.31 ^{+ 0.05 }_{- 0.04 }$  &  0.76  &  0.55  &  \ref{fig:gas_photometry} \\
$M_{gas, R25}$--$\nu L_{\nu, g, B}$  	 &  88  &  3  &  $ 1.07 ^{+ 0.10 }_{- 0.10 }$  &  $ -1.24 ^{+ 0.04 }_{- 0.04 }$  &  $ 0.36 ^{+ 0.04 }_{- 0.03 }$  &  0.76  &  0.55  &  \ref{fig:gas_photometry} \\
$M_{gas, R25}$--$\nu L_{\nu, r, A}$  	 &  47  &  1  &  $ 0.99 ^{+ 0.11 }_{- 0.11 }$  &  $ -0.47 ^{+ 0.05 }_{- 0.05 }$  &  $ 0.32 ^{+ 0.05 }_{- 0.04 }$  &  0.74  &  0.53  &  \ref{fig:gas_photometry} \\
$M_{gas, R25}$--$\nu L_{\nu, r, B}$  	 &  88  &  3  &  $ 1.00 ^{+ 0.10 }_{- 0.10 }$  &  $ -0.63 ^{+ 0.04 }_{- 0.04 }$  &  $ 0.37 ^{+ 0.04 }_{- 0.03 }$  &  0.74  &  0.53  &  \ref{fig:gas_photometry} \\
$M_{gas, R25}$--$\nu L_{\nu, I, A}$  	 &  47  &  1  &  $ 0.94 ^{+ 0.11 }_{- 0.11 }$  &  $ -0.05 ^{+ 0.05 }_{- 0.05 }$  &  $ 0.33 ^{+ 0.05 }_{- 0.04 }$  &  0.72  &  0.52  &  \ref{fig:gas_photometry} \\
$M_{gas, R25}$--$\nu L_{\nu, I, B}$  	 &  88  &  3  &  $ 0.95 ^{+ 0.09 }_{- 0.10 }$  &  $ -0.19 ^{+ 0.04 }_{- 0.04 }$  &  $ 0.38 ^{+ 0.04 }_{- 0.03 }$  &  0.73  &  0.51  &  \ref{fig:gas_photometry} \\
$M_{gas, R25}$--$\nu L_{\nu, z, A}$  	 &  47  &  1  &  $ 0.91 ^{+ 0.11 }_{- 0.11 }$  &  $ 0.27 ^{+ 0.05 }_{- 0.05 }$  &  $ 0.33 ^{+ 0.05 }_{- 0.04 }$  &  0.71  &  0.51  &  \ref{fig:gas_photometry} \\
$M_{gas, R25}$--$\nu L_{\nu, z, B}$  	 &  88  &  3  &  $ 0.92 ^{+ 0.09 }_{- 0.09 }$  &  $ 0.09 ^{+ 0.04 }_{- 0.04 }$  &  $ 0.38 ^{+ 0.04 }_{- 0.03 }$  &  0.72  &  0.51  &  \ref{fig:gas_photometry} \\
$M_{gas, R25}$--$\nu L_{\nu, J, A}$  	 &  47  &  1  &  $ 0.90 ^{+ 0.11 }_{- 0.11 }$  &  $ 0.52 ^{+ 0.05 }_{- 0.05 }$  &  $ 0.34 ^{+ 0.05 }_{- 0.04 }$  &  0.70  &  0.49  &  \ref{fig:gas_photometry} \\
$M_{gas, R25}$--$\nu L_{\nu, J, B}$  	 &  121  &  13  &  $ 0.82 ^{+ 0.07 }_{- 0.07 }$  &  $ 1.05 ^{+ 0.04 }_{- 0.04 }$  &  $ 0.42 ^{+ 0.04 }_{- 0.03 }$  &  0.68  &  0.50  &  \ref{fig:gas_photometry} \\
$M_{gas, R25}$--$\nu L_{\nu, H, A}$  	 &  47  &  1  &  $ 0.92 ^{+ 0.11 }_{- 0.11 }$  &  $ 0.24 ^{+ 0.05 }_{- 0.06 }$  &  $ 0.33 ^{+ 0.05 }_{- 0.04 }$  &  0.70  &  0.51  &  \ref{fig:gas_photometry} \\
$M_{gas, R25}$--$\nu L_{\nu, H, B}$  	 &  114  &  10  &  $ 0.85 ^{+ 0.08 }_{- 0.08 }$  &  $ 0.78 ^{+ 0.04 }_{- 0.04 }$  &  $ 0.39 ^{+ 0.04 }_{- 0.03 }$  &  0.64  &  0.45  &  \ref{fig:gas_photometry} \\
$M_{gas, R25}$--$\nu L_{\nu, Ks, A}$ 	 &  47  &  1  &  $ 0.93 ^{+ 0.11 }_{- 0.11 }$  &  $ 0.36 ^{+ 0.05 }_{- 0.05 }$  &  $ 0.32 ^{+ 0.05 }_{- 0.04 }$  &  0.71  &  0.51  &  \ref{fig:gas_photometry} \\
$M_{gas, R25}$--$\nu L_{\nu, Ks, B}$ 	 &  118  &  13  &  $ 0.85 ^{+ 0.08 }_{- 0.07 }$  &  $ 0.94 ^{+ 0.04 }_{- 0.04 }$  &  $ 0.40 ^{+ 0.04 }_{- 0.03 }$  &  0.69  &  0.48  &  \ref{fig:gas_photometry} \\
$M_{gas, R25}$--$\nu L_{\nu, W1, A}$ 	 &  47  &  1  &  $ 0.90 ^{+ 0.10 }_{- 0.10 }$  &  $ 1.15 ^{+ 0.05 }_{- 0.05 }$  &  $ 0.31 ^{+ 0.05 }_{- 0.03 }$  &  0.74  &  0.53  &  \ref{fig:gas_photometry} \\
$M_{gas, R25}$--$\nu L_{\nu, W1, B}$ 	 &  121  &  13  &  $ 0.81 ^{+ 0.06 }_{- 0.06 }$  &  $ 1.79 ^{+ 0.04 }_{- 0.04 }$  &  $ 0.40 ^{+ 0.03 }_{- 0.03 }$  &  0.72  &  0.53  &  \ref{fig:gas_photometry} \\
$M_{gas, R25}$--$\nu L_{\nu, W2, A}$ 	 &  47  &  1  &  $ 0.92 ^{+ 0.10 }_{- 0.10 }$  &  $ 1.21 ^{+ 0.05 }_{- 0.05 }$  &  $ 0.30 ^{+ 0.05 }_{- 0.03 }$  &  0.74  &  0.55  &  \ref{fig:gas_photometry} \\
$M_{gas, R25}$--$\nu L_{\nu, W2, B}$ 	 &  120  &  12  &  $ 0.79 ^{+ 0.07 }_{- 0.06 }$  &  $ 2.22 ^{+ 0.04 }_{- 0.04 }$  &  $ 0.41 ^{+ 0.03 }_{- 0.03 }$  &  0.71  &  0.52  &  \ref{fig:gas_photometry} \\
$M_{gas, R25}$--$\nu L_{\nu, W3, A}$ 	 &  47  &  1  &  $ 0.91 ^{+ 0.08 }_{- 0.08 }$  &  $ 1.23 ^{+ 0.04 }_{- 0.04 }$  &  $ 0.25 ^{+ 0.04 }_{- 0.03 }$  &  0.80  &  0.60  &  \ref{fig:gas_photometry} \\
$M_{gas, R25}$--$\nu L_{\nu, W3, B}$ 	 &  120  &  13  &  $ 0.75 ^{+ 0.06 }_{- 0.06 }$  &  $ 2.40 ^{+ 0.04 }_{- 0.04 }$  &  $ 0.37 ^{+ 0.03 }_{- 0.03 }$  &  0.74  &  0.57  &  \ref{fig:gas_photometry} \\
$M_{gas, R25}$--$\nu L_{\nu, W4, A}$ 	 &  47  &  1  &  $ 0.79 ^{+ 0.10 }_{- 0.10 }$  &  $ 2.32 ^{+ 0.05 }_{- 0.05 }$  &  $ 0.34 ^{+ 0.05 }_{- 0.04 }$  &  0.72  &  0.49  &  \ref{fig:gas_photometry} \\
$M_{gas, R25}$--$\nu L_{\nu, W4, B}$ 	 &  118  &  11  &  $ 0.68 ^{+ 0.06 }_{- 0.06 }$  &  $ 3.06 ^{+ 0.04 }_{- 0.04 }$  &  $ 0.40 ^{+ 0.03 }_{- 0.03 }$  &  0.68  &  0.47  &  \ref{fig:gas_photometry} \\
$M_{gas, R25}$--$\nu L_{\nu, I3, A}$ 	 &  47  &  1  &  $ 0.92 ^{+ 0.08 }_{- 0.08 }$  &  $ 1.13 ^{+ 0.04 }_{- 0.04 }$  &  $ 0.25 ^{+ 0.04 }_{- 0.03 }$  &  0.79  &  0.61  &  \ref{fig:gas_photometry} \\
$M_{gas, R25}$--$\nu L_{\nu, I3, B}$ 	 &  87  &  4  &  $ 0.75 ^{+ 0.07 }_{- 0.07 }$  &  $ 2.48 ^{+ 0.04 }_{- 0.04 }$  &  $ 0.38 ^{+ 0.04 }_{- 0.03 }$  &  0.76  &  0.58  &  \ref{fig:gas_photometry} \\
$M_{gas, R25}$--$\nu L_{\nu, I4, A}$ 	 &  47  &  1  &  $ 0.80 ^{+ 0.08 }_{- 0.08 }$  &  $ 1.97 ^{+ 0.05 }_{- 0.05 }$  &  $ 0.29 ^{+ 0.04 }_{- 0.03 }$  &  0.77  &  0.61  &  \ref{fig:gas_photometry} \\
$M_{gas, R25}$--$\nu L_{\nu, I4, B}$ 	 &  89  &  5  &  $ 0.74 ^{+ 0.07 }_{- 0.07 }$  &  $ 2.45 ^{+ 0.04 }_{- 0.04 }$  &  $ 0.39 ^{+ 0.04 }_{- 0.03 }$  &  0.75  &  0.57  &  \ref{fig:gas_photometry} \\
$M_{gas, R25}$--$\nu L_{\nu, 24, A}$ 	 &  47  &  1  &  $ 0.70 ^{+ 0.10 }_{- 0.10 }$  &  $ 3.08 ^{+ 0.06 }_{- 0.06 }$  &  $ 0.38 ^{+ 0.05 }_{- 0.04 }$  &  0.68  &  0.48  &  \ref{fig:gas_photometry} \\
$M_{gas, R25}$--$\nu L_{\nu, 24, B}$ 	 &  96  &  5  &  $ 0.57 ^{+ 0.07 }_{- 0.07 }$  &  $ 4.13 ^{+ 0.05 }_{- 0.05 }$  &  $ 0.46 ^{+ 0.04 }_{- 0.03 }$  &  0.65  &  0.46  &  \ref{fig:gas_photometry} \\
$M_{gas, R25}$--$\nu L_{\nu, 70, A}$ 	 &  47  &  1  &  $ 0.84 ^{+ 0.10 }_{- 0.10 }$  &  $ 1.28 ^{+ 0.05 }_{- 0.05 }$  &  $ 0.32 ^{+ 0.05 }_{- 0.04 }$  &  0.75  &  0.54  &  \ref{fig:gas_photometry} \\
$M_{gas, R25}$--$\nu L_{\nu, 70, B}$ 	 &  85  &  6  &  $ 0.74 ^{+ 0.08 }_{- 0.08 }$  &  $ 2.12 ^{+ 0.05 }_{- 0.05 }$  &  $ 0.43 ^{+ 0.04 }_{- 0.04 }$  &  0.75  &  0.50  &  \ref{fig:gas_photometry} \\
$M_{gas, R25}$--$\nu L_{\nu, 100, B}$	 &  47  &  1  &  $ 0.95 ^{+ 0.09 }_{- 0.09 }$  &  $ 0.25 ^{+ 0.04 }_{- 0.04 }$  &  $ 0.27 ^{+ 0.04 }_{- 0.03 }$  &  0.80  &  0.61  &  \ref{fig:gas_photometry} \\
$M_{gas, R25}$--$\nu L_{\nu, 100, B}$	 &  112  &  11  &  $ 0.82 ^{+ 0.06 }_{- 0.06 }$  &  $ 1.29 ^{+ 0.04 }_{- 0.04 }$  &  $ 0.35 ^{+ 0.03 }_{- 0.03 }$  &  0.73  &  0.57  &  \ref{fig:gas_photometry} \\
$M_{gas, R25}$--$\nu L_{\nu, 600, B}$	 &  47  &  1  &  $ 0.99 ^{+ 0.08 }_{- 0.08 }$  &  $ -0.13 ^{+ 0.04 }_{- 0.04 }$  &  $ 0.22 ^{+ 0.04 }_{- 0.03 }$  &  0.82  &  0.64  &  \ref{fig:gas_photometry} \\
$M_{gas, R25}$--$\nu L_{\nu, 160, B}$	 &  111  &  10  &  $ 0.84 ^{+ 0.06 }_{- 0.06 }$  &  $ 1.20 ^{+ 0.04 }_{- 0.04 }$  &  $ 0.35 ^{+ 0.03 }_{- 0.03 }$  &  0.77  &  0.59  &  \ref{fig:gas_photometry} \\
$M_{gas, R25}$--$\nu L_{\nu, 250, A}$	 &  47  &  1  &  $ 1.04 ^{+ 0.07 }_{- 0.07 }$  &  $ -0.01 ^{+ 0.03 }_{- 0.03 }$  &  $ 0.18 ^{+ 0.04 }_{- 0.03 }$  &  0.86  &  0.68  &  \ref{fig:gas_photometry} \\
$M_{gas, R25}$--$\nu L_{\nu, 250, B}$	 &  116  &  13  &  $ 0.85 ^{+ 0.05 }_{- 0.05 }$  &  $ 1.46 ^{+ 0.03 }_{- 0.03 }$  &  $ 0.30 ^{+ 0.03 }_{- 0.02 }$  &  0.82  &  0.66  &  \ref{fig:gas_photometry} \\
$M_{gas, R25}$--$\nu L_{\nu, 350, A}$	 &  47  &  1  &  $ 1.09 ^{+ 0.06 }_{- 0.06 }$  &  $ 0.11 ^{+ 0.03 }_{- 0.03 }$  &  $ 0.14 ^{+ 0.04 }_{- 0.03 }$  &  0.88  &  0.72  &  \ref{fig:gas_photometry} \\
$M_{gas, R25}$--$\nu L_{\nu, 350, B}$	 &  116  &  13  &  $ 0.90 ^{+ 0.05 }_{- 0.05 }$  &  $ 1.45 ^{+ 0.03 }_{- 0.03 }$  &  $ 0.28 ^{+ 0.03 }_{- 0.02 }$  &  0.85  &  0.69  &  \ref{fig:gas_photometry} \\
$M_{gas, R25}$--$\nu L_{\nu, 500, A}$	 &  47  &  1  &  $ 1.12 ^{+ 0.06 }_{- 0.06 }$  &  $ 0.43 ^{+ 0.03 }_{- 0.03 }$  &  $ 0.11 ^{+ 0.04 }_{- 0.03 }$  &  0.90  &  0.74  &  \ref{fig:gas_photometry} \\
$M_{gas, R25}$--$\nu L_{\nu, 500, B}$	 &  115  &  12  &  $ 0.94 ^{+ 0.05 }_{- 0.05 }$  &  $ 1.67 ^{+ 0.03 }_{- 0.03 }$  &  $ 0.25 ^{+ 0.03 }_{- 0.02 }$  &  0.87  &  0.72  &  \ref{fig:gas_photometry} \\
\hline
\end{tabular}
\\
\footnotesize {{\bf Notes.} Best-fit parameters and correlation coefficients. Columns: (1) Scaling relation; (2) Total number of objects with available $M_{gas, R25}$ and monochromatic luminosity; (3) Number of $M_{gas, R25}$ detections; (4), (5) and (6) Best-fit parameters, namely slope $\alpha$, normalization $\gamma$, and intrinsic dispersion $\delta_{intr}$; (7) and (8) are $P$ and $K$; (9) Reference to figure.}
\end{table*}
\FloatBarrier 
%
\begin{figure*}
	\includegraphics[width =\textwidth, keepaspectratio=True]{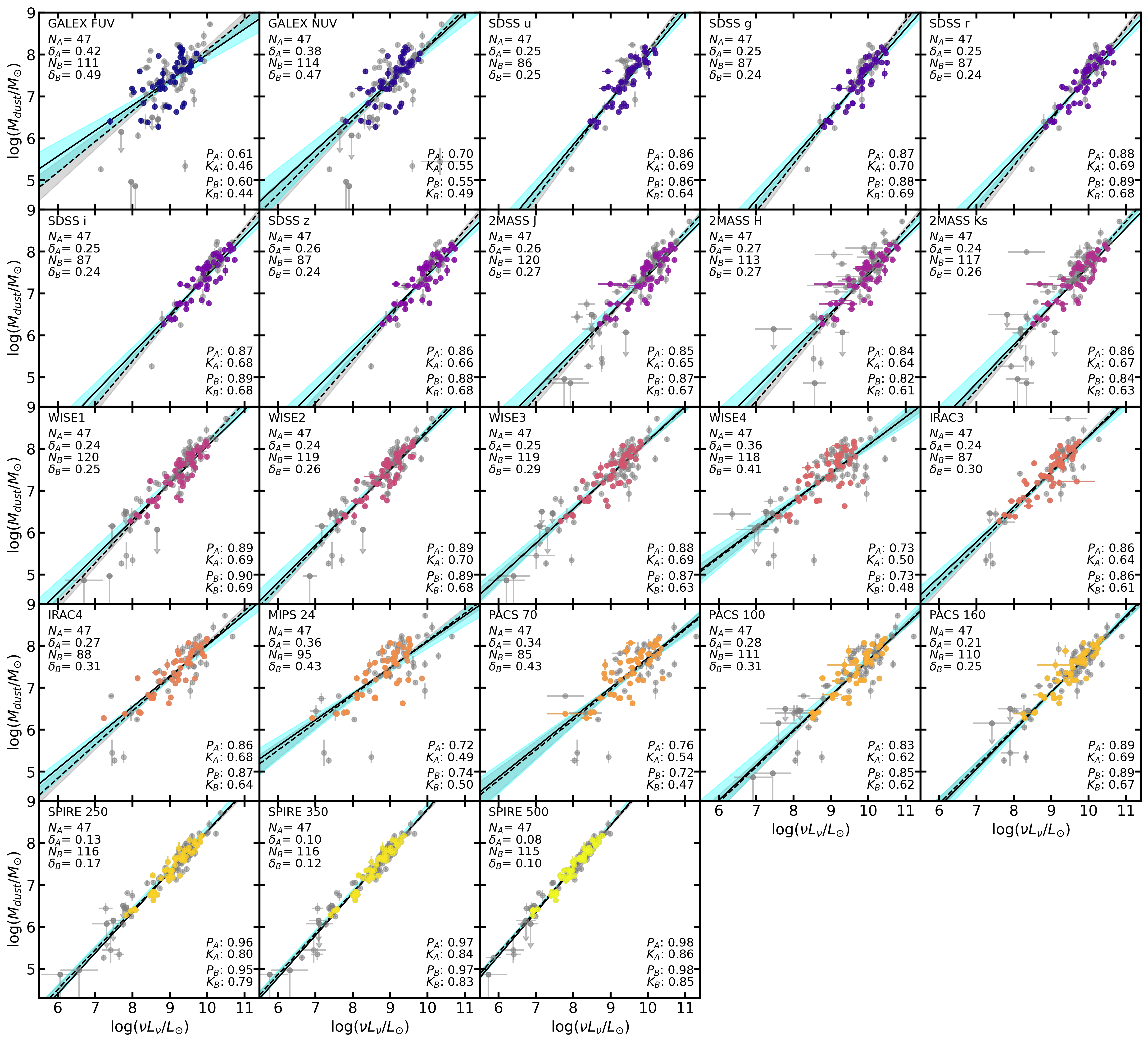}
	\caption{$M_{dust}$ versus monochromatic luminosities.
See the caption of Fig.~\ref{fig:photometry}.}
    \label{fig:dust_photometry}
\end{figure*}
\FloatBarrier 
\begin{table*}[ht]
\caption{$M_{dust}$ vs. monochromatic luminosities}
\label{table:corr_dust_phot}      
\centering          
\begin{tabular}[\textwidth]{l l l l l l l l l}     
\hline     
    Relation                &   Sample  & UL   & $\alpha$ & $\beta$ & $\delta_{intr}$ & P &   K  &   Fig. \\ 
    (1)                     &   (2)     & (3) &  (4)     &   (5)   &     (6)         &   (7) &   (8)  &  (9) \\ 
\hline
$M_{dust}$--$\nu L_{\nu, FUV, A}$	&  47  &  0  &  $ 0.61^{+0.12}_{-0.12 }$  &  $ 1.92^{+0.06}_{-0.06 }$  &  $ 0.42^{+0.05}_{-0.04 }$  &  0.43  &  0.31  &  \ref{fig:dust_photometry} \\
$M_{dust}$--$\nu L_{\nu, FUV, B}$	 &  112  &  5  &  $ 0.73^{+0.09}_{-0.08 }$  &  $ 0.83^{+0.05}_{-0.05 }$  &  $ 0.49^{+0.04}_{-0.03 }$  &  0.48  &  0.36  &  \ref{fig:dust_photometry} \\
$M_{dust}$--$\nu L_{\nu, NUV, A}$	 &  47  &  0  &  $ 0.82^{+0.13}_{-0.13 }$  &  $ -0.00^{+0.06}_{-0.06 }$  &  $ 0.38^{+0.05}_{-0.04 }$  &  0.55  &  0.42  &  \ref{fig:dust_photometry} \\
$M_{dust}$--$\nu L_{\nu, NUV, B}$	 &  115  &  6  &  $ 0.88^{+0.09}_{-0.09 }$  &  $ -0.62^{+0.04}_{-0.05 }$  &  $ 0.47^{+0.04}_{-0.03 }$  &  0.50  &  0.42  &  \ref{fig:dust_photometry} \\
$M_{dust}$--$\nu L_{\nu, u, A}$  	 &  47  &  0  &  $ 1.07^{+0.09}_{-0.09 }$  &  $ -2.68^{+0.04}_{-0.04 }$  &  $ 0.25^{+0.04}_{-0.03 }$  &  0.72  &  0.56  &  \ref{fig:dust_photometry} \\
$M_{dust}$--$\nu L_{\nu, u, B}$  	 &  87  &  0  &  $ 1.16^{+0.07}_{-0.07 }$  &  $ -3.54^{+0.03}_{-0.03 }$  &  $ 0.25^{+0.03}_{-0.02 }$  &  0.76  &  0.55  &  \ref{fig:dust_photometry} \\
$M_{dust}$--$\nu L_{\nu, g, A}$  	 &  47  &  0  &  $ 1.02^{+0.08}_{-0.09 }$  &  $ -2.61^{+0.04}_{-0.04 }$  &  $ 0.25^{+0.03}_{-0.03 }$  &  0.77  &  0.59  &  \ref{fig:dust_photometry} \\
$M_{dust}$--$\nu L_{\nu, g, B}$  	 &  88  &  0  &  $ 1.14^{+0.07}_{-0.06 }$  &  $ -3.70^{+0.03}_{-0.03 }$  &  $ 0.24^{+0.02}_{-0.02 }$  &  0.80  &  0.55  &  \ref{fig:dust_photometry} \\
$M_{dust}$--$\nu L_{\nu, r, A}$  	 &  47  &  0  &  $ 0.97^{+0.08}_{-0.08 }$  &  $ -2.25^{+0.04}_{-0.04 }$  &  $ 0.25^{+0.03}_{-0.03 }$  &  0.80  &  0.60  &  \ref{fig:dust_photometry} \\
$M_{dust}$--$\nu L_{\nu, r, B}$  	 &  88  &  0  &  $ 1.09^{+0.06}_{-0.06 }$  &  $ -3.39^{+0.03}_{-0.03 }$  &  $ 0.24^{+0.02}_{-0.02 }$  &  0.82  &  0.56  &  \ref{fig:dust_photometry} \\
$M_{dust}$--$\nu L_{\nu, I, A}$  	 &  47  &  0  &  $ 0.93^{+0.08}_{-0.08 }$  &  $ -1.88^{+0.04}_{-0.04 }$  &  $ 0.25^{+0.03}_{-0.02 }$  &  0.81  &  0.60  &  \ref{fig:dust_photometry} \\
$M_{dust}$--$\nu L_{\nu, I, B}$  	 &  88  &  0  &  $ 1.05^{+0.06}_{-0.06 }$  &  $ -3.02^{+0.03}_{-0.03 }$  &  $ 0.24^{+0.02}_{-0.02 }$  &  0.82  &  0.56  &  \ref{fig:dust_photometry} \\
$M_{dust}$--$\nu L_{\nu, z, A}$  	 &  47  &  0  &  $ 0.91^{+0.08}_{-0.08 }$  &  $ -1.63^{+0.04}_{-0.04 }$  &  $ 0.26^{+0.03}_{-0.03 }$  &  0.81  &  0.58  &  \ref{fig:dust_photometry} \\
$M_{dust}$--$\nu L_{\nu, z, B}$  	 &  88  &  0  &  $ 1.03^{+0.06}_{-0.06 }$  &  $ -2.84^{+0.03}_{-0.03 }$  &  $ 0.24^{+0.02}_{-0.02 }$  &  0.81  &  0.56  &  \ref{fig:dust_photometry} \\
$M_{dust}$--$\nu L_{\nu, J, A}$  	 &  47  &  0  &  $ 0.90^{+0.08}_{-0.08 }$  &  $ -1.53^{+0.04}_{-0.04 }$  &  $ 0.26^{+0.03}_{-0.03 }$  &  0.80  &  0.57  &  \ref{fig:dust_photometry} \\
$M_{dust}$--$\nu L_{\nu, J, B}$  	 &  121  &  6  &  $ 0.99^{+0.05}_{-0.05 }$  &  $ -2.40^{+0.03}_{-0.03 }$  &  $ 0.27^{+0.02}_{-0.02 }$  &  0.87  &  0.61  &  \ref{fig:dust_photometry} \\
$M_{dust}$--$\nu L_{\nu, H, A}$  	 &  47  &  0  &  $ 0.90^{+0.09}_{-0.09 }$  &  $ -1.48^{+0.04}_{-0.04 }$  &  $ 0.27^{+0.04}_{-0.03 }$  &  0.79  &  0.56  &  \ref{fig:dust_photometry} \\
$M_{dust}$--$\nu L_{\nu, H, B}$  	 &  114  &  4  &  $ 1.02^{+0.06}_{-0.06 }$  &  $ -2.73^{+0.03}_{-0.03 }$  &  $ 0.27^{+0.03}_{-0.02 }$  &  0.81  &  0.53  &  \ref{fig:dust_photometry} \\
$M_{dust}$--$\nu L_{\nu, Ks, A}$ 	 &  47  &  0  &  $ 0.92^{+0.08}_{-0.08 }$  &  $ -1.51^{+0.04}_{-0.04 }$  &  $ 0.24^{+0.03}_{-0.03 }$  &  0.83  &  0.59  &  \ref{fig:dust_photometry} \\
$M_{dust}$--$\nu L_{\nu, Ks, B}$ 	 &  118  &  6  &  $ 1.00^{+0.05}_{-0.05 }$  &  $ -2.31^{+0.03}_{-0.03 }$  &  $ 0.26^{+0.02}_{-0.02 }$  &  0.83  &  0.59  &  \ref{fig:dust_photometry} \\
$M_{dust}$--$\nu L_{\nu, W1, A}$ 	 &  47  &  0  &  $ 0.87^{+0.07}_{-0.07 }$  &  $ -0.61^{+0.04}_{-0.04 }$  &  $ 0.24^{+0.03}_{-0.02 }$  &  0.86  &  0.65  &  \ref{fig:dust_photometry} \\
$M_{dust}$--$\nu L_{\nu, W1, B}$ 	 &  121  &  6  &  $ 0.96^{+0.04}_{-0.04 }$  &  $ -1.43^{+0.02}_{-0.02 }$  &  $ 0.25^{+0.02}_{-0.02 }$  &  0.90  &  0.67  &  \ref{fig:dust_photometry} \\
$M_{dust}$--$\nu L_{\nu, W2, A}$ 	 &  47  &  0  &  $ 0.89^{+0.07}_{-0.07 }$  &  $ -0.50^{+0.04}_{-0.04 }$  &  $ 0.24^{+0.03}_{-0.02 }$  &  0.87  &  0.67  &  \ref{fig:dust_photometry} \\
$M_{dust}$--$\nu L_{\nu, W2, B}$ 	 &  120  &  5  &  $ 0.94^{+0.04}_{-0.04 }$  &  $ -0.91^{+0.03}_{-0.03 }$  &  $ 0.26^{+0.02}_{-0.02 }$  &  0.91  &  0.69  &  \ref{fig:dust_photometry} \\
$M_{dust}$--$\nu L_{\nu, W3, A}$ 	 &  47  &  0  &  $ 0.82^{+0.07}_{-0.07 }$  &  $ 0.08^{+0.04}_{-0.04 }$  &  $ 0.25^{+0.03}_{-0.02 }$  &  0.92  &  0.77  &  \ref{fig:dust_photometry} \\
$M_{dust}$--$\nu L_{\nu, W3, B}$ 	 &  120  &  6  &  $ 0.81^{+0.04}_{-0.04 }$  &  $ 0.07^{+0.03}_{-0.03 }$  &  $ 0.29^{+0.02}_{-0.02 }$  &  0.91  &  0.79  &  \ref{fig:dust_photometry} \\
$M_{dust}$--$\nu L_{\nu, W4, A}$ 	 &  47  &  0  &  $ 0.66^{+0.10}_{-0.09 }$  &  $ 1.46^{+0.05}_{-0.05 }$  &  $ 0.36^{+0.05}_{-0.03 }$  &  0.81  &  0.64  &  \ref{fig:dust_photometry} \\
$M_{dust}$--$\nu L_{\nu, W4, B}$ 	 &  118  &  4  &  $ 0.67^{+0.05}_{-0.05 }$  &  $ 1.37^{+0.04}_{-0.04 }$  &  $ 0.41^{+0.03}_{-0.03 }$  &  0.87  &  0.68  &  \ref{fig:dust_photometry} \\
$M_{dust}$--$\nu L_{\nu, I3, A}$ 	 &  47  &  0  &  $ 0.84^{+0.07}_{-0.07 }$  &  $ -0.12^{+0.04}_{-0.04 }$  &  $ 0.24^{+0.03}_{-0.02 }$  &  0.87  &  0.66  &  \ref{fig:dust_photometry} \\
$M_{dust}$--$\nu L_{\nu, I3, B}$ 	 &  87  &  1  &  $ 0.88^{+0.06}_{-0.05 }$  &  $ -0.51^{+0.03}_{-0.03 }$  &  $ 0.30^{+0.03}_{-0.02 }$  &  0.92  &  0.71  &  \ref{fig:dust_photometry} \\
$M_{dust}$--$\nu L_{\nu, I4, A}$ 	 &  47  &  0  &  $ 0.74^{+0.07}_{-0.07 }$  &  $ 0.62^{+0.04}_{-0.04 }$  &  $ 0.27^{+0.03}_{-0.03 }$  &  0.90  &  0.75  &  \ref{fig:dust_photometry} \\
$M_{dust}$--$\nu L_{\nu, I4, B}$ 	 &  89  &  0  &  $ 0.80^{+0.05}_{-0.05 }$  &  $ 0.05^{+0.03}_{-0.03 }$  &  $ 0.31^{+0.03}_{-0.02 }$  &  0.92  &  0.79  &  \ref{fig:dust_photometry} \\
$M_{dust}$--$\nu L_{\nu, 24, A}$ 	 &  47  &  0  &  $ 0.62^{+0.09}_{-0.09 }$  &  $ 1.84^{+0.05}_{-0.05 }$  &  $ 0.36^{+0.05}_{-0.04 }$  &  0.79  &  0.62  &  \ref{fig:dust_photometry} \\
$M_{dust}$--$\nu L_{\nu, 24, B}$ 	 &  96  &  0  &  $ 0.65^{+0.06}_{-0.06 }$  &  $ 1.60^{+0.04}_{-0.05 }$  &  $ 0.43^{+0.04}_{-0.03 }$  &  0.84  &  0.67  &  \ref{fig:dust_photometry} \\
$M_{dust}$--$\nu L_{\nu, 70, A}$ 	 &  47  &  0  &  $ 0.71^{+0.09}_{-0.10 }$  &  $ 0.55^{+0.05}_{-0.05 }$  &  $ 0.34^{+0.05}_{-0.03 }$  &  0.84  &  0.66  &  \ref{fig:dust_photometry} \\
$M_{dust}$--$\nu L_{\nu, 70, B}$ 	 &  85  &  0  &  $ 0.72^{+0.08}_{-0.08 }$  &  $ 0.48^{+0.05}_{-0.05 }$  &  $ 0.43^{+0.04}_{-0.03 }$  &  0.87  &  0.69  &  \ref{fig:dust_photometry} \\
$M_{dust}$--$\nu L_{\nu, 100, B}$	 &  47  &  0  &  $ 0.83^{+0.08}_{-0.08 }$  &  $ -0.59^{+0.04}_{-0.04 }$  &  $ 0.28^{+0.04}_{-0.03 }$  &  0.88  &  0.70  &  \ref{fig:dust_photometry} \\
$M_{dust}$--$\nu L_{\nu, 100, B}$	 &  112  &  5  &  $ 0.85^{+0.05}_{-0.05 }$  &  $ -0.81^{+0.03}_{-0.03 }$  &  $ 0.31^{+0.03}_{-0.02 }$  &  0.89  &  0.76  &  \ref{fig:dust_photometry} \\
$M_{dust}$--$\nu L_{\nu, 600, B}$	 &  47  &  0  &  $ 0.91^{+0.07}_{-0.07 }$  &  $ -1.29^{+0.03}_{-0.03 }$  &  $ 0.21^{+0.03}_{-0.02 }$  &  0.93  &  0.79  &  \ref{fig:dust_photometry} \\
$M_{dust}$--$\nu L_{\nu, 160, B}$	 &  111  &  3  &  $ 0.89^{+0.04}_{-0.04 }$  &  $ -1.09^{+0.03}_{-0.03 }$  &  $ 0.25^{+0.02}_{-0.02 }$  &  0.92  &  0.80  &  \ref{fig:dust_photometry} \\
$M_{dust}$--$\nu L_{\nu, 250, A}$	 &  47  &  0  &  $ 0.97^{+0.04}_{-0.04 }$  &  $ -1.32^{+0.02}_{-0.02 }$  &  $ 0.13^{+0.02}_{-0.02 }$  &  0.94  &  0.81  &  \ref{fig:dust_photometry} \\
$M_{dust}$--$\nu L_{\nu, 250, B}$	 &  116  &  6  &  $ 0.94^{+0.03}_{-0.03 }$  &  $ -1.13^{+0.02}_{-0.02 }$  &  $ 0.17^{+0.02}_{-0.01 }$  &  0.96  &  0.83  &  \ref{fig:dust_photometry} \\
$M_{dust}$--$\nu L_{\nu, 350, A}$	 &  47  &  0  &  $ 1.01^{+0.04}_{-0.04 }$  &  $ -1.16^{+0.02}_{-0.02 }$  &  $ 0.10^{+0.02}_{-0.01 }$  &  0.93  &  0.79  &  \ref{fig:dust_photometry} \\
$M_{dust}$--$\nu L_{\nu, 350, B}$	 &  116  &  6  &  $ 0.98^{+0.02}_{-0.02 }$  &  $ -1.03^{+0.01}_{-0.01 }$  &  $ 0.12^{+0.01}_{-0.01 }$  &  0.94  &  0.80  &  \ref{fig:dust_photometry} \\
$M_{dust}$--$\nu L_{\nu, 500, A}$	 &  47  &  0  &  $ 1.04^{+0.03}_{-0.03 }$  &  $ -0.88^{+0.01}_{-0.01 }$  &  $ 0.08^{+0.02}_{-0.01 }$  &  0.91  &  0.76  &  \ref{fig:dust_photometry} \\
$M_{dust}$--$\nu L_{\nu, 500, B}$	 &  115  &  5  &  $ 1.01^{+0.02}_{-0.02 }$  &  $ -0.69^{+0.01}_{-0.01 }$  &  $ 0.10^{+0.01}_{-0.01 }$  &  0.93  &  0.78  &  \ref{fig:dust_photometry} \\
$M_{dust}$--$M_{H2, A}$	 &  47  &  0  &  $ 0.72 ^{+0.06}_{-0.06}$  &  $ 0.98 ^{+0.03}_{-0.03}$  &  $ 0.21 ^{+0.02}_{-0.01}$  &  0.88  &  0.70  &   \\
$M_{dust}$--$M_{HI, R25, A}$	 &  47  &  0  &  $ 0.68^{+0.07}_{-0.07}$  &  $ 1.53 ^{+0.05}_{-0.04}$  &  $ 0.31 ^{+0.02}_{-0.02}$  &  0.88  &  0.70  &   \\
\hline
\end{tabular}
\\
\footnotesize {{\bf Notes.} Best-fit parameters and correlation coefficients. Columns: (1) Scaling relation; (2) Total number of objects with available $M_{dust}$ and monochromatic luminosity; (3) Number of $M_{dust}$ detections; (4), (5) and (6) Best-fit parameters, namely slope $\alpha$, normalization $\gamma$, and intrinsic dispersion $\delta_{intr}$; (7) and (8) are $P$ and $K$; (9) Reference to figure.}
\end{table*}
\FloatBarrier 

\end{appendix}

\end{document}